\documentclass[numberedappendix,twocolappendix]{emulateapj}

\newcommand{\noprint}[1]{}

\shorttitle{Faraday rotation in parsec-scale AGN jets}
\shortauthors{Hovatta et al.}

\begin{document}


\title{MOJAVE: Monitoring of Jets in Active Galactic Nuclei with VLBA Experiments. VIII. Faraday rotation in parsec-scale AGN jets}


\author{Talvikki Hovatta}
\affil{Department of Physics, Purdue University, 525 Northwestern Ave. West Lafayette, IN 47907, USA}
\affil{Cahill Center for Astronomy \& Astrophysics, California Institute of Technology, 1200 E. California Blvd, Pasadena, CA 91125, USA}
\email{thovatta@caltech.edu}

\author{Matthew L. Lister}
\affil{Department of Physics, Purdue University, 525 Northwestern Ave. West Lafayette, IN 47907, USA}

\author{Margo F. Aller and Hugh D. Aller}
\affil{Department of Astronomy, University of Michigan, 817 Dennison Building, Ann Arbor, MI 48109-1042, USA}

\author{Daniel C. Homan}
\affil{Department of Physics and Astronomy, Denison University, Granville, OH 43023, USA}

\author{Yuri Y. Kovalev}
\affil{Astro Space Center of Lebedev Physical Institute,
  Profsoyuznaya 84/32, 117997 Moscow, Russia}
\affil{Max-Planck-Institut f\"ur Radioastronomie, Auf dem
  H\"ugel 69, 53121 Bonn, Germany}

\author{Alexander B. Pushkarev}
\affil{Pulkovo Observatory, Pulkovskoe Chaussee 65/1, 196140
  St. Petersburg, Russia}
\affil{Crimean Astrophysical Observatory, 98409 Nauchny, Crimea, Ukraine}
\affil{Max-Planck-Institut f\"ur Radioastronomie, Auf dem
  H\"ugel 69, 53121 Bonn, Germany}

\author{Tuomas Savolainen}
\affil{Max-Planck-Institut f\"ur Radioastronomie, Auf dem
  H\"ugel 69, 53121 Bonn, Germany}


\begin{abstract}
We report observations of Faraday rotation measures for a sample of 191 extragalactic radio jets observed within the 
Monitoring Of Jets in Active galactic nuclei with VLBA Experiments (MOJAVE) program. Multifrequency 
Very Long Baseline Array (VLBA) observations were carried out over twelve epochs in 2006 at four frequencies between 8 and 15\,GHz. 
We detect parsec-scale Faraday rotation measures in 149 sources and find the quasars to have larger rotation measures 
on average than BL~Lac objects. The median core rotation measures are significantly higher than in the jet components. 
This is especially true for quasars where we detect a significant negative correlation between the magnitude of the rotation measure 
and the de-projected distance from the core. We perform detailed simulations of the observational errors of total intensity, polarization and Faraday rotation, 
and concentrate on the errors of transverse Faraday rotation measure gradients in unresolved jets. Our simulations show that the finite image restoring
beam size has a significant effect on the observed rotation measure gradients, and spurious gradients can occur due to noise in the 
data if the jet is less than two beams wide in polarization. 
We detect significant transverse rotation measure gradients in four sources (0923+392, 1226+023, 2230+114 and 2251+158). 
In 1226+023 the rotation measure is for the first time seen to change 
sign from positive to negative over the transverse cuts, which supports the presence of a helical magnetic field in the jet. 
In this source we also detect variations in the jet rotation measure over a time scale of three months, which are difficult to explain 
with external Faraday screens and suggest internal Faraday rotation.
By comparing fractional polarization changes in jet components between the four frequency bands 
to depolarization models we find that an external purely random Faraday screen viewed through only a few lines of sight can explain most of our 
polarization observations but in some sources, such as 1226+023 and 2251+158, 
internal Faraday rotation is needed.

\end{abstract}


\section{Introduction}
Polarimetric observations of active galactic nuclei (AGN) jets
enable studies of the magnetic field structure in the outflows. 
If the jets are launched from the rotating black hole or 
accretion disk, it is natural to expect that 
the magnetic field structure in the jets is helical
\citep[e.g.,][]{blandford77,meier01,vlahakis04,mckinney07}. On the other 
hand, it is not known whether the helical structure persists 
parsecs down from the central engine or if it becomes tangled 
due to re-collimation shocks or interaction with external medium 
\citep[e.g.,][]{marscher08}. Very Long Baseline Interferometry (VLBI) 
can be used to study the electric vector orientation in the 
parsec-scale jets of AGN. In the optically thin part of the jet, 
the magnetic field orientation is perpendicular to the 
electric vector position angles (EVPAs). 
Thus, observations of EVPAs parallel to the jet direction 
have resulted in claims of toroidally dominated magnetic fields \citep[e.g.,][]{gabuzda04}.
One should note that relativistic effects make the situation more complicated and 
when viewed at small angles a toroidally dominated magnetic field can appear 
poloidal in the observer's frame \citep{lyutikov05}.
Alternatively the observed magnetic field orientation can be 
accounted for by shocks compressing the magnetic field perpendicular 
to the jet \citep{laing80, hughes89}. 

Polarized waves are affected 
by Faraday rotation when propagating through non-relativistic plasma within or external 
to the source \citep[e.g.,][]{burn66}.  This effect can both diminish the 
observed degree of polarization and rotate the intrinsic EVPAs so that in order to 
study the intrinsic magnetic field orientation in the jets, the effect must be removed. 
In the case of external rotation, the effect can be described by a linear dependence between the 
observed EVPA ($\chi_\mathrm{obs}$) and wavelength squared ($\lambda^2$)
by 
\begin{equation}\label{eq:RM}
\chi_\mathrm{obs} = \chi_0 + \frac{e^3\lambda^2}{8\pi^2\epsilon_0m^2c^3}\int n_e \mathbf{B} \cdot \mathbf{\mathrm{d}l} = \chi_0 + \mathrm{RM}\lambda^2,
\end{equation}
where $\chi_0$ is the intrinsic EVPA and RM is the rotation measure, related to the electron density $n_e$ and 
the magnetic field component $\mathbf{B}$ parallel to the line of sight. The constant in the equation consists of the charge 
of the electron $e$, vacuum permittivity $\epsilon_0$, mass of the electron $m$, and speed of light $c$. The RM 
can thus be estimated by observing the EVPA at several frequencies.

If the Faraday rotation is internal to the jet it means that either the thermal plasma causing the rotation is intermixed 
with the emitting plasma, or the relativistic particle spectrum extends to low energies.
For total rotations larger than 45$^\circ$ internal Faraday rotation is expected to
cause severe depolarization \citep{burn66} which is not often seen \citep{zavala04}, but 
for smaller total rotations the internal rotation can appear linear and follow Eq.~\ref{eq:RM}.
External Faraday rotation could be caused by a screen very close to the jet itself 
where it can also interact with the jet (e.g., a sheath) or by a more distant screen 
such as the broad or narrow line regions, or even intergalactic and Galactic plasma. 
Distinguishing between the different alternatives can be very difficult, especially 
in the case of small rotations if additional information is not available \citep[e.g.,][]{homan09}.

Over the past few decades there have been numerous studies of 
Faraday rotation in parsec-scale jets associated with active galaxies. One of the largest is 
by \cite{taylor98, taylor00} and \cite{zavala02,zavala03,zavala04} who report Faraday rotation measures (RMs)
for a sample of 40 AGN. They find the typical absolute core RMs to 
be in a range of 500 to several thousand rad~m$^{-2}$ in the observer's frame. Additionally they report variability 
in several RMs over a time span of months to years, ruling out 
the narrow line region as the origin for the Faraday rotation. Instead they suggest 
that the Faraday rotation is caused by a screen close to the jet.
Similar conclusions about the screen were drawn by \cite{asada02} who detected 
a transverse rotation measure gradient in the jet of 3C 273. They 
interpreted the gradient as a signature of a helical magnetic field in the sheath 
surrounding the jet. Several other claims of transverse gradients 
have been published \citep[e.g.,][]{gabuzda04,asada08b,gomez08,mahmud09,croke10} but 
due to some uncertainties regarding transversely unresolved jets the issue remains controversial \citep{taylor10}.

Due to the complex nature of the sources and their surroundings, the situation 
is often more complicated. In addition to a helical field, interactions 
with the surrounding intergalactic medium will cause distinct RM 
structures \citep{gomez08}. Additionally, beam effects can severely complicate 
the interpretation of the maps as shown by simulations \citep{broderick10}.
Starting from general relativistic magnetohydrodynamical simulations, \cite{broderick10} 
created a canonical jet model and calculated RM maps which 
they convolved with different image restoring beam sizes to create unresolved and resolved jets.
They showed that within one beam width from the  optically thick core, 
any gradient seen in the RM map is generally unreliable,  and only at resolution obtained with 43\,GHz VLBA observations are in 
agreement with expected values, although still suppressed in magnitude. 
In the case of an optically thin jet, it could be possible to detect a gradient 
if the jet is surrounded by a helical field  even if the jet is unresolved (above 8\,GHz VLBA resolution), 
but the magnitude of the gradient may be suppressed due to the beam effects. 

In this paper we study the statistical properties of Faraday rotation 
in AGN by using a large sample of objects which are part of the MOJAVE 
(Monitoring of Jets in AGN with VLBA Experiments) survey \citep{lister09}.
Our goal was to create a set of RM maps in which all potential sources of error in the data 
processing have been accounted for. Therefore we have performed extensive simulations 
of the errors in polarization and Faraday rotation maps and have assessed when a RM gradient can be 
called significant. These simulations show that the finite beam size of VLBI observations 
has a large effect on the observed Faraday rotation, and caution needs to be taken when interpreting the 
maps. Neighboring pixels in beam-convolved images are not independent for a typical VLBI pixel size 
and the RM maps generally consist only of a few independent measurements.

We describe our observations and the detailed data
analysis process in Section \ref{sect:data}. The results of our statistical 
study are reported in Section \ref{sect:results}. We discuss our results in light 
of depolarization models, observed RM gradients and time variability in Section \ref{sect:discussion}. 
Our conclusions are summarized in Section \ref{sect:conclusions}. In Appendix A we discuss the 
effects of relative image alignment on RM maps, and in Appendices B and C we discuss observational errors in 
polarization and RM images.
Throughout the 
paper we use a cosmology where $H_0 = 71~\mathrm{km}\mathrm{s}^{-1}\mathrm{Mpc}^{-1}$,
$\Omega_M = 0.3$, and $\Omega_\Lambda = 0.7$ \citep[e.g.,][]{komatsu09}. 

\section{Observations and Data Reduction}\label{sect:data}

Our sample consists of 191 AGN observed within the MOJAVE VLBA survey \citep{lister09}. 
It includes 134 sources of the complete flux density-limited MOJAVE-1 sample, for which 
we monitor total intensity and polarization changes of 135 AGN jets
above declination $-$20$^\circ$ which have exceeded 
15\,GHz flux density 1.5\,Jy (2\,Jy at $\delta < 0^\circ$) at any epoch between 
1994 and 2002. The rest of the sources belong to the MOJAVE-2 sample\footnote{http://www.physics.purdue.edu/astro/MOJAVE/allsources.shtml}, which 
includes sources from the 2\,cm survey \citep{kellermann04}, gamma-ray blazars, and other sources 
with unusual jet properties.

The sources were observed with the VLBA in 2006 over 12 epochs with about 
monthly separation, each epoch containing 18 sources (except for epoch 2006-Feb-12 which 
included only 14 sources and epoch 2006-Apr-28 which included 17 sources). 
The observations were made in dual polarization mode using 
frequencies centered at 8.104, 8.424 (X-band), 12.119 and 15.369\,GHz (U-band). This setup was chosen 
because VLBA observes the gain centered at 8.4\,GHz, and 8.1\,GHz was chosen as the low frequency end of the X-band.
The bandwidths were 16 and 32\,MHz for the X and U-bands, respectively. The observations were 
recorded with a bit rate of 128 Mbits s$^{-1}$. In the X-bands the observations consist of 2 IFs in both frequencies and in the U-bands 
4 IFs. All ten VLBA antennas were 
observing except at epoch 2006-Aug-09 when Pie Town was not included. The sources 
and their observing epochs are listed in Table \ref{rmtable}.  A total of twenty sources were observed 
twice during the year. 

\begin{table*}[ht!]\scriptsize
\begin{center}
\caption{\label{rmtable}Sources and their rotation measure properties}  
\begin{tabular}{lccccrrrrr}
\tableline\tableline
IAU Name & Other name & z & Opt. Cl. & $\beta\mathrm{app}$ & Epoch & Gal. RM & med. RM & med. core RM & med. jet RM\\  
 & & & & (c) & & (rad~m$^{-2}$) & (rad~m$^{-2}$) & (rad~m$^{-2}$) & (rad~m$^{-2}$)\\
(1) & (2) & (3) & (4) &  (5) & (6) & (7) & (8) & (9) & (10)\\
\tableline
0003$-$066 & NRAO 005 & 0.3467 & B & 8.4 & 2006-07-07 & 4.5 & -20.7 & -34.8 & 130.0  \\ 
0003+380 & S4 0003+38 & 0.229 & Q & \nodata & 2006-03-09 & -90.0 & \nodata & \nodata & \nodata  \\ 
0003+380 & S4 0003+38 & 0.229 & Q & \nodata & 2006-12-01 & -90.0 & 6053.3 & 6053.3 & \nodata  \\ 
0007+106 & III Zw 2 & 0.0893 & G & 1.2 & 2006-06-15 & -3.4 & 604.2 & 604.2 & \nodata  \\ 
0010+405 & 4C +40.01 & 0.256 & Q & \nodata & 2006-04-05 & -77.8 & \nodata & \nodata & \nodata  \\ 
0010+405 & 4C +40.01 & 0.256 & Q & \nodata & 2006-12-01 & -77.8 & \nodata & \nodata & \nodata  \\ 
0016+731 & S5 0016+73 & 1.781 & Q & 8.1 & 2006-08-09 & -9.1 & 264.9 & 264.9 & \nodata  \\ 
\tableline
\end{tabular}
\tablecomments{Columns are as follows: (1) IAU Name (B1950.0); (2) alternate name; (3) redshift; (4) optical classification where Q = quasar, B = BL Lac object, G = active galaxy, and U = unidentified; (5) Apparent speed used in viewing angle calculation of Fig. \ref{fig:rmdist}; (6) epoch of the RM observation; (7) Galactic RM correction taken from \cite{taylor09}; (8) median RM over the source; (9) median RM over the core area; (10) median RM over the jet. Table 1 is published in its entirety in the electronic edition of the Journal, a portion is shown here for guidance regarding its form and content.}
\end{center}
\end{table*}

\subsection{Data reduction}\label{sect:datared}
The initial data reduction and calibration were performed following the standard procedures described 
in the AIPS cookbook\footnote{http://www.aips.nrao.edu}. All the frequency bands were 
treated separately throughout the data reduction process. The imaging and self-calibration
were done in a largely automated way using the Difmap package \citep{shepherd97}. 
For more details see \cite{lister09} for the standard data reduction and imaging process and 
\cite{lister05} for the calibration of the polarization data.

All the maps were modelfit with circular or elliptical Gaussian components using 
the standard procedure in the Difmap package. The 15\,GHz maps were modelfit already 
as a part of the MOJAVE survey \citep{lister09b}. Since one of our goals was 
to use the optically thin components in the jets to align our images, 
we used these 15\,GHz modelfits as a starting point for the other bands 
and modified the fit if needed. 

As the ({\it u,v}) plane coverage differs in the bands with higher frequency maps 
resolving smaller structures, we can get spurious features in the 
rotation measure maps, especially near the core region where there can be  
many components blending within the beam at lower frequencies.
Therefore, in order to have comparable ({\it u,v}) 
coverage in all the bands, we flagged the long baselines from the 15 and 
12\,GHz maps and short baselines from the 8\,GHz maps. The resulting typical  ({\it u,v}) 
range in our data is 7.3 - 231 mega-$\lambda$.  We tested on individual sources the difference 
between flagging the baselines compared to tapering and found the differences to be minimal; there was 
no difference in the final RM map area, and the differences in RM values were a fraction of 
the error bars.
Additionally, we restored all the maps to the 
beam size of our lowest frequency (8.1\,GHz).
All these steps were carried out in Difmap  after the initial data reduction and self-calibration 
which were done using the full ({\it u,v}) data.

\subsection{Absolute EVPA calibration}\label{sect:EVPAcal}

The absolute electric vector position angle (EVPA) offset is an instrumental
quantity that must be determined and applied to every VLBA polarization observation.
To calibrate the EVPAs of our data, we used Very Large Array (VLA), University of 
Michigan Radio Astronomy Observatory (UMRAO) and 15\,GHz VLBA data, and instrumental leakage term (D-term) phases.
The 15 GHz observations were previously calibrated as part of the MOJAVE project using the D-term 
calibration method \citep{gomez02,lister05}. Therefore we only had to calibrate the 8 and 12\,GHz bands.
For five epochs we were able to use the VLA/VLBA polarization calibration 
database\footnote{http://www.aoc.nrao.edu/$\sim$smyers/calibration/} to find polarization observations within a week of our 
epoch and including one or two of our sources. For those epochs, we also calculated the distribution of differences  
between the calibrated 15\,GHz EVPAs and other bands, and UMRAO 8 and 15\,GHz EVPAs versus our 
8 and 15 GHz EVPAs. Usually these difference histograms showed a peak at an angle similar to that determined from the VLA observations.
The typical errors in the VLA EVPAs range from 1 to 3 degrees and in the UMRAO data from 1 to 10 degrees but these cancel when 
multiple sources are used.

By using these five epochs, we were able to find D-term phases on various antennas that were stable enough over the 12 month period
to enable the use of D-term phases in the calibration of the EVPAs of the remaining epochs. 
The EVPA corrections for all the epochs are shown in Table \ref{table:evpa} where in column (1) we give the observing code 
of the epoch and list the epochs that were used to anchor the D-terms. The epoch of observations is listed in 
column (2), and the reference antenna used in the calibration in column (3). The EVPA corrections at 15.4, 12.1, 8.4 and 8.1\,GHz are given in columns (4)-(7).
Since we are using five different anchoring epochs with different VLA calibration sources and additionally the 
UMRAO data, the main source of error in our calibration method should be the scatter in the measured
D-term phases. By calculating the standard deviation of the mean for the scatter in each 
right-hand or left-hand phase and taking the maximum value over the frequency band as a conservative error 
estimate, we determine the absolute EVPA calibration errors to be 3$^\circ$, 2$^\circ$ and 4$^\circ$ 
at 15, 12 and 8\,GHz bands, respectively. The total error in the EVPAs is a quadrature sum of the 
calibration error and statistical error in the EVPA, with the latter being derived from the rms values in Q and U maps.

The error in the final rotation measure images is highly dependent on the error of the EVPA. We have 
performed detailed simulations to verify that our error estimate, derived with error propagation from the 
rms in Q and U images, is correct. These simulations are described in detail in Appendix \ref{app:polerr}, where 
we also give the equations used in the error calculation.

\begin{table*}[ht!]
\begin{center}
\caption{EVPA calibration corrections for all the epochs in degrees.\label{table:evpa}}
\begin{tabular}{llrrrrrrrrrrr}
\tableline\tableline
Obscode & Epoch & Ref. Ant. & 8.1 GHz & 8.4 GHz & 12.1 GHz & 15.4 GHz\tablenotemark{b}\\
\tableline
BL137A & 2006  Feb 2 & PT & $-$15.9 & $-$16.7 & $-$16.1 & $-$18.8\\
BL137B\tablenotemark{a} & 2006 Mar 6 & PT & $-$17.7 & $-$20.0 & $-$18.0 & $-$14.5\\
BL137C & 2006  Apr 5 & KP & 19.2 & 13.1 & 22.5 & 30.9\\
BL137D & 2006 Apr 28 & FD & 12.4 & 6.7 & $-$42.9 & $-$53.7\\
BL137E & 2006 May 24 & FD & $-$12.8 & $-$17.4 & $-$16.7 & $-$47.3\\
BL137F\tablenotemark{a}  & 2006 Jun 15 & FD &$-$42.2 & $-$47.3 & $-$47.1 & $-$49.2\\
BL137G\tablenotemark{a}  & 2006  Jul 7 & FD & $-$46.3 & $-$47.9 & $-$47.9 & $-$49.1\\
BL137H & 2006 Aug 9 & FD & $-$45.0 & $-$47.8 & $-$47.0 & $-$48.4\\
BL137I\tablenotemark{a}  & 2006 Sep 6 & PT & $-$10.0 & $-$10.0 & $-$14.3 & $-$14.9\\
BL137J & 2006  Oct 6 & FD & $-$45.3 & $-$46.8 & $-$45.6 & $-$48.3\\
BL137K\tablenotemark{a}  & 2006 Nov 10 & FD & $-$44.0 & $-$45.0 & $-$45.0 & $-$46.6\\
BL137L & 2006 Dec 1 & FD & $-$44.2 &  $-$47.1 & $-$47.4 & $-$50.5\\
\tableline
\end{tabular}
\tablenotetext{1}{Epoch used to anchor the D-term calibration}
\tablenotetext{2}{Calibrated as part of the MOJAVE project}
\end{center}
\end{table*} 

\subsection{Image alignment}

During the initial data reduction process the absolute coordinate position of the source is lost and 
the center of the image is shifted to the phase center of the map. This may not be the 
same position on the sky for different frequency bands, and therefore an extra step is needed to align the images.
This can be done using bright components in the optically thin part of the jet, 
whose position should not depend on the observing frequency \citep[e.g.][]{lobanov98,marr01,kovalev08, sokolovsky11}. This approach works well for knotty 
jets but is unreliable or impossible to use for faint or smooth jets. A solution 
is to use a 2D cross-correlation algorithm to look for the best alignment based on correlation of 
the optically thin parts of the jets at different bands \citep[e.g.][]{walker00, croke08}. 

We used both methods whenever possible, and concluded that the results matched 
very well when using bright optically thin components.  Similarly to e.g., \citet{marr01} and \citet{kovalev08}, all the shifts were 
verified by examining the spectral index maps before and after the alignment. In shifted maps 
the spectral index gradient along the jet was typically smoother, and any optically thin 
regions apparently upstream of the core disappeared.
The absolute shifts between 15\,GHz and other bands varied between 0\,mas and 2.02\,mas with a 
median value of 0.11\,mas. This is comparable to the pixel size of 0.1\,mas used in the 
RM images. The extreme value of 2.02\,mas is for the source 2134+004 between 15\,GHz and 
12\,GHz, where a different component is the brightest feature in the two maps. 
This illustrates the importance of correct alignment for the data analysis.
The small median shift, however, shows that in the majority of the sources the change is not extreme 
as is to be expected for bright, core-dominated objects. These shifts are determined in part by the 
frequency-dependent core shifts, which will be studied in our sample in \cite{pushkarev12}, although 
other effects can also contribute in some cases, such as in 2134+004 described above.

For 35 sources we were not able to find a reliable alignment due 
to the compactness of the source or the faintness of a featureless jet. 
In these cases we aligned the images based on the fitted core component position 
at each band. The median shift values for these sources were less than 
0.03\,mas. We used spectral index maps to verify that our alignments 
were reasonable. The spectral index maps of all the sources will be 
presented and discussed in a separate paper (T. Hovatta et al. 2012 in prep.).

Additionally we did several tests, described in Appendix \ref{app:align}, to study the 
effect of false alignment on spectral index and rotation measure maps. 
Based on the tests we conclude that even if our image alignment is off 
by 0.15\,mas between 15\,GHz and any other frequency band, it should not affect the results from our rotation measure maps, 
especially as we are not using the edge or low signal-to-noise regions to 
make conclusions about the RM structure.  We verify that the spectral index map is a 
good indicator of the image alignment because the effect of small fake shifts can readily be seen in the 
structure of the spectral index map. 

\subsection{Rotation measure maps}

For the calculation of the rotation measure (RM) maps we wrote a Perl Data Language (PDL) script that does 
the calculation semi-automatically for our large sample of sources. We verified the 
performance of the script by using the RM task manually in AIPS for several sources. 
In our calculations we blanked all the pixels which had polarized flux density less than 
three times the polarization error, defined in Appendix \ref{app:polerr}, at any of the frequency 
bands. Our script chooses the best $\lambda^2$-fit based on a 
$\chi^2$ criterion and blanks all the pixels where it is not met.  We calculate the $\chi^2$ of the fit 
using the standard formulae
\begin{equation}
\chi^2 = \sum_{i=1}^{N}\frac{(O_i - E_i)^2}{\sigma_i^2},
\end{equation}
where N is the number of data points, $O_i$ are the observed data, $E_i$ are the expected data based on the model and $\sigma_i$ is the 
measurement error of the individual data point \citep[e.g.][]{press92}. Due to the dependence on the errors of the 
data points, blanking of low signal-to-noise regions is essential to prevent small $\chi^2$ values simply due to 
large error bars. In general care must be taken in determining the EVPA errors because too small errors will prevent good fits 
while too large errors will result in too small $\chi^2$ values. Our EVPA errors are estimated by adding in quadrature an rms 
error using error propagation from Q and U images (see Appendix \ref{app:polerr} for details) and an absolute calibration error 
defined in Sect.~\ref{sect:EVPAcal}. As we are fitting a two-parameter model to four data points we have two degrees of 
freedom and from a $\chi^2$ distribution the corresponding 95\% confidence limit is $\chi^2 < 5.99$.

The EVPA is ambiguous for changes of 180$^\circ$ and in the calculation of the RM we need to solve for these 
n$\pi$-wraps. We first assumed that there are no n$\pi$-wraps between our frequency bands and 
calculated the RM fit. If the $\chi^2$ of the fit met our criterion, we accepted the RM value without any 
wraps. If the $\chi^2$ criterion is not met, we solved for all possible n$\pi$-wraps up to 
$3.3\times10^4$ rad~m$^{-2}$ and chose the fit with the smallest wrap meeting our $\chi^2$ criterion. 
The upper limit was primarily introduced to keep the computing time reasonable but also because 
based on earlier studies \citep[e.g.,][]{zavala03,zavala04}, we did not expect to resolve RMs larger than this with our frequency setup. 
If none of the wraps resulted in acceptable fits, we blanked the pixel.
By blanking the poor $\lambda^2$-fit regions, we prevent interpretations based on noisy data 
and identify regions with non-$\lambda^2$-law behavior. 

The error of the RM is calculated from the variance-covariance matrix of the least squares fit 
in each pixel. Our typical errors range between 70 and 150 rad~m$^{-2}$ depending on the 
signal-to-noise ratio of the total intensity in the jet, thus in the fainter jet edges the RM errors are 
larger. We verified that our error estimates are correct by performing detailed simulations 
described in Appendix \ref{app:rmerr}.

In order to study the distribution of the intrinsic, redshift-corrected, RM values, the Galactic Faraday 
rotation contribution must be taken into account. We used 
the averaged Galactic RM image of \cite{taylor09} and subtracted the value at the source 
location from each map. We list the values used for each source in Table \ref{rmtable}.
In the majority of sources the Galactic Faraday rotation is very small (the median absolute value 
for the sample is 12.3 rad~m$^{-2}$) but there are 14 sources for which the absolute value is 
more than 70 rad~m$^{-2}$, thereby exceeding our minimum error in the RM values. The largest absolute Galactic 
Faraday rotation values are observed 
for 2021+317 ($-$173 rad~m$^{-2}$) and 2200+420 ($-$156 rad~m$^{-2}$). For the majority of sources in 
our sample the values from \cite{taylor09} agree very well with previously published values 
\citep[][]{rudnick83,rusk88,wrobel93,pushkarev01}. However, we note that since we are using an averaged image, some 
Galactic RM values may be underestimated because small regions of high Galactic RM get smoothed out 
(e.g., 0235+164, 1749+096, 1803+784, 2200+420).

\section{Results}\label{sect:results}
\begin{figure*}[ht!]
\begin{center}
\includegraphics[scale=0.7]{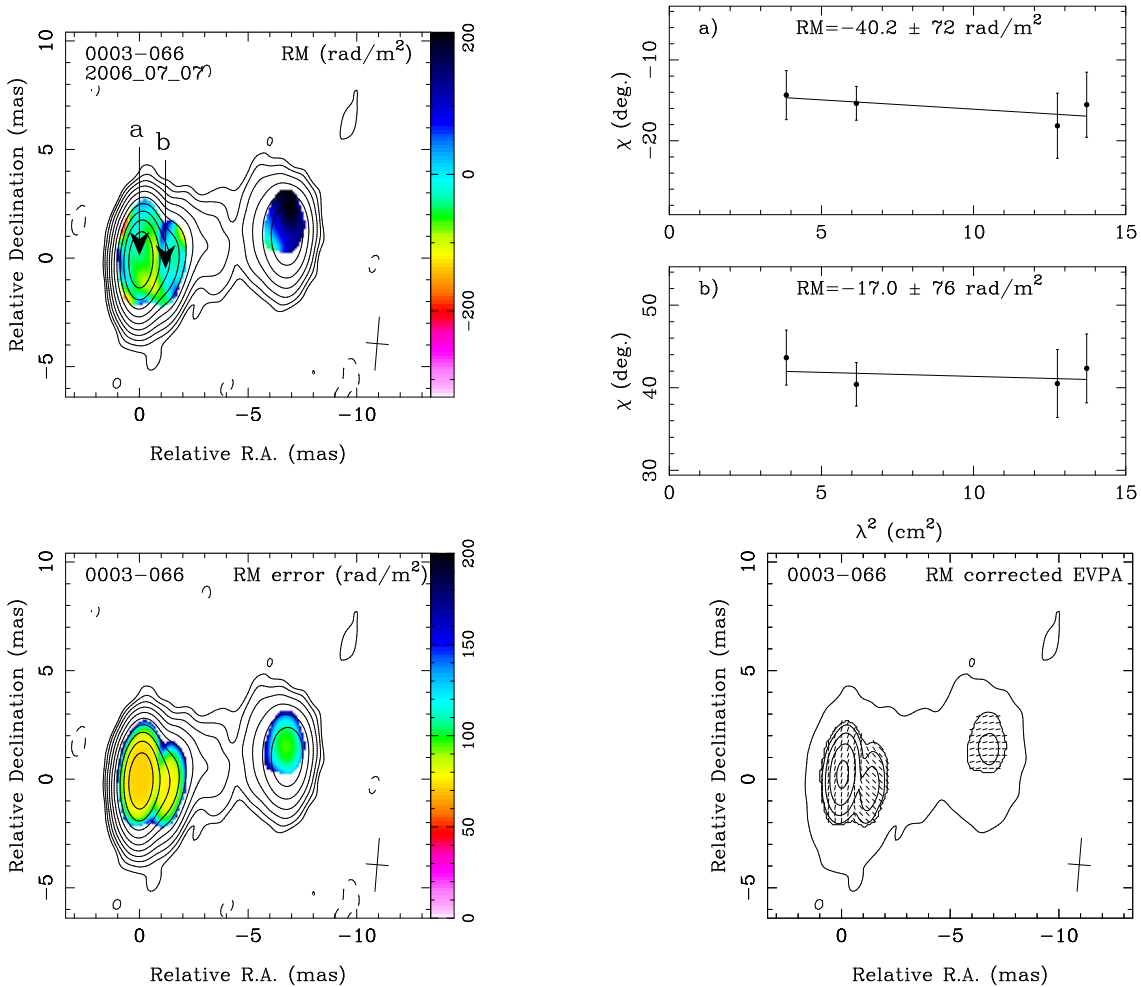}
\caption{Rotation measure maps of all the sources. RM map (top left) with $\lambda^2$-fits at one or two locations in the jet (top right). Error map (bottom left) and map of rotated EVPA values (bottom right). Figures 1.1-1.159 are available in color in the electronic edition of the Journal.\label{fig1}}
\end{center}
\end{figure*}

RM maps are shown in Fig.~\ref{fig1} for the 159 cases where we detect enough polarization to get a RM value for at least a few
pixels. We show the RM values in color scale overlaid on the 15\,GHz total intensity contours and examples of the 
$\lambda^2$ fits in two locations of the jet, chosen to be at the polarization peaks of the map. These locations 
are typically at least one beam width apart. In sources where clear polarization peaks were not seen, we chose 
the location to be in the middle of the RM region. 
Additionally we show the error of the RM in color and the intrinsic, RM corrected, 15\,GHz polarization vectors overlaid on 
the 15\,GHz polarization contours. All the RM maps in Fig.~\ref{fig1} and later in the paper are corrected for Galactic Faraday rotation.
In some cases, there appear to be pixels with very high RM values of over $\pm$ $2\times10^4$ rad~m$^{-2}$.
In most of the sources these coincide with edge pixels and/or regions of complex polarization 
structure. Our simulations of the RM error (see 
Appendix \ref{app:rmerr}), show that it is possible to have spurious high-RM pixels 
in the maps purely due to random noise in the polarization images. These were always 
more than $\pm$ $2\times10^4$ rad~m$^{-2}$ and in our real maps the high-RM regions resembled the 
simulated maps very well. Therefore we have blanked these 
extreme values in Fig.~\ref{fig1}.

\subsection{Extreme RM values}
In some sources we see very high RM regions around the core where we may expect more Faraday rotating material and stronger magnetic fields. 
\cite{udomprasert97} report an intrinsic RM (RM$_\mathrm{int}$) of $4\times10^4$ rad~m$^{-2}$ in the high-redshift quasar OQ~172 ($z = 3.52$). 
In the observed frame, defined as $\mathrm{RM}_\mathrm{obs} = (1+z)^{-2}\mathrm{RM}_\mathrm{int}$, this corresponds to a RM$_\mathrm{obs}$ 
of $\sim$ 2000 rad~m$^{-2}$ in the core. If the intrinsic 
value is correct, we might expect to observe extremely high RMs in some nearby objects. \cite{attridge05} report 
a difference of $\sim 3.2\times10^4$ rad~m$^{-2}$ between two components in the core of 3C~273 in observations 
at 43 and 86\,GHz. It is, however, very difficult at our observing frequencies with much less resolution
to distinguish true extremely high RM$_\mathrm{obs}$ values from the spurious ones due to noise and blending of components. 
For example, in 3C~273 we observe extreme RM$_\mathrm{obs}$ values of $\sim 2.3\times10^4$ rad~m$^{-2}$ around the core in the March epoch. In 
our June epoch, we do not detect these high values but instead see values of $\sim -2.9\times10^4$ rad~m$^{-2}$ around the 
same region. Similar behavior is seen in the cores of 3C~279 and 3C~454.3. In 2200+420 we do not find good $\lambda^2$-fits in the 
core in our April epoch but in November we detect extreme RM$_\mathrm{obs}$ values of $\sim +2.9\times10^4$ rad~m$^{-2}$, never seen before in 
this source \citep{mutel05, zavala03} including observations of \cite{osullivan09a} in July 2006. Other sources where we detect extreme 
RM$_\mathrm{obs}$ values in larger areas near the core include 
0149+218, 0420$-$014, 0605$-$085, 1038+064 and 2145+045. Out of these, 0420$-$014 was observed by \cite{zavala03} who do not 
see these extreme values although they too do not find a good $\lambda^2$ fit at the core component position.
0605$-$085 was observed by \cite{zavala04} who do not detect any extreme RM values and find the RM in the core 
to follow the $\lambda^2$-law. In 2145+067 the polarization structure at 15\,GHz is extremely complex with four 
separate components seen within the innermost jet, while at 8\,GHz only one component is seen. 
Therefore we do not believe that these extreme observed core RMs are real in our maps, but instead are due to multiple polarized components 
blending in the finite beam or due to different opacity properties at the frequency bands. This is further supported by continuing MOJAVE observations of 2200+420 which show a new component emerging from the core in February 2007\footnote{http://www.physics.purdue.edu/astro/MOJAVE/sourcepages/2200+420.shtml}. \cite{algaba11} observe high RMs in the cores of several sources in 
their study of eight sources between 12 and 43\,GHz. They observe a RM$_\mathrm{obs}$ of $2.2\times10^4$ rad~m$^{-2}$ in the core of 1633+382, for which we observe 
a small RM$_\mathrm{obs}$ of $-244$ rad~m$^{-2}$. Their result is based on a large difference in the EVPAs of the 22 and 24\,GHz observations 
which require a large RM. It is possible that they are able to resolve structures not seen in our maps due to 
their higher resolution. In some sources they also find that they need to divide the frequency range into high- and low-frequency 
parts to obtain acceptable $\lambda^2$-fits, which is further indication of different frequencies probing different regions in the core and 
multiple components blending within the beam in the lower frequency maps. 

The blending of components in the core region can also affect our $\lambda^2$-fits so that 
the $\chi^2$ criterion is not met and no RM values are shown in the maps. Another cause 
for this could be internal Faraday rotation, which could play a significant role in the AGN core 
regions. We also see non-$\lambda^2$ patches in the jets of some sources, sometimes 
due to the faintness of the jet emission but at other times also due to depolarization of the lower 
frequencies, a sign of internal Faraday rotation. The effects of internal Faraday rotation and 
other depolarization mechanisms are discussed in more detail in Section \ref{sect:depol}.

\subsection{Median RM distribution}
\begin{figure}[ht!]
\begin{center}
\includegraphics[scale=0.5]{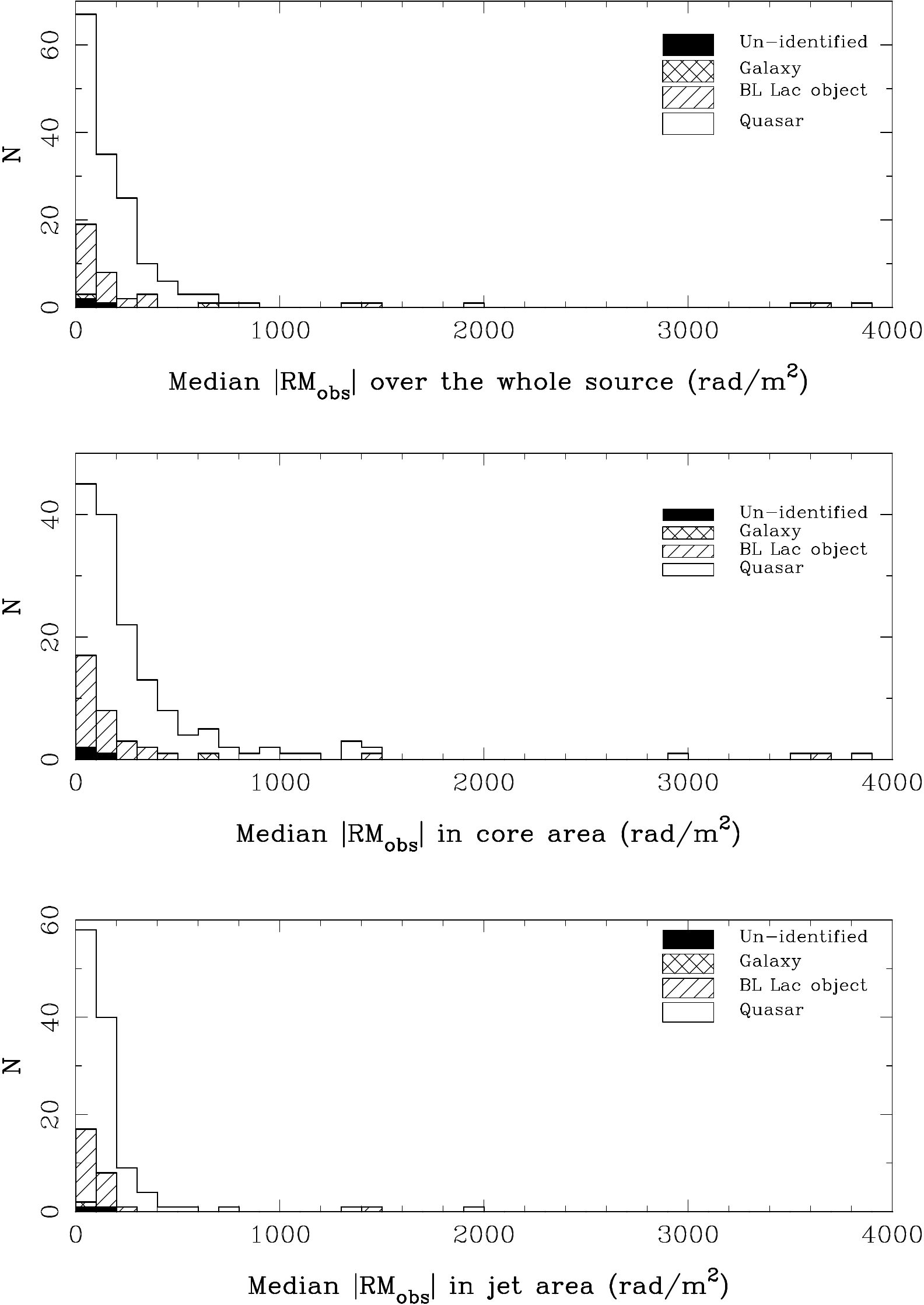}
\caption{Distribution of median $|$RM$_\mathrm{obs}|$ over the whole source (top), over the core region (middle), and over the jet region (bottom). Quasars are shown in white, BL~Lacs in hatched, galaxies in cross-hatched and un-identified sources in black. These plot excludes quasars 0003+380 and 2008-159 which have RM values over 6000 rad~m$^{-2}$.\label{fig:ave}}
\end{center}
\end{figure}

We were able to determine the median RM$_\mathrm{obs}$ for 159 maps, which are shown in the top panel of Fig.~\ref{fig:ave}
and given in Table \ref{rmtable}, where column (1) gives the B1950-name of the source and 
column (2) an alternative alias name. The redshift and the optical classification of the source are listed 
in columns (3) and (4). Apparent speed used for calculation of the de-projected distance in Sect.~\ref{sect:dist} is given in 
column (5) and the observing epoch is listed in column (6). The value used for Galactic 
Faraday rotation correction, taken from \cite{taylor09} is listed in column (7). The median RM$_\mathrm{obs}$ 
value, taken as the median of all the pixels in the source where RM is detected and not blanked, is listed in column (8). 
 Columns (9) and (10) give the median RM over the core and the jet regions, respectively.
We calculate the median instead of 
the average to lessen the effect of individual, possibly spurious, high-RM values. 
The vast majority of sources have a median RM$_\mathrm{obs}$ of  less than 1000 rad~m$^{-2}$, but the 
distribution has a tail to RM$_\mathrm{obs}$ values of 6500 rad~m$^{-2}$. The highest value shown in the plot, 6457 rad~m$^{-2}$, 
is for the source 2008$-$159, which only shows RM values in a small region  of less than half the beam size. At the redshift 1.18 of the source, 
this would result in an extremely high intrinsic RM$_\mathrm{int}$ of over $3\times10^4$ rad~m$^{-2}$ in the source frame. 
As the region over which we detect the high RM$_\mathrm{obs}$ value is so small and does not coincide with any total intensity component locations, it is difficult to say if 
this is a true RM of the source or due to blending of multiple components within the core region.

\subsection{Core vs jet distributions}\label{sect:corejet}
To study the difference between core and jet RM values we  1) divided the source into core and jet regions 
by defining the core region to be everything within a beam width from the center of the 15\,GHz core component position and jet region to be 
everything else and 2) took the 15\,GHz modelfit components \citep[see ][for details on the modelfitting]{lister09b} 
and divided the source into core and jet components. In the first approach the division is determined by a line perpendicular to the 
jet direction at one beam width away from the core. Median values over the pixels within the regions were calculated and are
given in Table~\ref{rmtable} columns (9) and (10) and shown in the bottom two panels of Fig.\ref{fig:ave}.

In the second approach, we calculated the average RM over 
the 9 contingent pixels around the component position to avoid basing conclusions 
on single pixel values. This corresponds to 10-30\% of the restoring beam width depending on the declination of the source.
The component locations and their RM$_\mathrm{obs}$ values are given 
in Table \ref{comptable}. The I.D. number of the component is listed in column (2) where 0 indicates a
core component. Columns (4) and (5) give the component distance and position angle from the phase 
center of the map. The RM$_\mathrm{obs}$ and its error are given in column (6).
Because the pixels are not independent (i.e., they cover a region smaller than the FWHM of the restoring beam), 
 it is not straightforward to estimate the error on the average carried out over 9 pixels. We define the error as 
the average of RM errors in the 9 individual pixels; this approach is conservative, and may overestimate the true error somewhat.
Column (7) indicates whether a jet component is isolated 
(see below). Most of the sources in the MOJAVE sample are core dominated, with a bright compact core that is 
optically thick at centimeter wavelengths and a fainter jet. In most of our sources, we identify the core as a bright, 
stationary feature in the jet, typically at one extreme end of the jet. In a few sources (especially with two-sided jets) 
the identification is not as simple and these are discussed separately in \cite{lister09b}. 

\begin{table*}[ht!]
\begin{center}
\caption{\label{comptable}Modelfit components at 15.3 GHz and their RM$_\mathrm{obs}$ values}  
\begin{tabular}{lccccccccc}
\tableline\tableline
Source & I.D. & Epoch & r & P.A. & RM & Isolated\\  
 & & & (mas) &  (deg) & (rad~m$^{-2}$) & \\ 
(1) & (2) & (3) & (4) & (5) & (6) & (7) & \\
\tableline
0003$-$066 & 0 & 2006-07-07 & 0.71 & -168.6 & -78 $\pm$ 73 &   \\ 
0003$-$066 & 1 & 2006-07-07 & 0.66 & -71.6 & -88 $\pm$ 75 &   \\ 
0003$-$066 & 4 & 2006-07-07 & 6.86 & -81.1 & 113 $\pm$ 98 & Y  \\ 
0003$-$066 & 5 & 2006-07-07 & 0.09 & 2.9 & -42 $\pm$ 72 &   \\ 
0003$-$066 & 6 & 2006-07-07 & 1.28 & -102.7 & -14 $\pm$ 77 &   \\ 
0007+106 & 1 & 2006-06-15 & 0.39 & -66.6 & 627 $\pm$ 130 &   \\ 
0016+731 & 0 & 2006-08-09 & 0.01 & -52.2 & 278 $\pm$ 74 &   \\ 
0016+731 & 2 & 2006-08-09 & 0.20 & 129.2 & 266 $\pm$ 74 &   \\ 
\tableline
\end{tabular}
\tablecomments{Columns are as follows: (1) IAU Name (B1950.0); (2) I.D. of the component (0 = core); (3) Observing epoch; (4) Component distance from the phase center of the I map; (5) Position angle of the component from the phase center; (6) Component RM; (7) Flag for isolated jet components. Table 3 is published in its entirety in the electronic edition of the Journal; a portion is shown here for guidance regarding its form and content.}
\end{center}
\end{table*} 

\begin{figure}[htp]
\begin{center}
\includegraphics[scale=0.5]{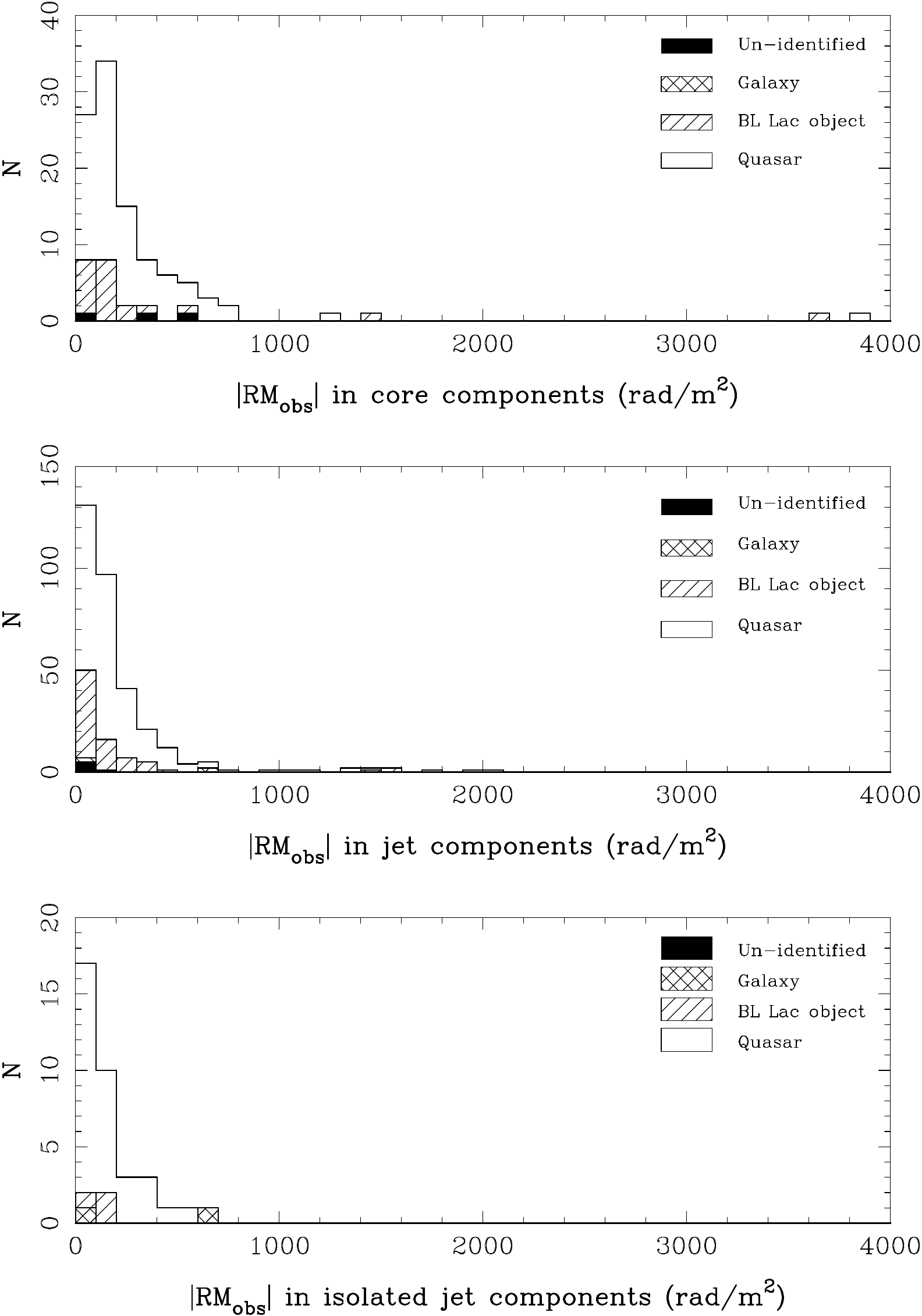}
\caption{Distributions of $|$RM$_\mathrm{obs}|$ in the model fit core (top) and  jet (middle) components. Jet components which have less than 30\% total intensity contribution from other components in the source are plotted in the bottom panel (see text for details). Quasars are shown in white, BL~Lacs in hatched, galaxies in cross-hatched and un-identified sources in black.\label{fig:corejetcomp}}
\end{center}
\end{figure}

The distribution of the RM$_\mathrm{obs}$ values in the components 
is shown in Fig.~\ref{fig:corejetcomp}.  We were able to determine the core component RM in 
104 maps (101 sources) and the jet RM in 324 components (121 sources).
From the distributions it is clear that the core component values have a tail to higher RM$_\mathrm{obs}$ values,
but there are also some jet components with high RM$_\mathrm{obs}$ values. In most of those cases
the jet component is within 1 mas of the core component and often still in the 
optically thick or self-absorbed region of the jet. To distinguish the jet components which are away from the bright core region, 
for each component we calculated the combined contribution of all the other jet components in the map at the component's peak intensity position. 
If this sum was less than 30\% of the component's total intensity, we considered the component to be isolated.
In this way we determined that the polarization and RM of the component were not 
affected by nearby bright components.  Out of all the jet components, 36 in 24 sources are listed as isolated.
The distribution of RM$_\mathrm{obs}$ values of these isolated components is plotted in the bottom panel of Fig.~\ref{fig:corejetcomp}. 
As can be seen, none of the RM$_\mathrm{obs}$ values greater than 700 rad~m$^{-2}$ are isolated.
The median component RM$_\mathrm{obs}$ in the whole sample is 
171 rad~m$^{-2}$ for the cores, 125 rad~m$^{-2}$ in all the jet components and 104 rad~m$^{-2}$ in the isolated jet components.
According to an Anderson-Darling (A-D) two sample test \citep[e.g.,][]{press92}, which 
is more sensitive to distribution tails than the Kolmogorov-Smirnov test,
the core and jet components have less than a 1\% probability of coming from the same parent population. Comparison 
of core and isolated jet components gives a probability of less than 2\% due to the smaller number of jet 
components.
In all our tests we consider the result significant if the probability is less than 
5\%.

 A similar trend is seen when comparing the core and jet regions in Fig.~\ref{fig:ave}. The median RM for the core regions 
is 187 rad~m$^{-2}$ and for the jets is 102 rad~m$^{-2}$, a result very similar to that for the isolated jet components. According to the A-D test the 
probability for these distributions to come from the same parent population is less than 0.001\%.

\subsection{Optical Subclasses}
Our sample can also be divided into subclasses based on the optical classification of the source.
These are also shown in Figs.~\ref{fig:ave} and \ref{fig:corejetcomp}.
The number of galaxies and optically un-identified sources in our 
sample is so small that they cannot be included in any statistical comparisons. 
The quasars and BL~Lac objects, however, can be compared, and it is clear that the high-RM tail in 
the distributions consists mainly of quasars.  The median absolute RM$_\mathrm{obs}$ value in 
quasars is 144 rad~m$^{-2}$ and in the BL~Lacs 79 rad~m$^{-2}$.

If we look at the core and jet components individually, there is less than 0.1\% probability that the 
jet components of quasars and BL~Lacs are drawn from the same population. The median jet 
RM$_\mathrm{obs}$ for quasars is 141 rad~m$^{-2}$ while for the BL~Lacs it is 71 rad~m$^{-2}$.  However,
this difference is affected by the components within one beam width of the core in quasars because the 
median values in the jet regions of quasars is 116 rad~m$^{-2}$ and in BL~Lacs it is 76 rad~m$^{-2}$. According 
to the A-D test we cannot reject the null hypothesis that the distributions come from the same parent population.

In the cores differing results are also obtained when core components and regions are compared.
In quasars the median core component value is 183 rad~m$^{-2}$ compared to 134 rad~m$^{-2}$ in BL~Lacs, 
and we cannot reject the null hypothesis that they come from the same parent population.
In the core regions, however, the median for quasars is 200 rad~m$^{-2}$ and for BL Lacs it is 105 rad~m$^{-2}$, 
and there is less than 1\% probability that these come from the same distribution. 

These results can be compared to \cite{zavala04}, who saw a difference in the 
core RMs for quasars and BL~Lacs but not in the jet values. Our results on the jet RMs agree if we look at the 
jet regions which are not contaminated by components near the core. We cannot verify if there is a difference 
between the cores in quasars and BL~Lacs because of the differing results depending on if we look at the core 
components or core regions. However, 
the higher median values observed 
in quasars than in BL~Lacs are in accordance with the
standard models in which BL~Lac objects have less material around them, resulting in 
dimmer and narrower emission lines than in quasars.

When examining the intrinsic, redshift corrected, RM$_\mathrm{int}$ values in the components the difference between BL~Lacs and quasars 
is more significant. This is mainly due to the BL~Lacs in our sample having smaller 
redshifts than the quasars (median redshift 0.31 vs. 1.12), which is enhanced in the correction defined as $\mathrm{RM}_\mathrm{int} = \mathrm{RM}_\mathrm{obs}(1+z)^2$, where $z$ is the redshift.

The absolute RM$_\mathrm{int}$ values in the cores of quasars range from 4.8 to 6436 rad~m$^{-2}$ with a median 
of 798 rad~m$^{-2}$. In BL~Lacs the range of the core RM$_\mathrm{int}$ values is from 13 to 3873 rad~m$^{-2}$, similar to
the quasars, but the median is significantly smaller (274 rad~m$^{-2}$) and according to an A-D test 
the probability for the two distributions to come from the same parent population is less than 3\%. 
The median RM$_\mathrm{int}$ value for the jet components of BL~Lacs (range from 0.8 to 1937 rad~m$^{-2}$ 
with a median of 148 rad~m$^{-2}$ ) is significantly smaller than that of the cores even though the range is similar. 
In quasars they range from 1 to 8975 rad~m$^{-2}$ with a median of 563 rad~m$^{-2}$  but the probability
to reject the null hypothesis in the case of the intrinsic jet and core components of quasars is only 6.8\% and therefore not 
significant. Similarly to the case of core components, the difference between quasars and BL~Lacs is significant also in the 
intrinsic jet components.

\section{Discussion}\label{sect:discussion}
One of the main scientific motivations for the multifrequency survey of the MOJAVE 
sources was to determine the effects of Faraday rotation on the observed polarization 
structure of the sources at 15\,GHz. Based on the first epoch MOJAVE data, \cite{lister05} showed 
that in BL~Lac objects the distribution of EVPAs with respect to the local jet direction appears bimodal. The effect of Faraday 
rotation was not taken into account and therefore these results could be affected by sources with 
high RM values. The RM distribution of Fig.~\ref{fig:ave} shows that in over 80\% of our sources 
the RM$_\mathrm{obs}$ values are less than 400 rad~m$^{-2}$, which will rotate the 15\,GHz electric vectors by 
about 10 degrees. This means that the results of a large sample of \cite{lister05} should approximately reflect the true 
distribution at 15\,GHz. However, when studying some individual sources, the Faraday rotation must 
be taken into account as a rotation measure of 2000 rad~m$^{-2}$ (seen in the median RM distributions for 
a few individual sources) can rotate the 15\,GHz EVPAs by 40 degrees. For example, in 0429+415 we detect 
RM$_\mathrm{obs}$ of $\sim$ 1900 rad~m$^{-2}$ in several jet components 40 mas from the core, similarly to 
\cite{mantovani10}. In 1101+384 and 1725+044 we detect core component RM$_\mathrm{obs}$ as high as 3800 rad~m$^{-2}$, although 
it must be noted that in these two sources we detect RM only in a very small region around the core and therefore 
cannot be sure if it is true RM or due to blending of multiple polarized components within the finite beam.

\subsection{Distance dependence}\label{sect:dist}
\begin{figure*}[htp]
\begin{center}
\includegraphics[scale=0.5]{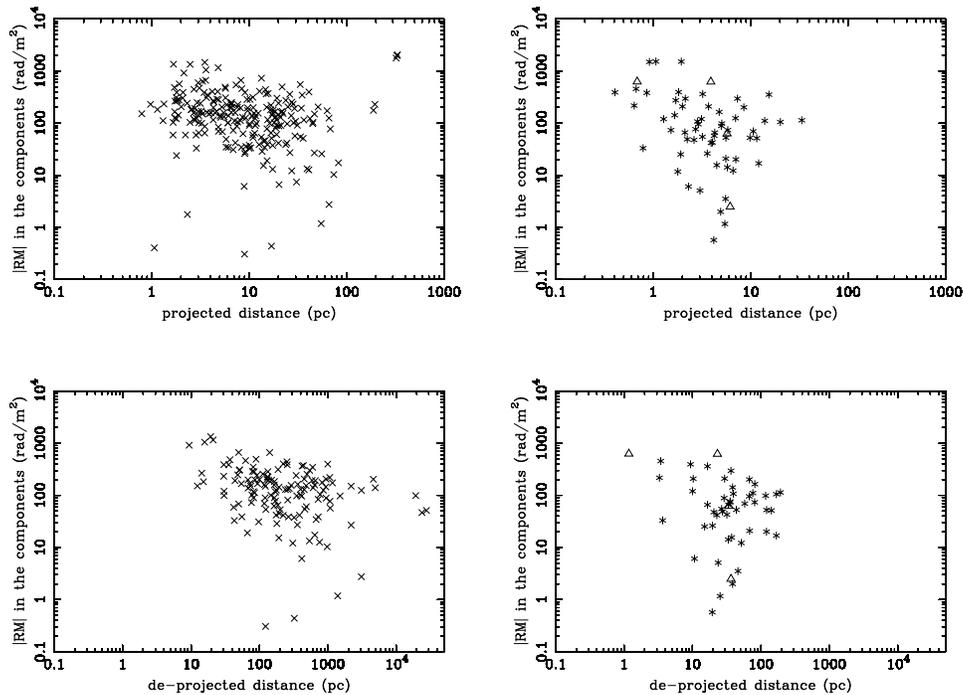}
\caption{$|$RM$_\mathrm{obs}|$ against the projected distance of the component from the core (top panel) and against the de-projected distance taking the viewing angle of the source into account (bottom panel). Left panels show quasars (crosses), and right panels BL~Lac objects (asterisks), and galaxies (triangles).\label{fig:rmdist}}
\end{center}
\end{figure*}
In Fig.~\ref{fig:rmdist} (top panel) we show the RM$_\mathrm{obs}$ versus projected distance from the core for all the modelfit jet components.
The dependence for the total sample is not very clear although according to a non-parametric Kendall's tau correlation test \citep[e.g.,][]{press92} 
there is a significant negative correlation ($\tau=-0.13$, p=0.00058). When quasars and BL~Lacs are studied 
separately it can be seen that the correlation in quasars is stronger ($\tau=-0.17$, p=$6.3\times10^{-5}$) while 
for the BL~Lacs alone $\tau = -0.21$, p=0.02. The picture is, however, more complicated 
as the true distance from the core depends on the viewing angle of the source. We have therefore de-projected 
the distances using viewing angles $\theta = \tan^{-1}[(2\beta_\mathrm{app})/(\beta_\mathrm{app}^2+\mathrm{D_{var}}^2-1)]$ 
determined with Doppler factors D$_\mathrm{var}$ from \cite{hovatta09} and apparent speeds $\beta_\mathrm{app}$ 
from \cite{lister09b}. Some of the speeds have been updated since \cite{lister09b}, and all the speeds used are tabulated in 
Table \ref{rmtable}. Both values were available for 138 components in quasars, 47 components in BL~Lacs and 
4  jet components in galaxies. The RM$_\mathrm{obs}$ against the de-projected distance is shown in Fig.~\ref{fig:rmdist} (bottom panel).
The negative correlation in quasars remains significant despite the smaller number of sources ($\tau=-0.23$, p=$4.8\times10^{-5}$) while in BL~Lacs 
the correlation vanishes ($\tau=-0.01$, p=0.92). However, we do not detect jet components as far away from the core 
in BL~Lacs as in quasars which could affect the correlation. The correlation seen in quasars supports the results from the 
simple core and jet component comparison in Sect. \ref{sect:corejet} showing that the amount of Faraday rotating material 
diminishes as a function of distance from the core. 
\begin{figure}[htp]
\begin{center}
\includegraphics[scale=0.5]{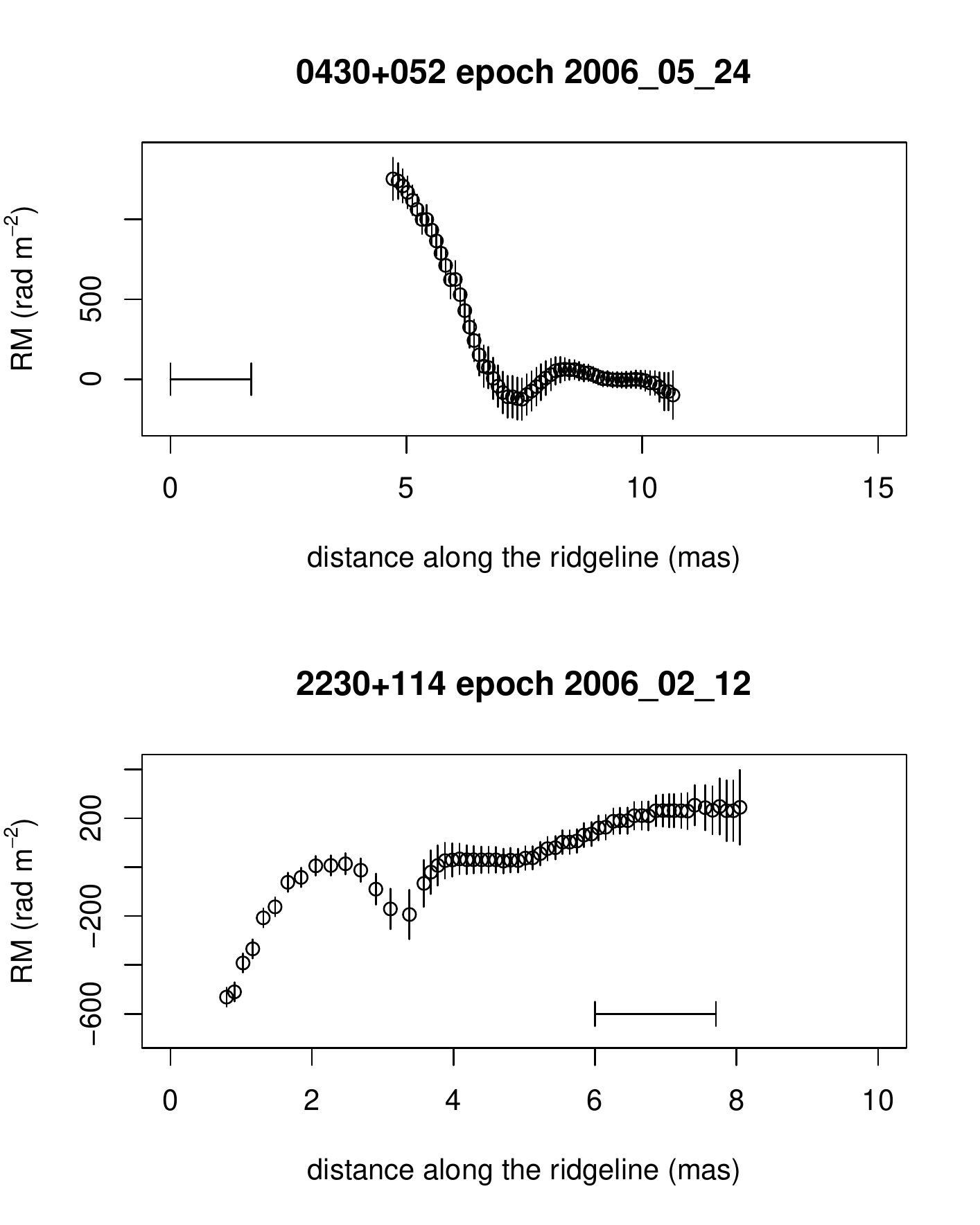}
\caption{RM$_\mathrm{obs}$ along the total intensity ridge line in 0430+052 (top) and 2230+114 (bottom).  The average FWHM beam size along the jet is shown in both plots as a scale bar.\label{fig:ridgeline}}
\end{center}
\end{figure}

 The above approach is a simplification of a more complex behavior in the RM values as a function of distance and even if 
the general trend shows a decline in RM along the jet, individual sources may deviate from this trend and show complex structures.
Another way to study the distance dependence is to calculate the RM values along the total intensity ridge line of the jet. 
Two examples are shown in Fig.~\ref{fig:ridgeline} where the top panel shows the RM along the ridge line in 0430+052 where 
the core is depolarized (Fig.~1.28) but further along the jet RM declines very sharply, in accordance with the simple scenario. 
In 2230 +114 (Fig.~1.151) a more complex structure can be seen along the jet, with the RM changing sign along the jet. 
This example shows that with better resolution along the jet, the situation may not be as simple as 
a linear dependence along the jet but more complex regions are seen. The recent sensitivity upgrades of the VLBA (e.g., higher bit rate observations) 
will allow us to detect more polarization further down from the core and help to study this in more sources.

\subsection{Faraday depolarization}\label{sect:depol}
Faraday rotation can cause different amounts of depolarization depending on the nature of the 
rotating screen and if it is internal or external to the jet \citep{burn66}. By applying a very simple Burn-type internal 
Faraday depolarization model to the core components of 40 AGN, \cite{zavala04} concluded
that internal Faraday rotation alone cannot explain the steep decline in fractional polarization 
as the magnitude of the rotation measure increases. The equations they used are valid only in 
the optically thin regime and therefore not applicable to the core regions of AGN at 15\,GHz. We 
explore the viability of possible models by directly fitting individual isolated jet components for depolarization and 
comparing the results to our observed RM values.

\begin{figure*}[htp]
\begin{center}
\includegraphics[scale=0.6]{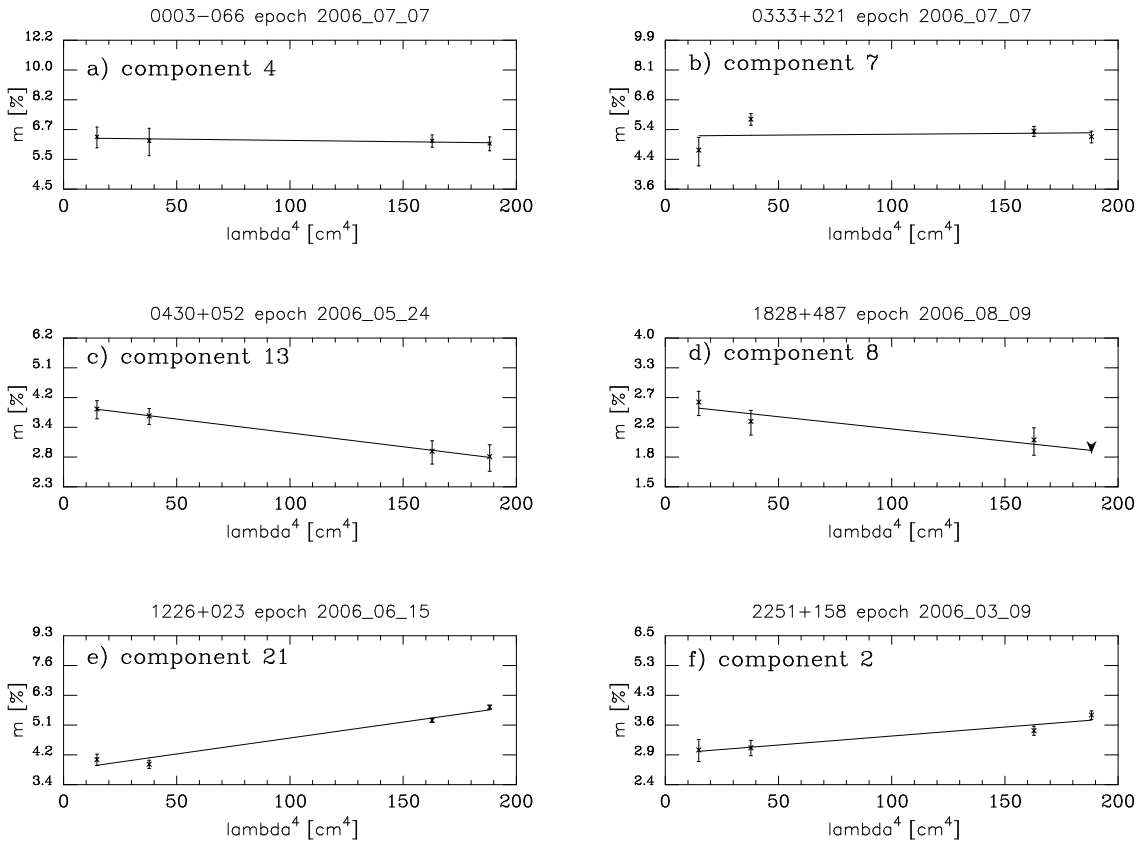}
\caption{Examples of fits of depolarization curves to isolated components. a) No apparent depolarization. b) Ambiguous behavior. c) and d) Depolarization. e) and f) inverse depolarization. See text for details. Upper limits are shown in arrows. Note the scale of the vertical axis which is not linear but in natural logarithm. Fits for all isolated components are available in the online edition of the Journal.\label{fig:depolfit}}
\end{center}
\end{figure*}

\begin{figure}[htp]
\begin{center}
\includegraphics[scale=0.65]{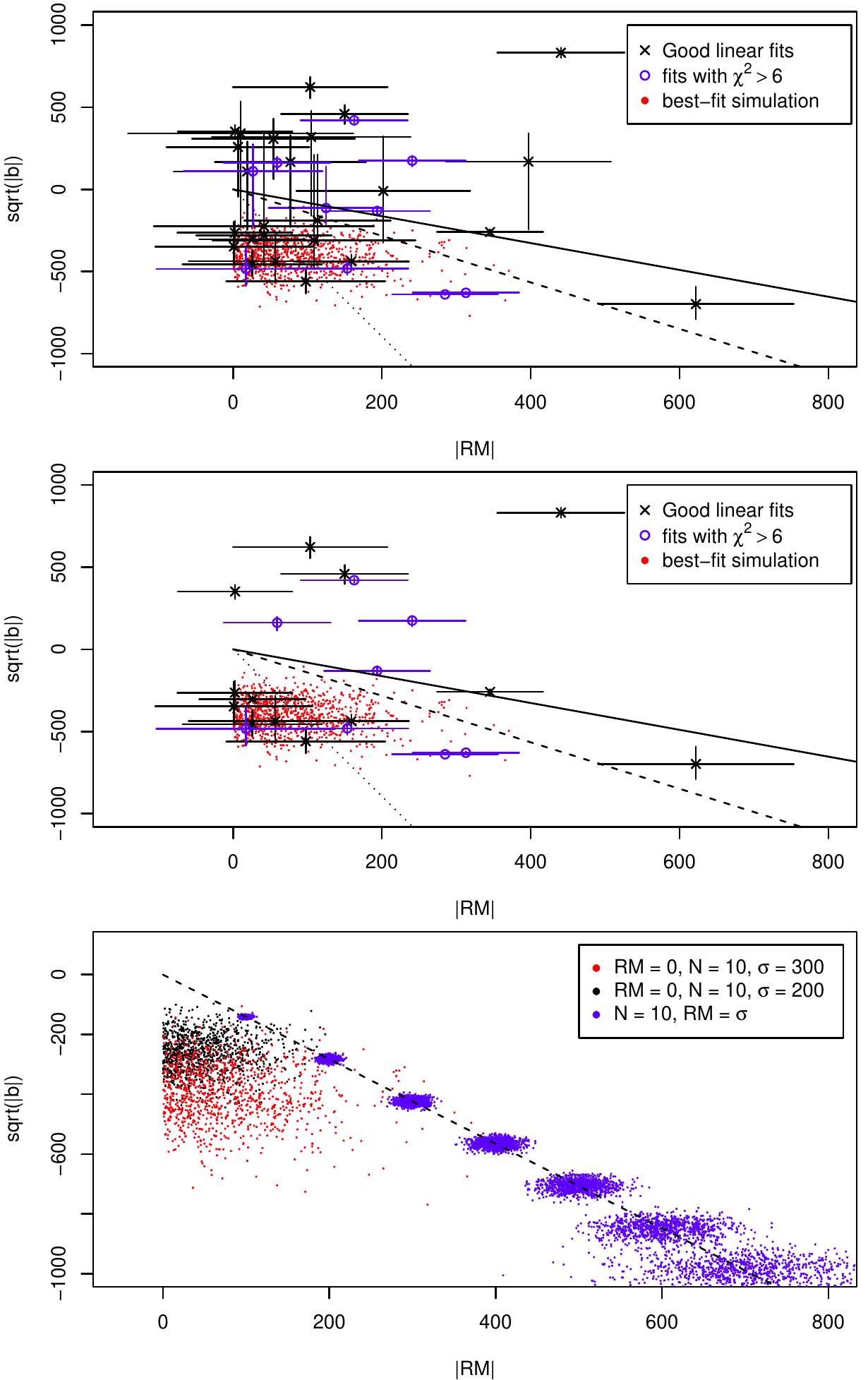}
\caption{Top: Fitted slope b from Fig.~\ref{fig:depolfit} against the $|$RM$_\mathrm{obs}|$ in the isolated jet components. Crosses (black in the online version) have good fits with $\chi^2 < 6$, open circles (blue) have $\chi^2 > 6$ and may not be well described by our simple model. Grey dots (red in the online version) are our best-fit simulation case from the bottom panel. The solid line shows the expected amount of internal Faraday depolarization and dashed and dotted lines the expected amount of external Faraday depolarization for different parameters (see text for details). Middle: Same as top panel but showing only components that are more than 2$\sigma$ from $b = 0$. Bottom: Simulations for external depolarization over varying average RM, number of lines of sight N and dispersion $\sigma$ in the RM. The dashed line shows the expected amount of external Faraday depolarization when average RM = $\sigma$. Dark gray (blue in the online version) dots show the same from several simulations using N=1000. Black dots have average RM = 0, N = 10, $\sigma = 200$, light gray (red) dots have average RM = 0, N = 10, $\sigma = 300$. \label{fig:rm_b}}
\end{center}
\end{figure}

For internal depolarization assuming a uniform magnetic field and the optically thin regime we have
\begin{equation}\label{eq:depol}
\mathrm{m}_\mathrm{obs} (\%)= \mathrm{m}_\mathrm{max} \left| \frac{\sin (2 \lambda^2\mathrm{RM})}{2 \lambda^2\mathrm{RM} }\right|,
\end{equation}
where m$_{\mathrm{obs}}$ is the observed 
fractional polarization, $\mathrm{m}_\mathrm{max}$ is the maximum fractional polarization in the 
specific magnetic field configuration \footnote{Note that $m_\mathrm{max} \simeq 70$\% for a pure uniform magnetic field \citep {pacholczyk70}; however, we do not assume a value for $m_\mathrm{max}$ in our analysis and include it as a free parameter, $m_0$, in our fits.}, RM is the 
observed rotation measure and $\lambda$ is the observing wavelength \citep{burn66, homan09}. 
In case of external depolarization 
\begin{equation}\label{eq:extpol}
\mathrm{m}_\mathrm{obs} (\%)= \mathrm{m}_\mathrm{max} e^{-2\sigma^2\lambda^4},
\end{equation} 
where $\sigma$ is the standard deviation of the RM fluctuations and the rest of the parameters are as in Eq.~\ref{eq:depol} \citep{burn66}. Here we assume 
that the component is optically thin and homogeneous (not a combination of multiple components),
and also that the angular scale of RM variations is much less than the angular resolution of our observations.
The functional forms of depolarization in Eqs. \ref{eq:depol} and \ref{eq:extpol} are similar over the range of $|$RM$_\mathrm{obs}|$ 
we observe in the isolated jet components (up to 800 rad~m$^{-2}$) and both follow the functional form $m = m_0e^{b\lambda^4}$ where 
$b$ is $-$2RM$^2$ in the case of internal depolarization and $-2\sigma^2$ in the case of external depolarization. 
We can linearize the formula to fit $\ln m = \ln m_0 + b\lambda^4$ to our observations. This way we will get an estimate 
of total depolarization $b$ from the slope of the fit  and the intercept, $\ln m_0$, gives the maximum
polarization for that component. We use the isolated jet components only to ensure that we are looking at 
homogeneous components in the optically thin part of the jet. The polarization values for the isolated components are given in 
Table \ref{depoltable} where the RM is given in column (6), columns (7) - (10) show the fractional polarization and 
its error at the different frequency bands, and column (11) the value $b$.
In Fig.~\ref{fig:depolfit} we show examples of the fits and in the top panel of 
Fig.~\ref{fig:rm_b} we show the square root of $|b|$ with its sign preserved to distinguish depolarization from 
inverse depolarization where polarization is higher at 8\,GHz than at 15\,GHz. 
In 16 out of 61 components our simple model does not fit the data well which may be an indication of a more complex behavior than 
the simple exponential model can explain. These components are clearly marked in Fig.~\ref{fig:rm_b} and Table \ref{depoltable}.

The solid line in the top panel of Fig.~\ref{fig:rm_b} gives the 
expected amount of internal Faraday depolarization for a given RM. If internal Faraday 
depolarization alone would account for our observations, most of the data points should fall 
on this line which is not the case. The majority of the depolarized components fall below the line indicating 
that the fractional polarization in these components falls faster than expected for internal Faraday rotation. This is 
clearer in the middle panel of Fig.~\ref{fig:rm_b} which shows only the components that are more than 2$\sigma$ away from 
$b = 0$. The same appears to be true for external depolarization when we assume that the dispersion in the RM values 
is proportional to the observed RM, i.e. $\sigma^2 = \mathrm{RM}^2$ (dashed line). Most of the components fall below this line, 
indicating that the dispersion in the Faraday screen is larger than the mean RM produced by that screen. This suggests that the 
Faraday screen is dominated by random RM fluctuations between independent lines of sight. For a random Faraday screen the 
observed average RM will approximately follow a 
relation $\mathrm{RM} \propto \sigma/\sqrt{\mathrm{N}}$ where N is the number of lines of sight. 
The dotted line in the top panel of Fig.~\ref{fig:rm_b} uses N = 10 for the calculation of $\sigma$ in Eq.~\ref{eq:extpol}, where 
we assume the angular scale of the RM dispersion to be much smaller than the beam so that $\sigma^2 = 10\mathrm{RM}^2$. 
This line fits our data much better than assuming $\sigma^2 = \mathrm{RM}^2$, but there is still a large scatter about the line. 
The line produced by Eq.~\ref{eq:extpol} assumes the scale of Faraday rotating cells to be much smaller than the beam size. 
This may not be true for high-angular-resolution observations such as these by the VLBA \citep{tribble91}.  
In order to take the number of lines of sight correctly into account, we directly simulate the expected depolarization and RM for 
a variety of $\sigma$, RM and N combinations.

In the simulations, we define $\lambda^2$ values in the range of 0 to $1.6\times10^{-3}$m$^{2}$ and initialize each frequency to have 
70\% fractional polarization and 0$^\circ$ EVPA. \footnote{Note that the initial values chosen here do not affect our results as we are 
interested only in the change in p and EVPA due to combination of multiple lines of sight within our beam.} We pass this initial polarization 
through N individual lines of sight. Each individual line of sight is simulated by adding an average RM, which is the same for all lines of sight, and a random 
contribution drawn from the Gaussian distribution of variance $\sigma^2$.  For each wavelength, we sum the contribution of 
N lines of sight drawn in a similar manner to obtain an average p and EVPA value. We treat these 
average values as our real observations and fit the p values for $b$ and EVPA values for RM. We repeated the simulation 1000 times 
to obtain a range of $b$ and RM values to compare with our observations. In Fig.~\ref{fig:rm_b} bottom panel we plot 
several of our simulations and in the top panel we show our best fit case overlaid with the real data points.  

 The blue dots in the bottom panel of Fig.~\ref{fig:rm_b} are from several simulations with a large number of lines of sight
(N=1000) where we have set average $\mathrm{RM} = \sigma$ (using values 100, 200, 300, 400, 500, 600, and 700 for RM and $\sigma$),
and these simulations are plotted with the dashed line produced by Eq.~\ref{eq:extpol}. As can be seen, these large N simulations follow the 
expected curve very well with increased scatter for the large sigma values. 
As described above, our data in the top panel of Fig. ~\ref{fig:rm_b} largely fall below this line, indicating that the Faraday screens may be random
screens with a small number of lines of sight.  To test this possibility, we set the average RM
applied to all lines of sight to be 0 in our simulation, the number of lines of sight N = 10, and either 
$\sigma=200$ rad~m$^{-2}$ (black dots) or $\sigma=300$ rad~m$^{-2}$ (red dots). The red dots cover almost the same
region as our data while the black dots produce too little depolarization. As one might expect for
a random-walk style Faraday screen,  $\sigma/\sqrt{\mathrm{N}}$ for the red dots is 95 rad~m$^{-2}$ which is in close
agreement with the observed median  $|$RM$|$ = 104 rad~m$^{-2}$ for our sample of isolated components.  Therefore we conclude that 
most of our observations can be explained with a completely random external foreground screen 
viewed through a small number of lines of sight. This implies that the linear size of the RM cells 
may not be too much smaller than our beam size. We plot the red dots of the bottom panel of 
Fig. ~\ref{fig:rm_b} in the top panel to show the good correspondence with our data.

Additional complications to note are that if the depolarization is 
much higher, we do not detect enough polarization at 8\,GHz to calculate the RM or the internal rotation causes non-$\lambda^2$-law 
behavior and we do not get good enough fit in our RM calculation. To study this further, we have included in Table \ref{depoltable} isolated jet components 
for which we detect fractional polarization at some of our four frequency bands but not necessarily all (15 components), and ten components for which we 
detect fractional polarization but no RM (a sign of non-$\lambda^2$ behavior). In the calculation of the slope $b$, we have assumed the upper limits to be detections. 
This way we get a lower limit estimate for the depolarization.

Based on the polarization behavior in Fig.~\ref{fig:rm_b} we can divide all our  
fits into four categories. Constant polarization over the frequency range is seen in 13 out of 60 components,
as the example case in Fig.~\ref{fig:depolfit}a. These components are within error bars of zero $b$ in Fig.~\ref{fig:rm_b}.
In 18 components, the fractional polarization did not follow a linear trend but 
was changing randomly between the frequencies, as seen in Fig.~\ref{fig:depolfit}b. In several of these the slope $b$ is consistent with zero 
within the error bars.
Depolarization is seen in 20 of the components, and two examples are given in 
Figs.~\ref{fig:depolfit}c and \ref{fig:depolfit}d. In 1828+487 we were not able to calculate a RM value because we only detect an upper limit in linear polarization at the 8.1\,GHz band.
From the slope of the fit we can estimate that the amount of internal Faraday rotation required to cause such depolarization would be 
970 rad~m$^{-2}$, higher than what we observe in any of the isolated components. There are four additional components which show slopes steeper than 
the typical range in Fig.~\ref{fig:rm_b}, and the depolarization in these sources could be produced by internal Faraday depolarization.

Additionally, we see nine components with 
inverse depolarization structure, where the fractional polarization at 8\,GHz is higher than at 15\,GHz.
In only five of these we detect RM as well and these are the most significant points above zero $b$ in Fig.~\ref{fig:rm_b}.
The other inverse depolarization components above zero are in the category where a linear fit did not describe the fractional polarization 
behavior well and the slope $b$ is not a good indicator of the depolarization. 
The nine components each show a significant rise in the fractional polarization as shown in Figs.~\ref{fig:depolfit}e and \ref{fig:depolfit}f. 
This is unexpected and cannot be easily explained with any standard external depolarization models. Interestingly, seven of the nine components (and all the five for which we 
have RM value) are in 3C~273 and in 3C~454.3, both of which show transverse RM gradients in their jets (see Sect. \ref{sect:grad}). The other two are in 1458+718
where the slope is still within 2$\sigma$ from zero, and in 1514$-$241 where the fractional polarization rises from an 8.6\% upper limit at 15 GHz to 
24\% at 8\,GHz. In this source we do not detect any RM values. Internal Faraday rotation together with helical or loosely tangled random 
magnetic field configurations could possibly explain the observed inverse depolarization and this model is investigated in detail
by \cite{homan12}.

\subsection{Transverse RM gradients}\label{sect:grad}
 If AGN jets are launched from a rotating black hole or accretion disk, it could be expected that the magnetic field 
around the jet has an ordered toroidal component \citep[e.g.,][]{mckinney07}. A signature of such a toroidal component 
(often interpreted as a component of a helical field) would be 
a rotation measure gradient transverse to the jet flow direction as the line-of-sight magnetic field changes its 
direction \citep[e.g.,][]{blandford93}. In this case, the gradient should be seen in multiple locations of the jet,  
which distinguishes it from isolated local gradients that arise from changes in the density of the Faraday rotating material.
The detection of such gradients is challenging due to the limited number of bright sources 
with polarized, well-resolved jets \citep{taylor10}. Furthermore, the jet structures can be 
very complex, and it is likely that both kind of gradients exist in the same sources, as in the case of the radio galaxy
3C~120 \citep{gomez11}. Therefore even if a transverse RM gradient is observed, it does not automatically 
indicate the presence of a helical magnetic field, and detailed modeling is needed to probe its nature.

In Appendix \ref{app:grad}, we perform 
simulations to investigate how large spurious transverse gradients can arise due to image noise and finite restoring beam size. 
Based on our simulations we conclude that the convolved jet should be at least 1.5 beams wide (but preferably more than 2) in polarization
along the direction of the gradient and that a gradient should exceed the
3$\sigma$ level to be considered significant.  We define $\sigma$ as the largest RM error at the edge of the jet when the 
systematic error due to absolute EVPA calibration, $\sim 60$~rad~m$^{-2},$ is first removed in quadrature. The significance 
of a gradient is then simply the total change in RM divided by the $\sigma$.
These criteria are similar to 
the ones described by \cite{broderick10} and \cite{taylor10}, although our simulations indicate a minimum transverse width of 1.5-2 rather 
than 3 beamwidths. \cite{broderick10} show that 
due to the complexity of AGN cores, gradients within one beam width of the core may be unreliable  at our resolution, and 
therefore we have not considered these regions in our study. \cite{murphy11} argue that a RM gradient 
due to helical magnetic field is significant even when the jet is not resolved, based on simulations where 
they convolve a simulated gradient with different beam sizes. However, their simulation does not take into account that 
a spurious gradient can arise due to noise in the data, which we show to be a major effect on VLBA observations of unresolved jets.

\begin{figure*}[ht!]
\begin{center}
\epsscale{0.9}
\plottwo{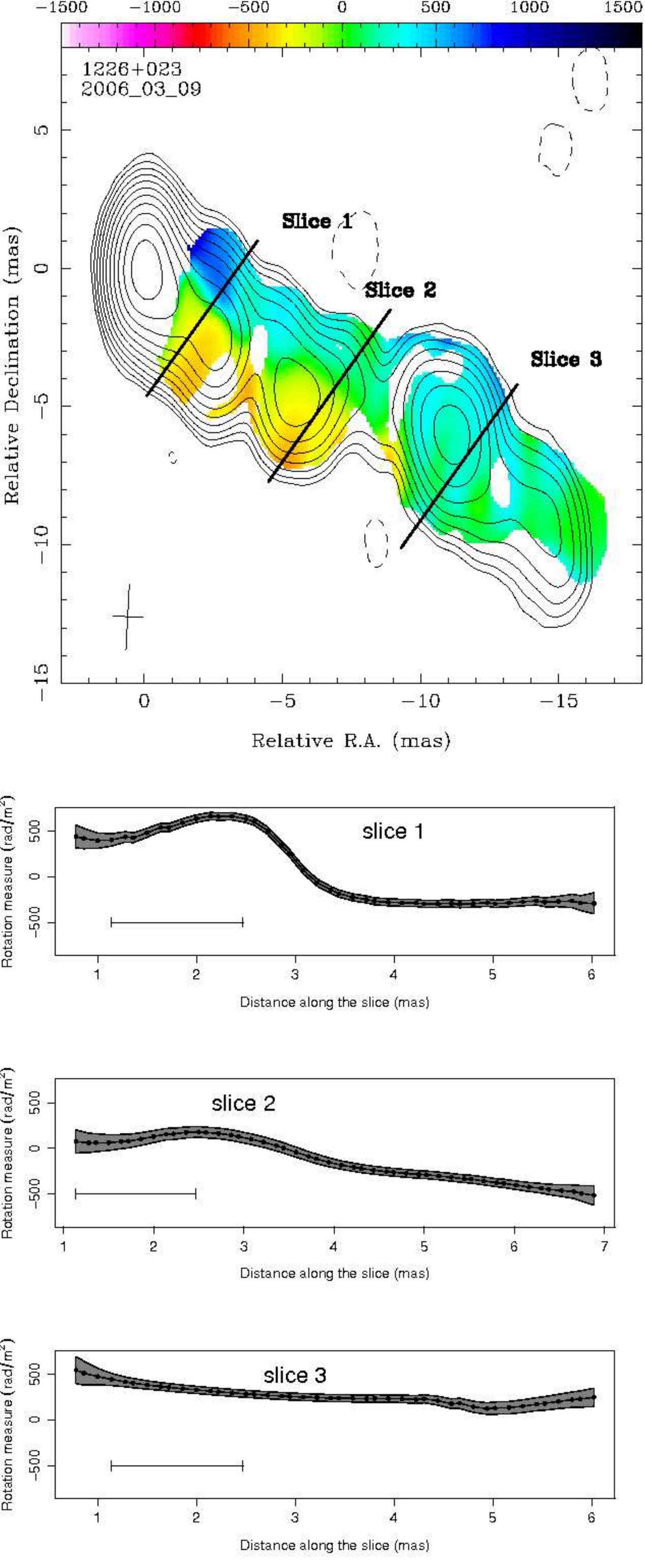}{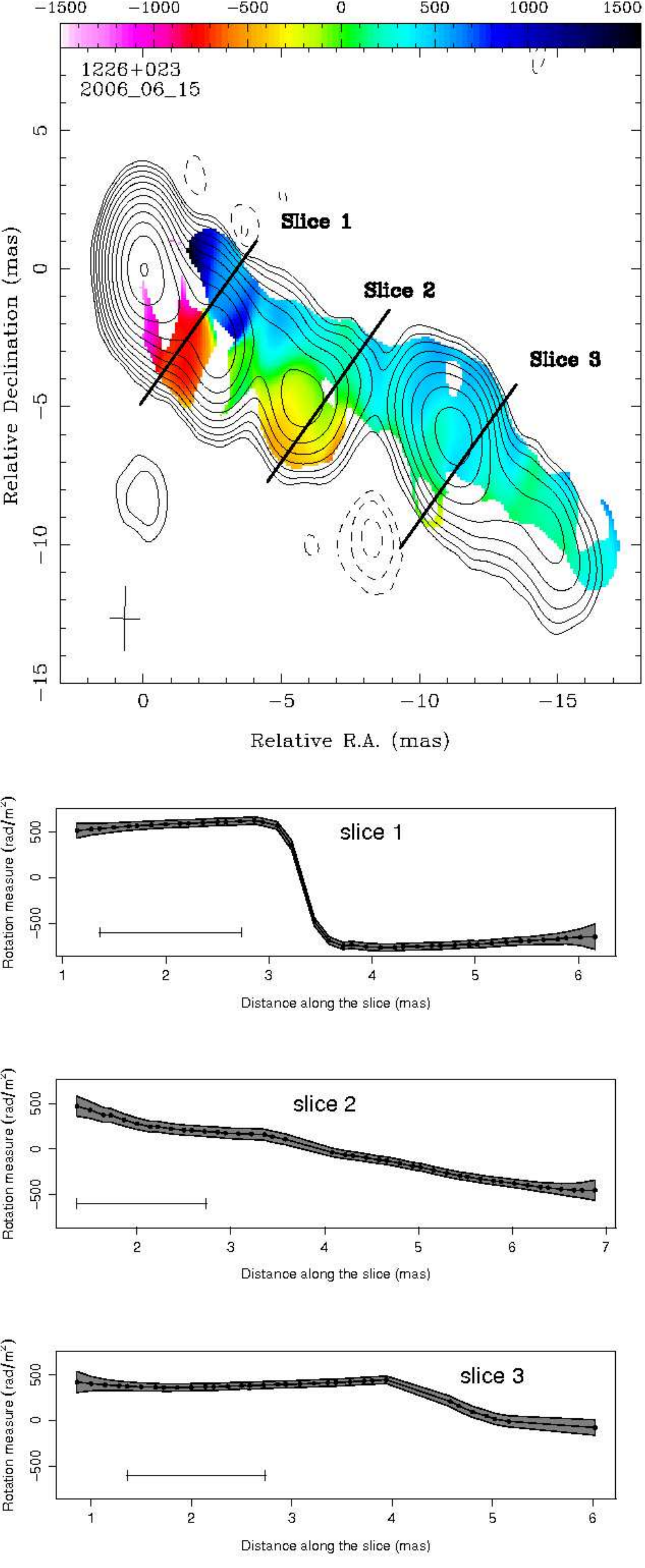}
\caption{Rotation measure maps of 3C 273 with high-RM pixels in the cores blanked for our March (left) and June (right) epochs. The black lines show slices transverse to 
the jet from which the gradients (bottom 3 panels) have been taken. The beam size along the slice is shown in each plot with a scale bar. A color version is available in the online edition of the Journal. \label{fig:1226grad}}
\end{center}
\end{figure*}

Following the above guidelines, we examined all our RM maps in detail.
Our observations show a clear gradient across the jet of 3C~273 (Fig.~\ref{fig:1226grad}), confirming the 
observations of \cite{asada02,asada08a} and \cite{zavala05}. The gradient is detected above the 3$\sigma$ 
level, and the jet is nearly 3 beams wide along the gradient direction.
For the first time the RM is seen to 
change sign over the gradient, which is a further indication of a helical field. We believe we are seeing this now due to a different part of the jet being 
illuminated in the earlier observations, similarly as seen in 3C~120 by \cite{gomez11}. In fact, if we compare the mean jet direction, calculated 
from modelfit components within 7 mas from the core, in our 2006 observations to the 2000 observations by 
\cite{zavala05}, we see a change of 10$^\circ$. We also do not detect as high positive RM values as \cite{zavala05}, who 
see values up to 2000 rad~m$^{-2}$, while our maximum values are near 500 rad~m$^{-2}$. The maximum gradient is 
detected about 3 - 7 mas from the core, where the RM changes from +500 rad~m$^{-2}$ to $-600$ rad~m$^{-2}$. Further 
down the jet the gradient becomes less pronounced and we also detect less polarized jet emission.

\cite{asada02} suggested that the gradient originates from a helical magnetic field in the jet. Based on the large RM values observed in their 
study, \cite{zavala05} preferred an external origin, possibly a sheath around the jet. The main argument they used was that internal Faraday rotation
values of 2000 rad~m$^{-2}$ should cause severe depolarization in a uniform magnetic field, which was not observed. However, combinations of 
different magnetic field configuration and number of lines of sight can possibly explain high RM and complex polarization structure 
(see Sect. \ref{sect:depol}).
\cite{asada08a} report variations in the transverse gradient of 3C 273 between their observations in 1995 and 2002. 
They present several calculations ruling out the narrow line region as the origin of the Faraday rotation due to 
the variability and suggest the variations are caused by the external slower moving sheath changing over time scale of several years. 
Our observations are only three months apart, but there are still differences between the maps, especially 
in the region 2 - 5 mas from the core on the South side of the jet as can be seen in the values of slice 1. In the first epoch the RM changes from 
+450 to $-300$ rad~m$^{-2}$ but in our second epoch the change is from +500 to $-760$ rad~m$^{-2}$. The large discrepancy in the 
negative values is also seen when examining component 14 in Table \ref{comptable}, located 2.7~mas from the core in Fig.~\ref{fig:1226grad}. The component has not moved 
more than 0.2 mas (0.6 pc) over the two epochs, and still the values differ by over 400 rad~m$^{-2}$ so the change cannot be caused by 
the component illuminating a different part of the Faraday screen. If the scale of RM variations in the screen were this small, we would not expect to see consistent 
RM values over the jet or well-defined gradients.

This component also shows clear depolarization between 15 and 8.1\,GHz as the fractional polarization drops from 9.2 to 2.8\% in the first epoch and 
from 8.4 to 1.4\% in the second. This component is not on our list of isolated jet components due to the proximity of the brighter (in total intensity)
component 12 and is therefore not shown in Fig.~\ref{fig:rm_b}. However, if we fit the polarization data similarly as in 
Sect. \ref{sect:depol}, we obtain values $-\sqrt{|b|}$ of $-$843 and $-$1001 for the first and second epoch, respectively. 
This would make the component the most depolarized in our sample. It is very difficult to explain such fast variations with external Faraday rotation, 
and therefore it is possible that we are seeing internal Faraday rotation in this case. 
Another alternative is that the variations we observe in the RM are due to interaction of 
the jet with a sheath. \cite{chen05} observed variations of comparable magnitude over similar time scales in 3C~273. He proposes
that the fast variations could be caused by expansion of the components compressing the surrounding medium increasing the 
magnetic field and electron density. However, it is difficult to explain the complex depolarization observations with this model. 

\begin{figure}[htp]
\begin{center}
\includegraphics[scale=0.6]{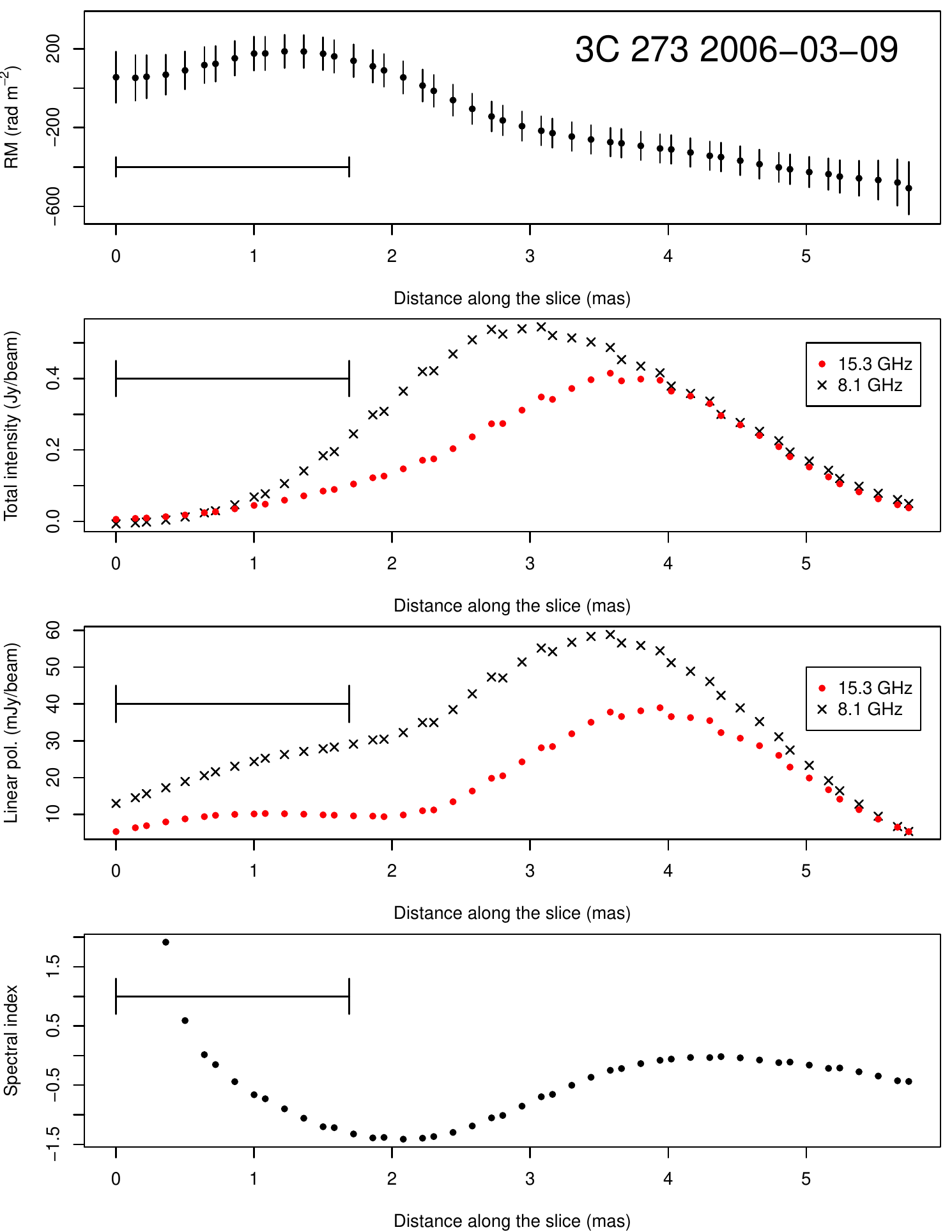}
\caption{Transverse slices of 3C~273 along slice 2 in Fig.~\ref{fig:1226grad} left panel for RM (top panel), total intensity (2nd panel from the top) linearly polarized intensity (3rd panel from the top), and spectral index between 15.3 and 8.1\,GHz (bottom panel). 15.3\,GHz observations are shown in (red) filled circles and 8.1\,GHz observations in (black) crosses. The rms error in total intensity is 1.2 mJy/beam and 4.0 mJy/beam at 15.3 and 8.1\,GHz, respectively. The rms error in linear polarization is  0.6 and 0.8 mJy/beam at 15.3 and 8.1\,GHz, respectively. The error in spectral index varies from 0.004 in the center of the jet to 0.3 at the edges. The beam size along the slice is shown in each plot with a scale bar. Color version is available in the online edition of the Journal.\label{fig:ipslice}}
\end{center}
\end{figure}

The observational signatures of large-scale helical magnetic fields were recently studied from a theoretical perspective by \citet{clausen11}. 
They suggest that the best way to distinguish signatures of helical 
fields from interaction with external medium is to look for correlated behavior in the total intensity, spectral index, and 
polarization profiles. In their model, the total intensity, polarization and spectral index should have skewed profiles, so that 
a tail in total intensity is found on the same side of the jet where the polarization is lower and where the spectral index is steeper. 
The skewness of the profiles depends on the Lorentz factor and viewing angle of the jet.
We have studied the total intensity, polarization and spectral index profiles at the locations of the gradients in 3C~273 and show them 
for slice 2 in our March 2006 epoch in Fig.~\ref{fig:ipslice}. It is obvious that the profiles are skewed in the 
way predicted by \citet{clausen11}, supporting models with helical magnetic fields in 3C~273. The spectral index gradient was also 
detected by \cite{savolainen08}. Unfortunately, such skewed signatures are in general difficult to detect due to beam effects and large 
errors in polarization towards the jet edges, and even in our observations the signature is not as clear in all jet locations. 
Our higher resolution and more sensitive VLBA follow-up observations will enable us to study this further.

\begin{figure*}[ht!]
\begin{center}
\epsscale{0.9}
\plottwo{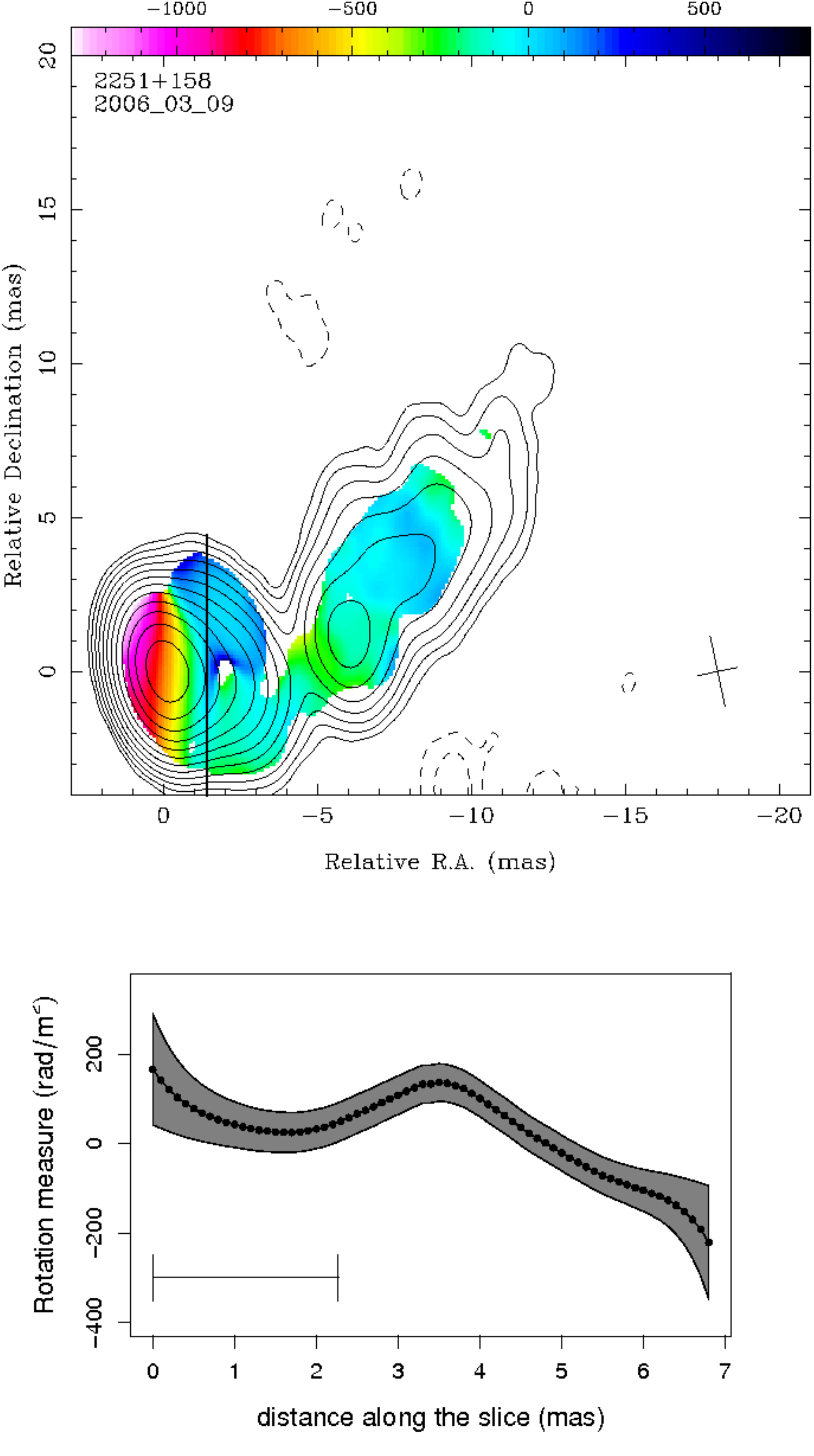}{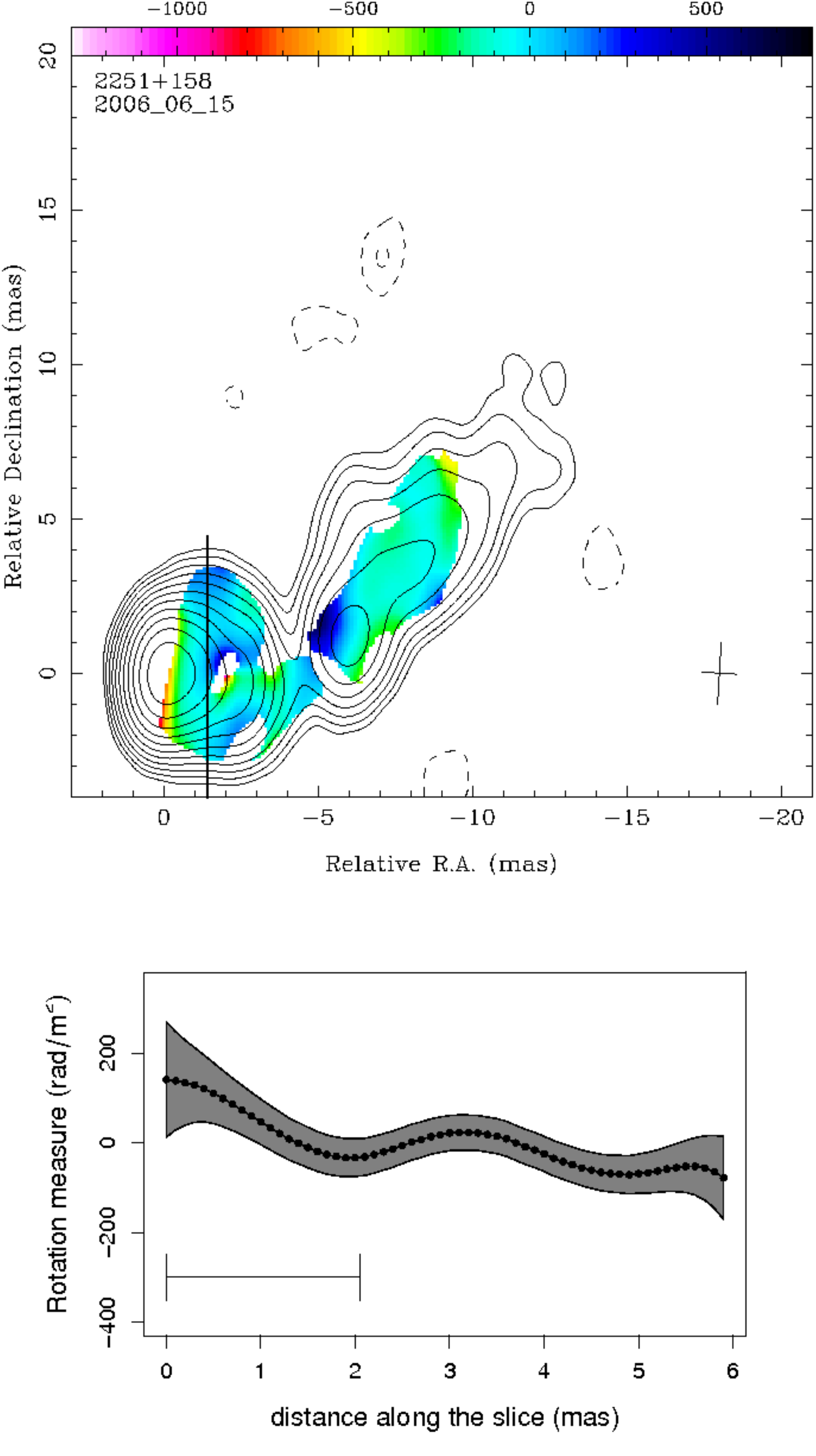}
\caption{Rotation measure maps of 3C 454.3 with high-RM pixels blanked for our March (left) and June (right) epochs. The beam size along the slice is shown in each plot with a scale bar. Color version is available in the online edition of the Journal.\label{fig:2251_grad}}
\end{center}
\end{figure*}

Another source which shows a significant transverse gradient in its March 2006 epoch is 3C~454.3, shown in Fig.~\ref{fig:2251_grad}. 
The gradient is seen between 1 - 3 mas from the core and exceeds 3$\sigma$. The magnitude of the gradient varies slightly depending on the 
chosen location with a maximum of about 63 rad~m$^{-2}$mas$^{-1}$. In the slice of Fig.~\ref{fig:2251_grad} it is about 57 rad~m$^{-2}$mas$^{-1}$ 
when the jet is 3 beams wide. In our second epoch in June 2006, the gradient is not as 
clear, but that can be attributed to lower data quality in the 8.1\,GHz band during that epoch, as a smaller region of the jet is visible above the noise level. Another complication arises from the 
bending of the jet because it is difficult to determine the transverse direction when the jet bends. In 3C~454.3 we have chosen the 
local jet direction when studying the gradient, but it is obvious that the gradient is no longer seen further down in the jet after it bends. 
We see variations in the RMs of the jet components 1 and 2 which are 8.6 and 6.1 mas from the core, respectively, as well as inverse depolarization in several 
components, and this could point towards internal Faraday rotation as seen in 3C~273. Interestingly, Fig. 27 in \cite{zavala03} seems to hint at a 
RM gradient in the same direction, although this was not reported by \cite{zavala03}, who were not concerned with the possible presence of transverse gradients in their RM maps. When our observations are combined 
with total intensity, polarization and spectral index observations (Zamaninasab et al. 2012, in prep.) they seem to follow a modification of the \citet{clausen11} model.
Details of this modeling will be presented in Zamaninasab et al. 

\begin{figure*}[ht!]
\begin{center}
\epsscale{0.9}
\plottwo{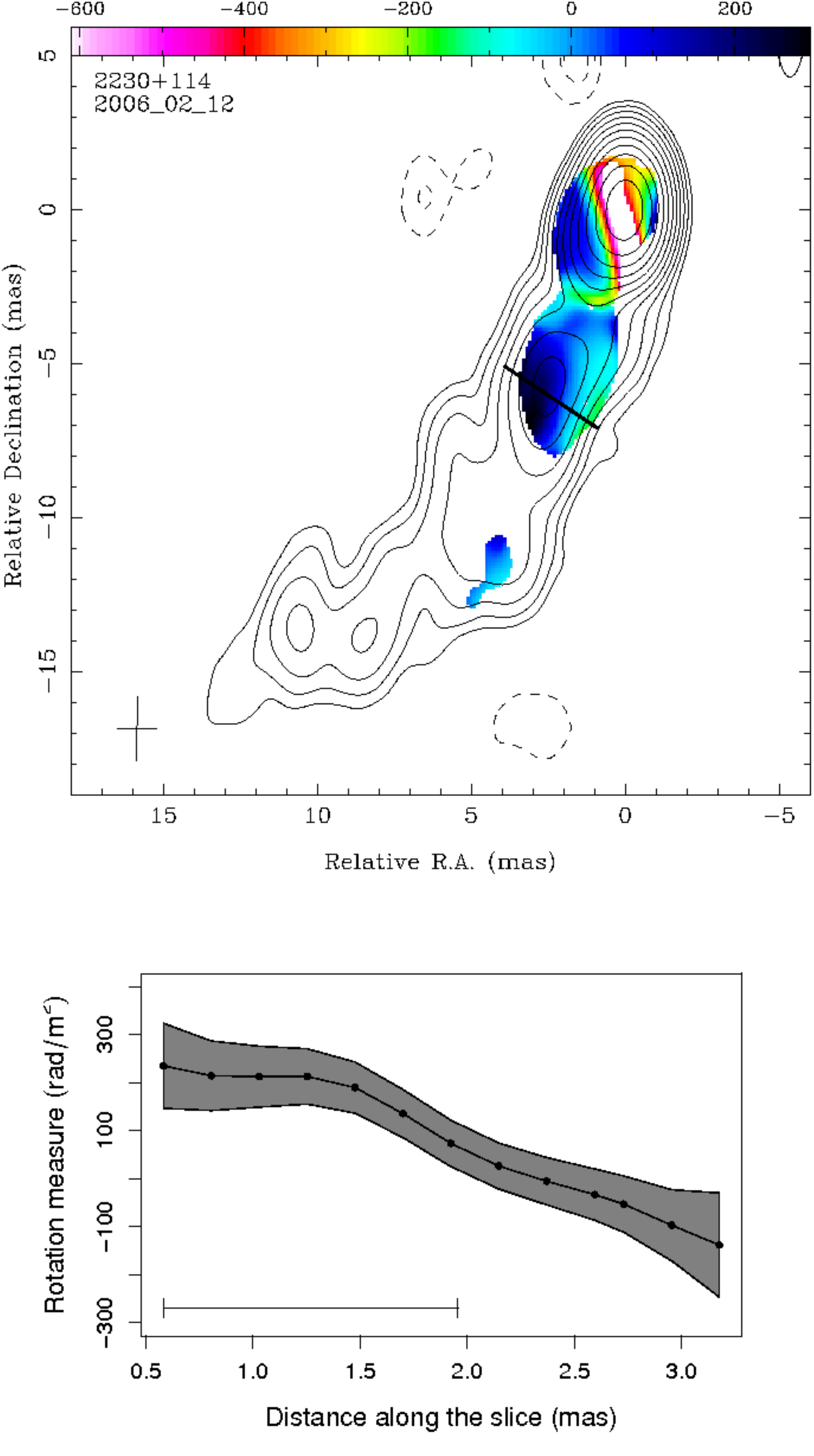}{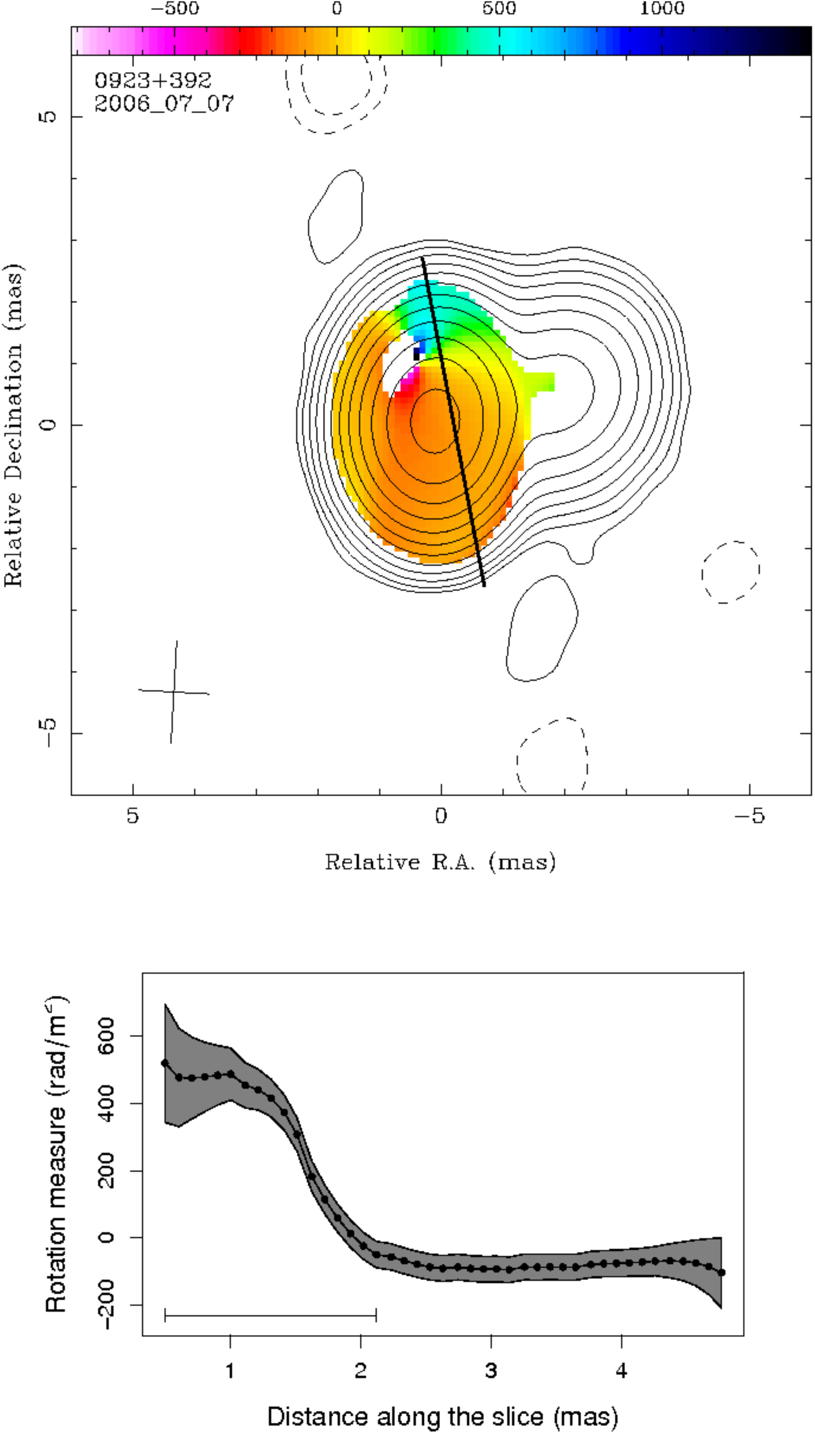}
\caption{RM maps and transverse RM slices of 2230+114 (left) and 0923+392 (right). Beam size along the slice is shown in each plot with a scale bar. Color version is available in the online edition of the Journal.\label{fig:2230_0923grad}}
\end{center}
\end{figure*}

Additionally, several other sources show interesting transverse RM structures. In 2230+114 we detect a gradient 
 of 144 rad~m$^{-2}$mas$^{-1}$ at 3$\sigma$ level about 7 mas from the core where the jet is 1.9 beams wide
(Fig.~\ref{fig:2230_0923grad}). Based on our criteria 
above, this can be considered as a significant gradient, but it is more difficult to tie it to any specific 
model because the region over which the gradient is detected is small. Therefore we have included the source 
in follow-up VLBA observations which are designed to give better sensitivity and resolution in hope of confirming the 
gradient and modeling it in more detail. The results of the follow-up observations will be presented in a separate paper. 

Similarly, 0923+392 shows a significant total gradient of over 624 rad~m$^{-2}$ where the polarized jet is 2.6 beams wide (Fig.~\ref{fig:2230_0923grad}). The 
gradient, however, is confined within one beam width and is about 385 rad~m$^{-2}$~mas$^{-1}$ and depends on the 
high RM region at the Northern side of the jet.
Interestingly, the high RM values are seen right where the jet is shown to bend outside the line-of-sight of the observer
\citep{alberdi00} and therefore could be a sign of the jet interacting with the intergalactic medium. The polarization structure we observe is 
also consistent with the observations of \citet{alberdi00} where the change in polarization over several years is 
shown to be consistent with a moving component interacting with a stationary feature where the 
jet bends. Another alternative could be that the three dimensional geometry of the jet is complex and the North and South side of the jet 
probe different regions along the jet so that the other is further downstream and we are seeing effects of different optical thickness.
With the present data it is difficult to distinguish between the alternative models. Therefore this source was also included in our follow-up VLBA observations.

We do see hints of transverse gradients in three other sources as well, but none of these fulfill all of our criteria.
For example, in 2134+004 (Fig.~\ref{fig1}.139) the change in the RM is more than 2$\sigma$, but it 
is detected only at a single location of the jet and is only ~ 65 rad~m$^{-2}$~mas$^{-1}$ where the jet is 1.8 beams wide, 
so we do not call it significant. In 0945+408 the RM map (Fig.~\ref{fig1}.67) shows clearly two different RM regions, and even though the 
jet is more than 2 beams wide, the gradient is within 2$\sigma$ errors. A similar gradient is seen in 1641+399 (Fig.~\ref{fig1}.110)
where it also is within 2$\sigma$ errors. Therefore we cannot call the gradients in these sources significant but have 
included them in the follow-up higher sensitivity observations for further study.

We have also compared our maps to other studies reporting 
transverse RM gradients in the sources in our sample. \cite{asada08b} and \cite{reichstein11} report a transverse 
gradient of a few hundred rad~m$^{-2}$ in the source 0333+321 in VLBA observations between 5 and 8\,GHz. 
We do see a gradient of similar magnitude (Fig.~\ref{fig1}.21), but 
it extends only one beam width across the jet, and it is also very much dependent on low S/N-jet edge pixels. 
Additionally, it does not extend over the whole length of the jet. Some of this may be due to our 
higher observing frequency which causes us to detect less polarized emission across the jet. Therefore we 
do not consider this a robust gradient in our images but merely suggestive of a possible gradient. 
\cite{reichstein11} also report gradients in both the core and jet of 1150+812 but in both 
cases the slices they take are less than two beams wide at the location of the gradient. We do see a change 
in RM values in a similar direction (Fig.~\ref{fig1}.79) but again it is very much dependent on unreliable edge pixels and also the jet is 
only one beam width across at our 8\,GHz resolution so that we do not consider the gradient to be robust.

In 3C~120 (Fig.~\ref{fig1}.28) we do not detect the transverse RM gradient that was seen by \cite{gomez08} in observations 
made in 2001 at 15, 22 and 43\,GHz. This it not unexpected because \cite{gomez11} demonstrate in their Fig.~9 how a different 
region of the jet in 3C~120 is seen in their 2001 and 2007 observations at 15\,GHz. Our observations are 
close to their latter epoch, and therefore we also see a different part of the jet and do not detect the 
gradient.

Other reported sources include 0836+710, for which \cite{asada10} find a possible gradient of $\sim$150 rad~m$^{-2}$ in 
observations between 4.6 and 8.6\,GHz. 
The gradient is at a location in which we were not able to obtain good $\lambda^2$-fits (Fig.~\ref{fig1}.59) and therefore we cannot 
confirm the result. In any case, the jet is not well-resolved transversely in our observations. Another source 
of interest is 1803+784, for which \cite{mahmud09} report RM gradients in four epochs observed with several different 
frequency setups from 4.6 to 43\,GHz. In our map we do not 
see any indication of a gradient (Fig.~\ref{fig1}.122) , but it must be noted that our map does not extend as far down the jet 
as some of theirs. This is because in their lower frequency maps they detect polarization further down the jet 
compared to our 15\,GHz observations.
\cite{osullivan09a} 
studied the Faraday rotation in six blazars observed in July 2006 between 4.6 and 43\,GHz, and detected a transverse gradient in two of them
in multiple frequency ranges. In 0954+658 
they detect a gradient which is mainly visible in the low-frequency RM map from 4.6 to 15.4\,GHz. Even in this 
case the jet is unresolved. In the 7.9 - 15.4\,GHz RM map, very close to our resolution, they see an indication of 
the gradient, but again the jet is unresolved. Our map (Fig.~\ref{fig1}.69) shows similar changes in the RM from positive to near 0 rad~m$^{-2}$ at the 
same location, but the jet is less than two beam widths wide and the gradient is also very much dependent on the edge pixels, 
which we have noted in the previous section to be unreliable.
We cannot compare the core RM values because we do not find good $\lambda^2$-fits in the core 
region. We also detect changes in the jet RM because in our observations the jet RM varies between 0 and 600 rad~m$^{-2}$ 
while in the same frequency range map in \cite{osullivan09a}  the values are between 0 and 200 rad~m$^{-2}$. This means that  
we are either seeing a different part of the jet, or the RM has changed over a time span of three months. 
The gradient they detect in 1156+295 is
highly dependent on a few unreliable edge pixels. Additionally it is within 
one beam width from the core and therefore unreliable as shown by the simulations of \cite{broderick10}. We 
do not detect any polarized jet emission in this source at our epoch (Fig.~\ref{fig1}.79) and therefore do not search for gradients. We do not obtain good $\lambda^2$-fits 
in the core but the few pixels where we obtain acceptable fits agree with the RM values of \cite{osullivan09a}.

\cite{contopoulos09} list 29 sources for which a gradient can be seen in RM maps collected from the literature.
Our sample includes 25 of these and we have looked for gradients in the 18 sources in which 
\cite{contopoulos09} report a gradient at least 2 mas from the core. These include 3C~273, 3C~454.3 and 2230+114 and two others which are in our list of 
sources showing hints of gradients. In the remaining 14 we either do not detect RM outside a beam width 
from the core (5 sources) or do not see any indication of a gradient in the jet at our epochs.  We note that some of these 
observations were done at longer wavelengths and therefore may have been more sensitive to polarization and Faraday rotation.

Based on our results of a very large sample of RM maps we conclude that robust mas-scale RM 
gradients are difficult to detect at $> 3\sigma$ level in the inner jets of AGN. The major cause for this is that the majority of the jets we observe are not 
transversely well-resolved and therefore we do not consider the gradients as robust based on our 
detailed simulations in Appendix \ref{app:grad}.  We note that multiepoch observations may help to confirm other gradients
because in principle, detections of 
three 2$\sigma$ gradients in the same source at the same location would give an overall significance of 3$\sigma$. However, 
this requires that the significance of the individual detections is determined using statistically correct error bars derived from 
error propagation of Q and U rms error and accounting for additional D-term and CLEAN errors as described in Appendix \ref{app:polerr}.
In addition to the four sources where we detect significant transverse gradients, our 
sample includes only five other sources which have polarized jets wider than 
two beam widths and show no sign of a transverse gradient. 
The reduced sensitivity and angular resolution makes the detection of gradients 
very difficult in objects at higher redshifts. For example, if we restore the maps of 3C~273 with a beam size corresponding to angular 
resolution of a $z=0.5$ object and reduce the flux density by a factor of 10 to achieve a typical flux density of a source at $z=0.5$, the 
gradient would no longer be significant. Higher sensitivity and better angular resolution observations 
would help to solve the problem. Given the existing constraints of the current VLBI arrays, this is a strong science motivation for 
high-sensitivity space-VLBI.

\subsection{Time variability}\label{sect:variability}
As was noted in the previous section, there is clearly variability in the RM values over time scales of 
years or even months. If the Faraday rotation is caused by an external screen very far away from the jet, 
for example in intergalactic clouds or even in the narrow line region of the AGN, the RMs observed in fixed 
locations (with respect to the core) in the jets of AGN should remain constant over time scales of years. In order 
for external clouds to cause variations over time scales of years, they should be so small in size that 
the number density is not large enough to cause significant variations in the observed RM. 
Therefore it is important to verify that the variations are seen in the same parts of the jet and are not due to changes in the 
jet position angle \citep{gomez11}. 

Twenty sources in our sample were observed twice during the 12 month period, and in ten AGNs we measured  
RM values in both of the epochs. Additionally, we can compare our observations to the 8-15\,GHz observations of \cite{taylor98, taylor00} and 
\cite{zavala02,zavala03, zavala04} where RM maps for 36 sources in our sample, mainly quasars, are shown (3C 279 is shown in three of them).
These observations were obtained between 1997 and 2000, and our comparisons are therefore affected by the long gap between 
our observations. Another factor that will affect at least some sources is that Zavala \& Taylor did not remove the contribution of 
Galactic Faraday rotation from their maps. This affects, for example, 2200+420 where \citet{zavala03} observe a jet RM of  
$-$287 rad~m$^{-2}$ which will reduce to $-$136 rad~m$^{-2}$ if the Galactic Faraday rotation is accounted for. This value is 
consistent within our error bars with our jet RM values of about 0 rad~m$^{-2}$. RM maps of five of our sources were obtained in
July 2006 by \cite{osullivan09a} which allows us to probe shorter time scale variations. They include RM maps 
between 7.9 and 15\,GHz for most of their sources, close to our resolution.
Out of these five, 0954+658 and 1156+295 were already discussed in the previous section.

\subsubsection{Variability in the core RMs}
In four sources which were observed twice during our program (0215+015, 0716+714, 0834$-$210, 0847$-$120), we detect RM values only in small regions or 
near the core and the values are all within 1$\sigma$ error bars in the two epochs. In 0219+428 and 2200+420 our 
observations agree very well. We do not detect good $\lambda^2$-fits in the cores of these two sources in 
either of the epochs. In the case of 2200+420 this agrees well with observations of \citet{mutel05} who 
observed 2200+420 over nine epochs between 1997 and 2002 at 15, 22 and 43\,GHz, and did not detect good $\lambda^2$-fits in 
four of their epochs. They showed that this is due to blending of multiple components in the core region. The 
components were seen as separated in their 22 and 43\,GHz observations but were blending together when 
restored with the 15\,GHz beam. This could also be the case in 1418+546, where four Gaussian components are required 
to fit the region 2 mas from the core at 15 GHz. \cite{osullivan09a} also observed this source in July 2006 between 4.6 and 43\,GHz. 
In our February 2006 epoch we obtain a similar core RM value as they do, although we do not 
detect the positive RM patch as seen in their 7.9--15.3\,GHz map. In our November 2006 map the core values change by
200 rad~m$^{-2}$, compared to our previous epoch and their map from July. In this epoch, we also detect the positive 
RM patch seen in their map.

\cite{osullivan09a} observed 1749+096 in July 2006 but they do not obtain any good $\lambda^2$ fits, which is surprising because we 
detect RM values in the core of 1749+096 in our June epoch, observed only two weeks earlier. They attribute the inability to obtain good fits to a 
possible flare in the source which may affect the polarization structure. Indeed, their peak flux in total intensity in 
the 7.9\,GHz band is 4.2 Jy/beam while in our 8.1\,GHz observation it is 3.5 Jy/beam so that a flare is probably 
on-going and could affect their July observations. In fact, a flare peaking in October 2006 is seen in the 15\,GHz MOJAVE observations\footnote{http://www.physics.purdue.edu/astro/MOJAVE/sourcepages/1749+096.shtml}. 
Additionally, our wavelength coverage is not as wide as theirs 
so it is possible that we do not detect the non-$\lambda^2$ behavior they see. 

Variations in the core RMs are also seen in comparison of our observations to those 
of \cite{taylor98, taylor00} and \cite{zavala03, zavala04}. In general, the core values between our maps and theirs differ significantly and there are 
only five sources where we detect similar core RM values. This can also be attributed to the blending of components within the 
finite beam. Alternatively, if the Faraday rotation is internal to the jet, changes in the particle density or magnetic field strength 
due to newly emerging components could cause the variations.  In six sources the core RM in our maps has a different sign than in their 
observations which indicates a change in the direction of the line-of-sight component of the magnetic field. However, due to the long 
gap between our observations and the complexity of the core regions in these sources, it is difficult to distinguish if this is due to a sign-reversal 
in an ordered magnetic field or due to changes in properties of new components.

\subsubsection{Variability in the jet RMs}
3C~273 and 3C~454.3 were seen to have large variations in the jet RMs over time scales of three month in our observations, which was 
discussed in the previous section. In some sources it is clear that we are seeing a different part of the jet at successive epochs
and sometimes we do not detect the polarized jet emission at all, or vice versa. This is seen, for example, in complex sources such 
as 1253$-$055 and 1418+456 where we do not see the exact same part of the jet because the same jet components 
are not detected or the components have moved significantly between the two epochs.  

In many sources we still see fairly similar jet RM values within the error bars. In the jet of 2200+420 we observe RM values consistent 
with 0 rad~m$^{-2}$, in agreement with \citet{mutel05}. This source was observed by \cite{osullivan09a} in July 2006 between 4.6 and 43\,GHz where 
they obtain slightly higher RM values in the jet than we do. However, these are still for the most part within the error bars 
of our observations. Only in 2230+114 do we see differing 
values in the jet RM, even though it looks like we are looking at the same portions of the jet. \cite{taylor00} detect 
a RM of $-$185 rad $m^{-2}$ in the jet of 2230+114 (component C in their paper) which is almost at the same location 
as our component 3 with RM of +173 rad~m$^{-2}$.  The different sign in the observed RMs indicates a change of direction in the 
line-of-sight component of the magnetic field. It will be interesting to compare these to our follow-up observations to determine if the 
sign has remained the same.

Even though at first glance it looks like the RM values in the jets of AGN change over time scales of 
several years, a detailed comparison taking the differing locations of the polarized components into account shows 
that we do not detect significant variations in the jet RMs of the majority of AGN. The fact that we detect non-zero RMs in the jets 
even after the Galactic Faraday rotation contribution is taken into account suggests that the RM is occurring outside our own Galaxy, 
for example in the narrow line region of the AGN \citep[e.g.][]{zavala03}. In three sources which show signs of transverse rotation measure gradients, 
3C~273, 2230+114 and 3C~454.3, RM time variations are seen in the jet as well.  In 3C~273 and 3C~454.3 the variations happen over 
time scales of three months, pointing to either internal Faraday rotation or 
interaction between the jet and the Faraday rotating material, possibly a sheath around the jet, as discussed in more detail in Sect.~\ref{sect:grad}. 
If the rotation is internal, the rotating plasma is either the low-energy end of the relativistic electron population or thermal plasma intermixed with 
the emitting plasma, and therefore fast changes in the magnetic field of the jet can cause fast variations in the observed RM.
If the jet is surrounded by a sheath, the sheath must be mildly relativistic because we do not detect counter-jet emission in these sources. Therefore 
over time scales of years, it is possible that the RM variations are caused by changes in the screen itself \citep[e.g.][]{asada08a}. This could explain the variability in 2230+114 but not in the two other sources where the gas in the sheath would not have moved sufficiently over the time scale of three months to cause significant variations. In sources with fast apparent superluminal motion, it may be possible to observe variations over time scales of 
months when the components illuminate a different part of the Faraday screen \citep[e.g.][]{gomez11}. In these sources the same RM 
should be seen when another component passes the exactly same part of the jet which could then be studied with frequent multiepoch observations. 
This seems not to be the case in 3C~273 or 3C~454.3 where the components have only moved about 0.1~mas between the two epochs.
We note that if the variations happen only in small parts of the jets, our ability to detect variability is easier for nearby sources with resolved jets. 

\subsection{Comparison to simulations}
\begin{figure*}[htp]
\begin{center}
\includegraphics[scale=0.97]{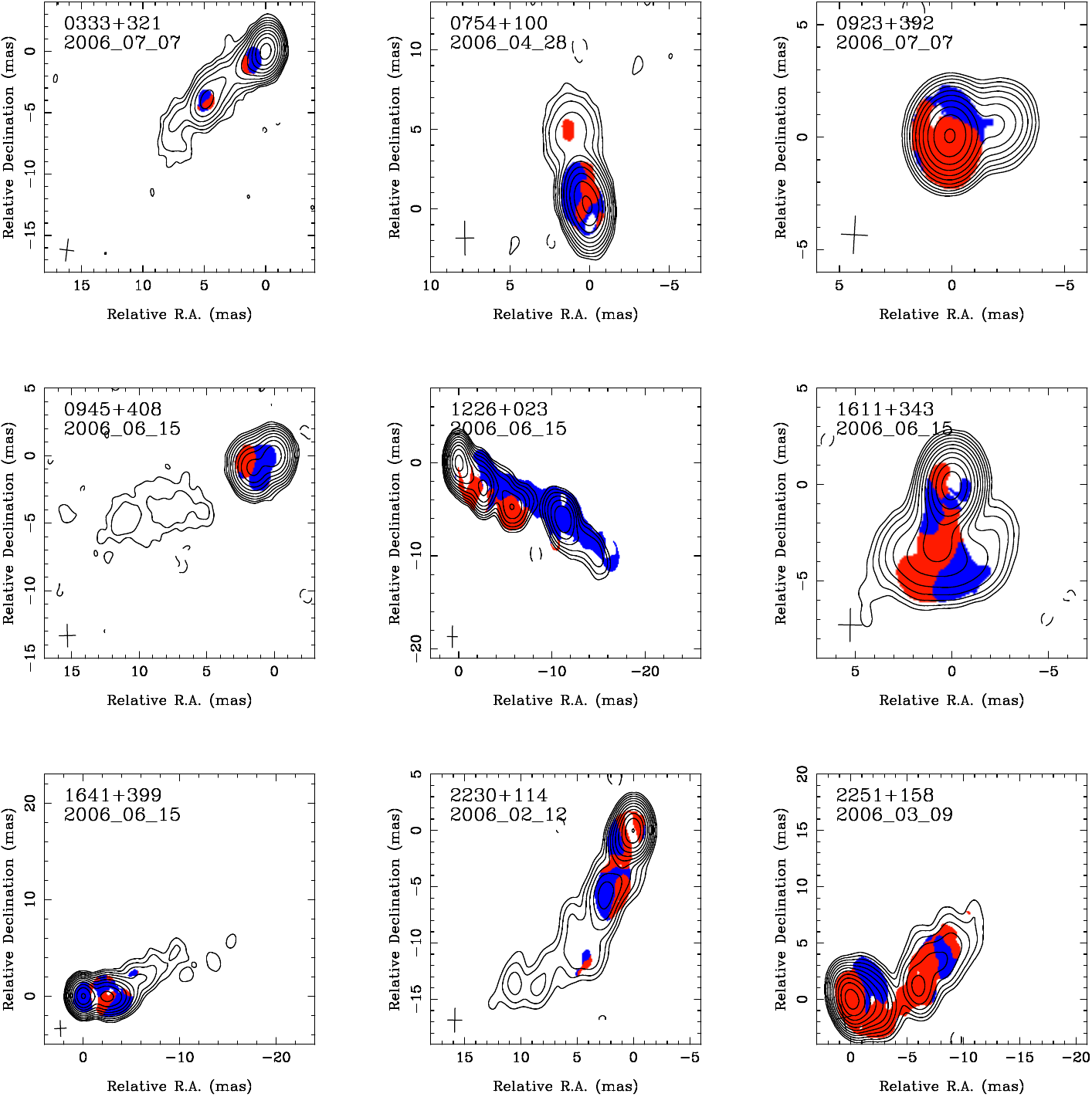}
\end{center}
\caption{Maps of the rotation measure sign in nine example objects showing bilateral structure (see text for details). Dark gray (blue) areas are positive RM and light gray (red) areas negative RM. Color version is available in the online edition of the Journal.\label{fig:signs_a}}
\end{figure*}

Recently, using GRMHD simulations, \cite{broderick10} produced simulated RM maps for AGN jets with 
large scale toroidally dominated magnetic fields. They produced maps with different resolutions and beam 
sizes to additionally study the effect of finite beam size on RM observations. Their results showed that any 
RM gradients observed within a beam width from the core are unreliable  at resolutions below 43\,GHz at VLBA. Further down the jet  
it should be possible to detect true RM gradients if the magnetic field is toroidal. Formally, with infinite sensitivity, even in unresolved 
jets it may be possible to detect gradients of the right sign, with the magnitude severely 
suppressed. They did not, however, take the effect of noise into account in their simulations, which we have shown in Appendix~\ref{app:grad} 
to have a major effect on the reliability of gradients in unresolved jets.
They also showed that if the toroidal component of the field is made to vanish, any significant 
gradients will vanish too. They conclude that linear, resolved transverse gradients are due to large-scale 
toroidally dominated magnetic fields within the Faraday screen. This provides us a way to compare our 
observations with their simulations. It is difficult to make quantitative comparisons, but \cite{broderick10} show that even maps of the 
RM sign should show the bilateral structure if the toroidal magnetic field component dominates. Their 
Fig.~12 shows examples of sign maps for different resolutions and other parameters, such as the 
jet viewing angle and black hole mass. Additionally they present a model in which the toroidal field is made to vanish, and a random 
foreground Faraday screen is inserted. 

\begin{figure*}[htp]
\begin{center}
\includegraphics[scale=0.97]{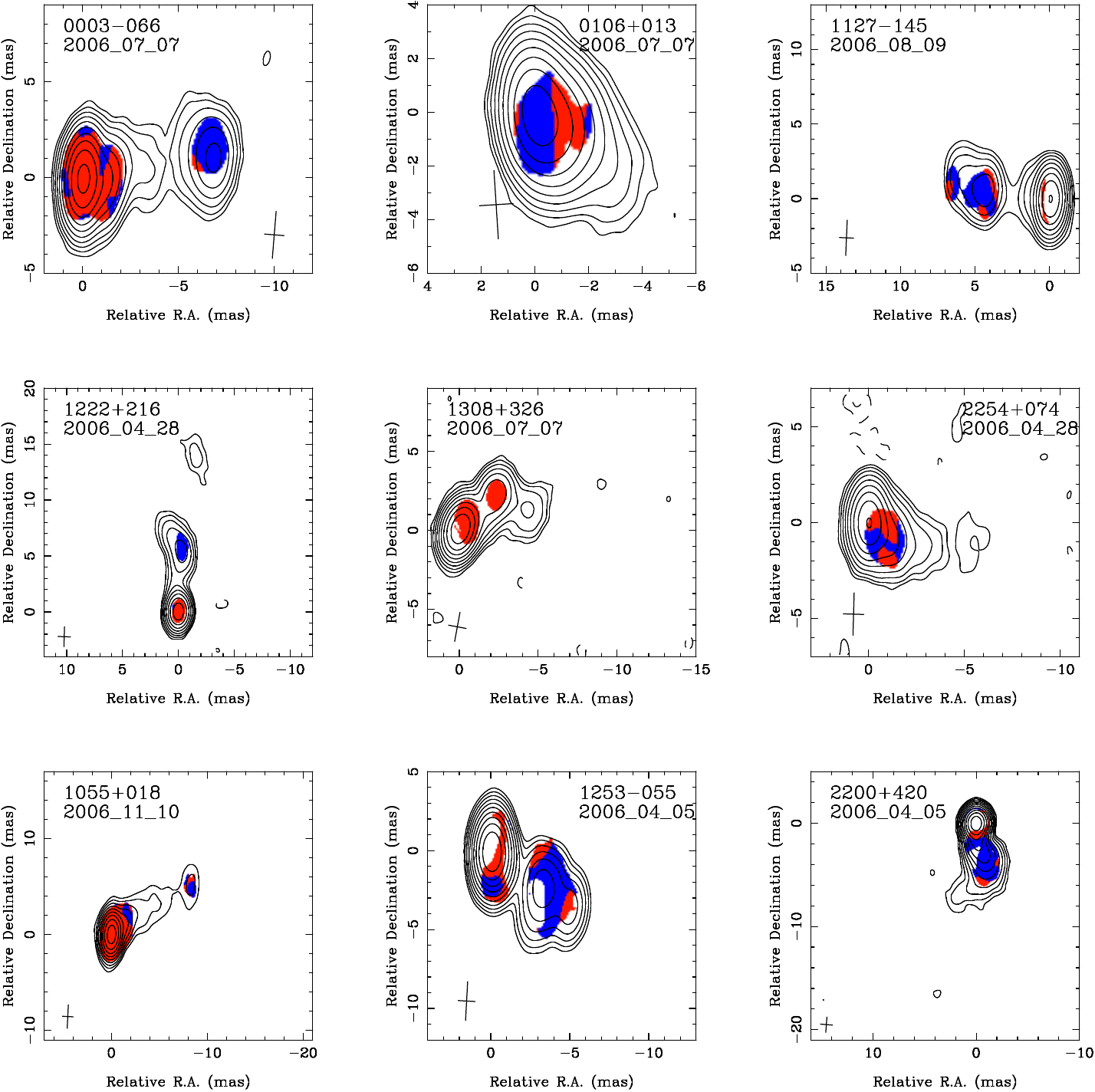}
\end{center}
\caption{Maps of the rotation measure sign in nine example objects that represent the typical sources in our sample (see text for details). Dark gray (blue) areas are positive RM and light gray (red) areas negative RM. Color version is available in the online edition of the Journal.\label{fig:signs_b}}
\end{figure*}

In nine sources  visual inspection of the sign maps reveals clear bilateral structures as shown in Fig.~\ref{fig:signs_a} . These include 3C~273, 3C~454.3, 
0923+392 and 2230+114 in which the gradient is significant based on our detailed analysis in Sect.~\ref{sect:grad}. 
In two other sources (0945+408 and 1641+399), in we which detect the gradient at a 2$\sigma$ level, the sign maps show different signs on 
both sides of the jet along the whole polarized region of the jet. These are candidate sources for toroidally dominated 
large-scale magnetic fields. However, one should note that for example in the case of 1611+343, even though the 
sign map shows a bilateral structure, any possible gradient is within the 1$\sigma$ error bars, and the jet is over 
four beams wide at its widest location. In 0333+321 and 0754+100 the jets are not sufficiently resolved for us to make 
any conclusions about the significance of the gradient.  If the jets are not viewed directly from the side in the jet frame, it may be 
that we are seeing only the positive or negative side of the possible toroidal field. Therefore we have produced sign maps where the 
median jet region RM is subtracted from the map to see if we can detect more sources with bilateral structures. There are only two 
such sources, 1150+812 and 2005+403. \cite{reichstein12} report a sign change centered around zero in 1150+812 in lower frequency 
data between 4.6 and 8.9\,GHz, while in our map the center is about 240 rad~m$^{-2}$. In our map the change in magnitude also 
happens nearly transverse to the jet while in their map it is more along the jet. The jet of 1150+812 is only one beam width 
wide in our observations and therefore we cannot make conclusions about the significance of the gradient. In 2005+403 the jet is 
1.5 beams wide in polarization and the values are centered around $\sim$110 rad~m$^{-2}$, but are all within one $\sigma$ of each other and 
therefore the gradient is not significant.

In Fig.~\ref{fig:signs_b} we show 9 example cases that we believe represent typical types of 
behavior in the sign maps. In many sources we see how the RM sign changes along the jet direction (e.g., 0003$-$066 and 
1222+216). These kind of structures look similar to simulations of unresolved jets with 
zero toroidal magnetic field component or random magnetic field. Multi-epoch studies of changes in the 
RM and polarization structure could give more insights into the nature of the screen and if it is internal or external to the jet.
In the bottom row we show examples of sources in which the jet is transversely resolved but the sign maps 
show a very complex structure similar to the simulated case with resolved jet and zero toroidal 
magnetic field or random foreground screens (e.g., 1253$-$055) or no change in the sign of the RM. The majority of the sources in our sample show sign maps 
similar to Fig.~\ref{fig:signs_b} or sign maps without any change in the sign. 
We note, however, that the majority of our jets are transversely unresolved and therefore it may be 
possible that we simply cannot detect the bilateral structure in our maps.

\section{Conclusions}\label{sect:conclusions}
We have observed a sample of 191 AGN jets (20 at two epochs) at 8.1, 8.4, 12.1, and 15\,GHz with the VLBA to determine the Faraday rotation measures 
in the parsec-scale jets of these sources. One motivation for the study was to find out how much Faraday rotation would 
affect the EVPAs of the MOJAVE program observations at 15\,GHz. The polarization is sufficiently strong to study Faraday rotation in 149 sources (159 maps), 
and in over 80\% of the sources the median observed RM is less than 400 rad~m$^{-2}$, which would rotate the 
15\,GHz EVPAs by less than 10 degrees. Additionally, we do not see significant variations in the jet RMs of most of the sources so that these 
results can be extrapolated to other MOJAVE epochs as well. However, there are several sources with high observed RMs of larger than 
1000 rad~m$^{-2}$ so that when studying polarization in individual sources (especially at frequencies lower than 15\,GHz), one needs to 
take the Faraday rotation into account. The highest intrinsic redshift-corrected absolute RM 8975 rad~m$^{-2}$ is detected in a jet component 0.5~mas from the 
core in the high-redshift quasar 0642+449 at $z=3.396$.

We find the quasars to have generally higher absolute RMs than the BL~Lac objects  (median $|$RM$_\mathrm{obs}|$ 144 rad~m$^{-2}$ vs. 79 rad~m$^{-2}$) and 
the core components in the sources to have higher RMs (median $|$RM$_\mathrm{obs}|$  171 rad~m$^{-2}$) compared to the jets (median $|$RM$_\mathrm{obs}|$ 125 rad~m$^{-2}$) which is also seen in both quasars and BL~Lacs separately. 
In quasars, we also detect a significant negative correlation with the magnitude of the RM and the de-projected distance of the component from the core.

We perform detailed simulations of the measurement error of RM, and focus especially on the effect of noise and finite restoring beam size 
on putative transverse RM gradients. Because of the finite beam size, neighboring pixels are not independent  and this has to be taken into account when interpreting radio interferometric images. For a typical jet there are only few independent measurements across a jet of apparently many 
pixels. Our simulations show that it is possible to obtain spurious, yet significant looking transverse RM gradients if the polarized region in the jet is 
less than two beams wide. We give several guidelines on how the errors in RM maps need to be taken into account to determine the 
reliability of a transverse gradient. The reliability is strongly dependent on the jet width, and the jet should preferably be two beams wide and 
more than 3$\sigma$ in significance to call the gradient reliable. The errors in the RM maps should be calculated from the variance-covariance matrix of the 
least-squares fit where the EVPA errors are determined using error propagation from Stokes Q and U rms errors, 
additionally accounting for effects due to the CLEAN procedure.
Following these guidelines, we detect significant transverse RM gradients in the flat spectrum radio quasars 
3C~273, 3C~454.3, 0923+392 and 2230+114. 
In 3C~273 the RM is for the first time seen to change sign along the transverse slice, giving further support for a helical magnetic field in the jet. 
The main reason why we are not able detect gradients in more sources is that the jets are insufficiently resolved at our lowest observing frequency of 8.1\.GHz. 
In addition to these four sources, there 
are only five which have wide-enough polarized jets where a gradient is not detected. The reduced sensitivity and angular resolution 
makes the detection of gradients very difficult in objects at higher redshifts.
Higher angular resolution observations of much greater sensitivity are needed 
to study this phenomenon further.

Comparison of our RM maps to earlier studies confirm that significant temporal RM variations in the core components of AGNs are common. This could 
due to multiple polarized components blending within the finite beam at different times or intrinsic changes in the magnetic field or particle density 
if the rotation is internal to the jet. Higher resolution observations at higher frequencies are required to 
uncover the true RMs in the cores of many of the sources. In almost all the cases where our jet RM values differ from earlier studies, it can be 
explained with different part of the jet being illuminated by the components at different times. Only in 3C~273, 3C~454.3, and 2230+114 do we detect 
variations which cannot be explained by moving components. In 2230+114 the variations happen on time scales of years, which could be possible 
if the Faraday rotation occurs in a mildly-relativistic sheath around the jet. In 3C~273 and 3C~454.3 we see variations over a time scale of three months which 
are difficult to explain with external Faraday rotation models, so more likely the rotation is internal to the jet. This is further supported by 
observations of inverse depolarization structures in the jet components of these two sources which show the 8\,GHz polarization to be higher than 
at 15\,GHz. This is the opposite trend with wavelength expected from
standard external depolarization and may be explained by combining
internal Faraday rotation with helical or loosely tangled random magnetic
fields \citep{homan12}. However, in the majority
of isolated jet components we studied, the relationship between depolarization and RM can be explained with a purely random 
external Faraday screen viewed through a small number of lines of sight. In these sources the rotation 
could be caused by intergalactic clouds or the narrow line region of the AGN.

\acknowledgments
We thank the anonymous referee for the many suggestions that helped to improve the paper.
The authors wish to acknowledge the other members of the MOJAVE team. We thank
D. Gabuzda, G. Taylor and K. Wiik for useful discussions.
The MOJAVE project is supported under National Science Foundation grant AST-
0807860 and NASA Fermi grant NNX08AV67G. Work at UMRAO has been supported 
by a series of grants from the NSF and NASA and by funds
for operation from the University of Michigan. T. Hovatta was supported in part by the
Jenny and Antti Wihuri foundation. D. Homan was funded by National Science Foundation grant AST-0707693. 
Part of this work was done when T. Savolainen and Y. Y. Kovalev were research 
fellows of the Alexander von Humboldt Foundation. Y. Y. Kovalev was supported in part by the Russian Foundation 
for Basic Research (grant 11-02-00368) and the Dynasty Foundation. The VLBA is a facility of the 
National Science Foundation operated by the National
Radio Astronomy Observatory under cooperative agreement with Associated Universities,
Inc. This research has made use of NASA's Astrophysics Data System, and the NASA/IPAC Extragalactic Database (NED). 
The latter is operated by the Jet Propulsion Laboratory, California Institute of Technology, under contract with the National 
Aeronautics and Space Administration.

{\it Facilities:} \facility{VLBA}, \facility{UMRAO}, \facility{VLA}

\appendix

\section{Effect of image alignment on RM maps}\label{app:align}
To study the effect of image alignment errors on the rotation measure maps, we  
used 12 sources for which we have a very good correspondence between the shifts from 
our 2D cross correlation and from optically thin component positions. In all these 
sources the difference between the two shifts is less than 0.1\,mas.
We assumed the shifts in these cases to be correct and introduced artificial shifts between 
the maps to study the effect on both spectral index and rotation measure maps. The shifts 
were applied to the maps of all the other frequency bands with respect to the 15.3\,GHz maps.
In the case of spectral index maps, the effect was large, with some sources 
showing clear differences with shifts as small as 0.03\,mas. In other sources, 
even a shift of 0.18\,mas, almost twice our median shift for the whole sample, did 
not cause major changes in the spectral index maps. For each of the test sources 
we calculated rotation measure maps for shifts that were seen to create spurious 
features in the spectral index maps. The effect on the rotation measure maps 
was much smaller, and shifts as large as 0.18\,mas did not affect the 
general structure. In most of the cases small differences could be seen 
in the edge pixels or close to patches of low signal to noise, which in 
any case are considered unreliable. In only 
two maps did we see appreciable differences in the jet rotation measure, and in these 
cases the wrong shifts were 0.09 and 0.15\,mas, but those were also in very 
complicated regions of the jet. 
Even in these two cases the general structure of the RM map did not change 
and our conclusions would be unaffected.
Therefore we conclude that even if our image alignment is off 
by 0.15\,mas, it should not affect the results from our rotation measure maps, 
especially as we are not using the edge or low signal-to-noise regions to 
make conclusions about the RM structure.
By using the spectral index map as an additional indicator of 
the goodness of the alignment we ensure that our rotation measure maps are 
not affected by false alignment between the different bands. 

\begin{figure*}[htp]
\begin{center}
\includegraphics[scale=0.6]{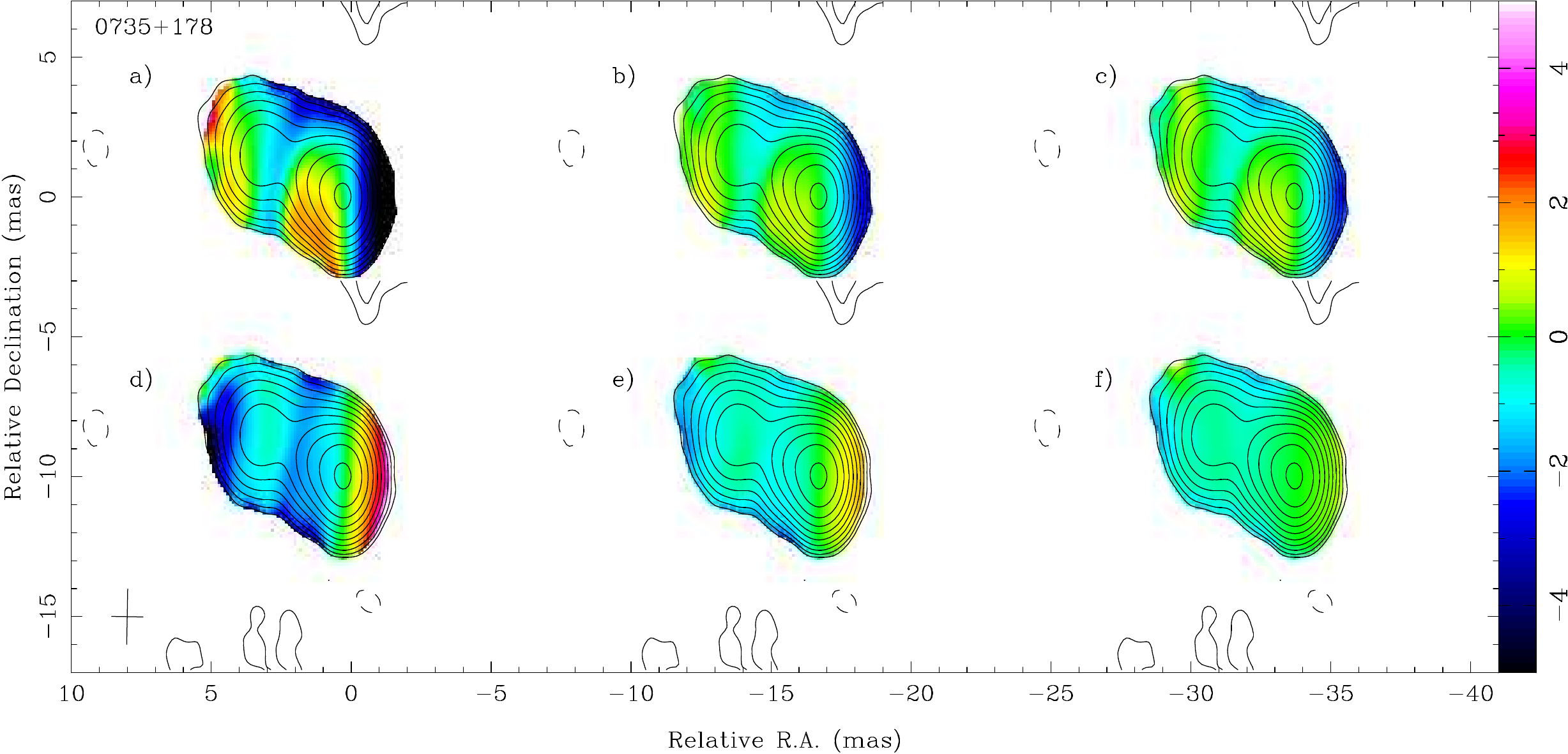}
\end{center}
\caption{Spectral index maps between 15 and 12\,GHz (a and d), 15 and 8.4\,GHz (b and e), 15 and 8.1\,GHz (c and f) aligned by map center (top panel a-c) and by using correct shifts from the 2D-cross correlation (bottom panel d-f). Contours correspond to the total intensity at 15\,GHz. A color version is available in the online edition of the Journal.\label{fig:alpha}}
\end{figure*}

An example of spectral index maps in the case where the alignment is very 
important is shown in Fig.~\ref{fig:alpha}. The figure shows spectral index maps 
without and with shifting between all the bands. The correct shifts were obtained using the 
2D-cross correlation and are 0.39 and 0.03 mas in right ascension and declination between 
the 15 and 12\,GHz bands, 0.45 and 0.03 mas between the 15 and 8.4\,GHz bands and 
0.33 and 0.03 mas between the 15 and 8.1\,GHz bands, and are shown in the bottom panel of 
Fig.~\ref{fig:alpha}. Analogous to results of \cite{kovalev08}, it is clear that the maps without shifting that have been aligned based on the phase center  
(top panel of Fig.~\ref{fig:alpha}) have spurious features, such as optically thin regions apparently upstream of the core 
and a spectral index gradient transverse to the jet, which all disappear when the images are properly aligned.

\section{Errors in polarized flux density, fractional polarization and EVPA}\label{app:polerr}
Errors in polarized flux density $\sigma_p$ are typically assumed to be the average rms error in the Q and 
U images. Additional sources of error are errors due to the CLEAN procedure which can be difficult to 
overcome \citep[e.g.,][]{lister01} and errors due to instrumental polarization (D-terms)  which are not evenly 
distributed across the images \citep{roberts94}.
We used simulations to examine whether the error estimates we use are consistent with 
the above error contributions to the data. The simulations were carried out in several steps:
\begin{enumerate}
\item A Stokes I model of the source was created using calibrated ({\it u,v}) data of a real source and CLEAN in Difmap. 
\item Stokes Q and U models were created by setting Q and U to be known fractions of Stokes I for each CLEAN component. 
Without any noise added, this corresponds to a uniform fractional polarization and EVPA across the source.
\item The original ({\it u,v}) data were loaded into AIPS and the task UVMOD was used to replace the real data with 
the values produced in the previous step. Additionally, random noise of the same order as 
seen in our real data was added. 
\item The UVMOD task was repeated 100 times to produce 100 simulated ({\it u,v}) data sets with random noise added. 
\item In the case of D-term simulations, additional random D-term error with a standard deviation of 0.002, determined from the scatter in the D-terms of our data
(see Section \ref{sect:EVPAcal} for details on the determination of the stable D-terms), was added to each set using the task SPLIT.
\item The simulated ({\it u,v}) data were then imaged in Difmap following the same procedure 
as for the real data to obtain 100 images in Stokes I, Q and U.
\item  The rms in each image was obtained by shifting the map by 1 arcsec and calculating the rms using the command 'imstat' in Difmap.
\end{enumerate}

\begin{figure}[htp]
\begin{center}
\includegraphics[scale=0.5]{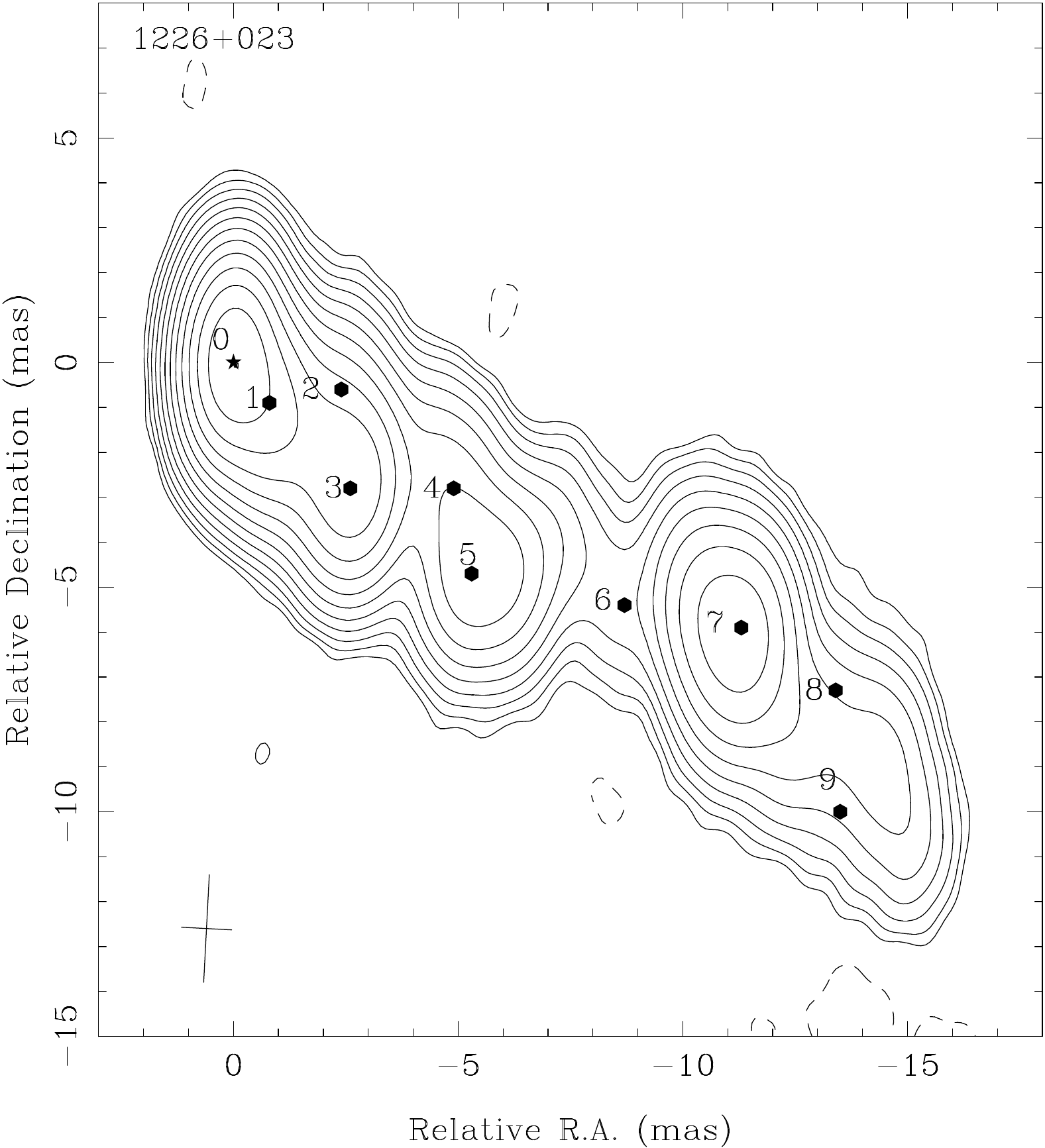}
\end{center}
\caption{Locations of the simulated distributions in the jet of 1226+023 overlaid on the 15 GHz total intensity contours. Star shows the component 
discussed in the error analysis.\label{fig:1226ipoints}}
\end{figure}

\begin{figure}[htp]
\begin{center}
\includegraphics[scale=0.5]{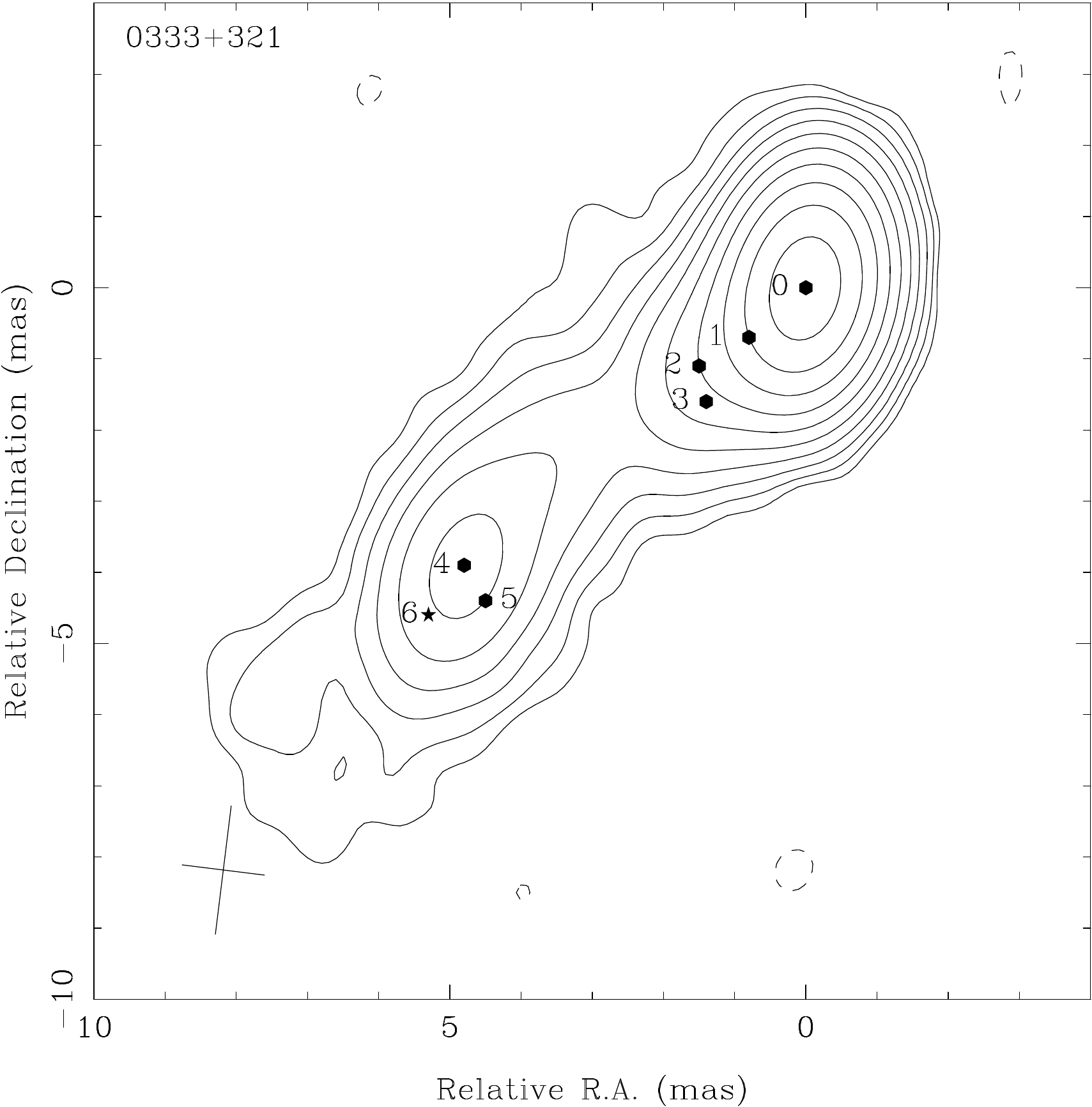}
\end{center}
\caption{Locations of the simulated distributions in the jet of 0333+321 overlaid on the 15 GHz total intensity contours. Star shows the component 
discussed in the error analysis.\label{fig:0333ipoints}}
\end{figure}

The process was repeated for two sources, 0333+321 and 1226+023, to verify that the errors 
are general and do not depend on the specific structure of a given source. In order to address the errors due to the CLEAN 
procedure and the clipping of the ({\it u,v}) coverage of our real data (see Sect. \ref{sect:datared}), the simulations were repeated 
for the original un-clipped 15\,GHz data and the data at all the frequency bands were treated with the same cutoffs as our real images.
We then studied the distributions of polarized flux density, fractional polarization and EVPA in individual jet locations 
shown in Figs.~\ref{fig:1226ipoints} and \ref{fig:0333ipoints}. We chose locations at both the brightest parts of the jet and at the jet edges to 
see how well the error formula reproduces the standard deviation of the simulations. 

For each location we determine the expected value and its expected error for all the parameters. For I, Q and U images the expected values are 
determined from low noise images which have been created using the same simulation procedure but adding only 1\% of the typical noise in the 
task UVMOD. Additionally, the last CLEAN step (6) was replaced by a simple restoration of the model components from step 1
to have comparison images which do not suffer from CLEAN errors.
The expected rms value is taken from the simulated maps with noise added at a location 1 arcsec from the map center. The expected 
p, EVPA and m values are calculated from the low noise images with p = $\sqrt{Q^2 + U^2}$, EVPA = $1/2\tan^{-1}(U/Q)$ and m = p/I. The 
expected error values are calculated using the following equations:
\begin{equation}\label{eq:p_err}
\sigma_p = \frac{\sigma_Q+\sigma_U}{2},
\end{equation}
\begin{equation}\label{eq:evpa_err}
\sigma_\mathrm{EVPA}=\frac{\sqrt{Q^2\sigma_U^2+U^2\sigma_Q^2}}{2(Q^2+U^2)} = \frac{\sigma_p}{2p},
\end{equation}
and
\begin{equation}\label{eq:m_err}
\sigma_m = \frac{\sigma_p}{I},
\end{equation}
where $\sigma_Q$ and $\sigma_U$ are the Q and U rms values in most of the simulations, and Q and U are the Stokes parameters in the 
given pixel. In \ref{eq:m_err} we assume that the $\sigma_I$ term can be neglected as $(p/I^2)^2$ will be very small for all our components.
For the simulations which take the 
instrumental polarization into account, an additional error $\sigma_\mathrm{Dterm}$ defined as
\begin{equation}\label{eq:dterm_err}
\sigma_\mathrm{Dterm} = \frac{0.002}{\sqrt{N_\mathrm{ant}*N_\mathrm{IF}*N_\mathrm{scan}}}\sqrt{I^2 + (0.3*I_\mathrm{peak})^2},
\end{equation}
where $N_\mathrm{ant}$ is the number of antennas, $N_\mathrm{IF}$ is the  number of IFs, $N_\mathrm{scan}$ is the number of scans 
with independent parallactic angles and $I_\mathrm{peak}$ is the peak total intensity of the map \citep{roberts94}. This additional error term is 
added in quadrature to the rms errors. The factor 0.002 in the equation is determined from the scatter of the 
D-terms in our data. In our data the number of antennas is 10, the number of IFs is 4 for 15.4 and 12.1\,GHz,
2 for 8.4 and 8.1\,GHz, and the number of independent scans is 4. The equation is defined and explained by 
\cite{roberts94}, where they study the effects of instrumental polarization with detailed simulations. For example, 
they show that the D-term errors scatter across the image (factor 0.3 in Eq.~\ref{eq:dterm_err}) which seems to be supported by our simulations. 
Therefore we include a contribution from the total intensity peak and the current location in Eq.~\ref{eq:dterm_err}. 

\begin{figure*}[htp]
\begin{center}
\includegraphics[scale=0.6]{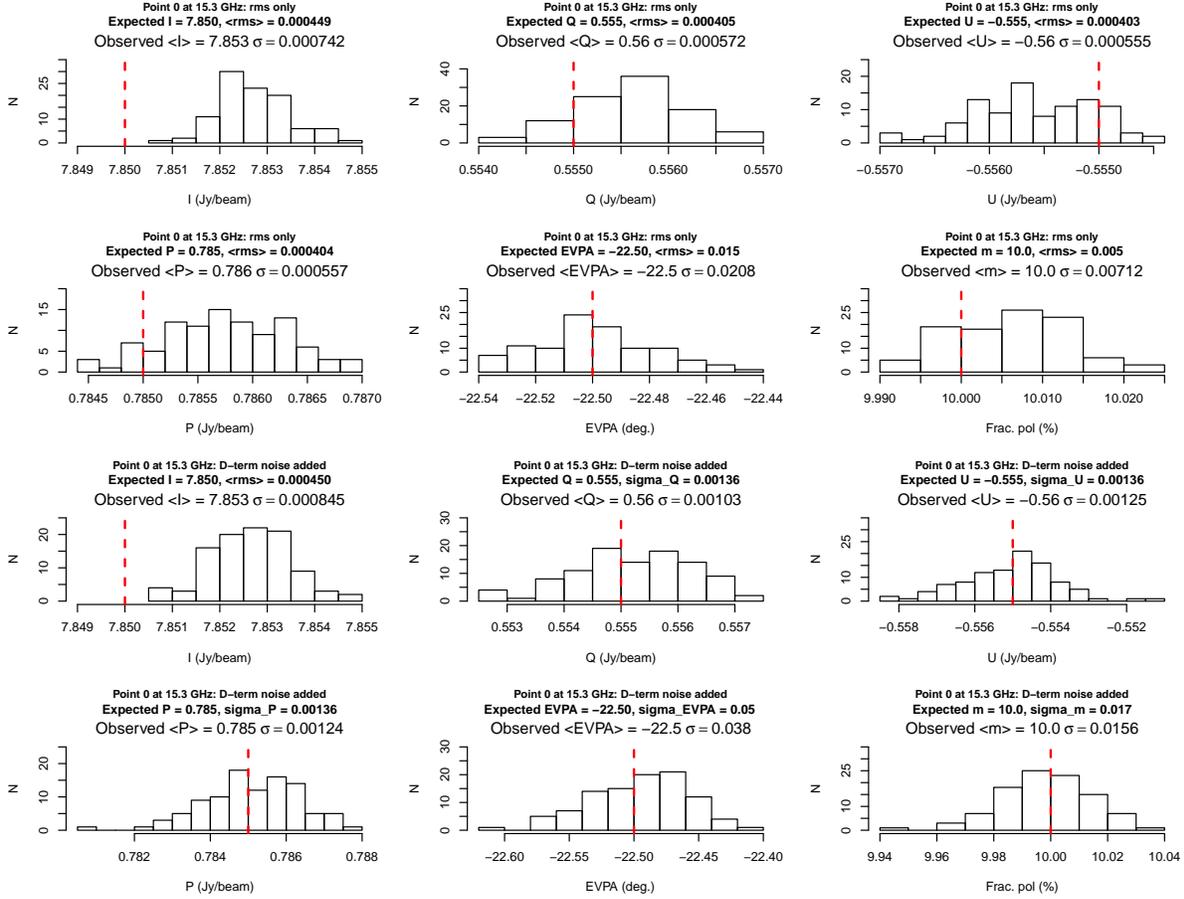}
\end{center}
\caption{Distributions of the 100 simulated values of total intensity, Stokes Q and U, polarized flux density, EVPA and fractional polarization for 1226+023 from jet location 0 shown in Fig.~\ref{fig:1226ipoints} at 15 GHz. Top six panels show simulations including the rms noise only and bottom six panels with D-term noise added. The expected value is shown by a dashed line (red in the online version). See text for information on the expected error values.}\label{fig:1226isim}
\end{figure*}

 The D-term error is strongly dependent on the Stokes I and is not distributed evenly across the images. Therefore 
its effect is largest on the bright locations of the source, especially near the core and it is not accounted for in the rms errors. 
To demonstrate the effect, in Fig.~\ref{fig:1226isim} we show distributions of total intensity I, Stokes parameters Q and U, polarized flux density p, 
EVPA and fractional polarization m for 1226+023 at point 0  (shown by a star) of Fig.~\ref{fig:1226ipoints}. In this simulation Q and U were set to 
be 0.0707 $\times$ I in each component. Without any noise added, this results in uniform fractional polarization $m=0.1$ and EVPA = $-$22.5 degrees.
Above each histogram we give the expected value for each parameter and its expected error value calculated using Eqs.~\ref{eq:p_err}, \ref{eq:evpa_err}, and \ref{eq:m_err}.
The top six panels show 100 simulations without adding the D-term error in step 5 in the simulations. As can be seen 
from the expected and observed values, the observed standard deviations of Q and U are slightly larger than the measured $Q_\mathrm{rms}$ and $U_\mathrm{rms}$ and 
the distributions are not peaking at the expected values, which can be attributed to the CLEAN errors.  
In the bottom six panels, the simulation is repeated with additional D-term noise added, which is also included in the 
expected error value using Eq.~\ref{eq:dterm_err}. Now the expected and observed rms values correspond very well to each other. 
In this bright location of the source, the contribution from D-term errors to the error in polarized flux density is twice as large as the 
rms error (1.1 mJy compared to 0.56 mJy). The distributions of other components are not shown here but they follow the same pattern, except 
that the D-term error contribution diminishes with decreasing Stokes I so that in most of the components of 3C~273 the rms errors 
and D-term errors are of the same order. In fainter sources, such as 0333+321 the D-term errors are typically negligible in the 
jet components and smaller than rms errors even in the core. This effect is correctly accounted for by using Eq. \ref{eq:dterm_err}.

\begin{figure*}[htp]
\begin{center}
\includegraphics[scale=0.6]{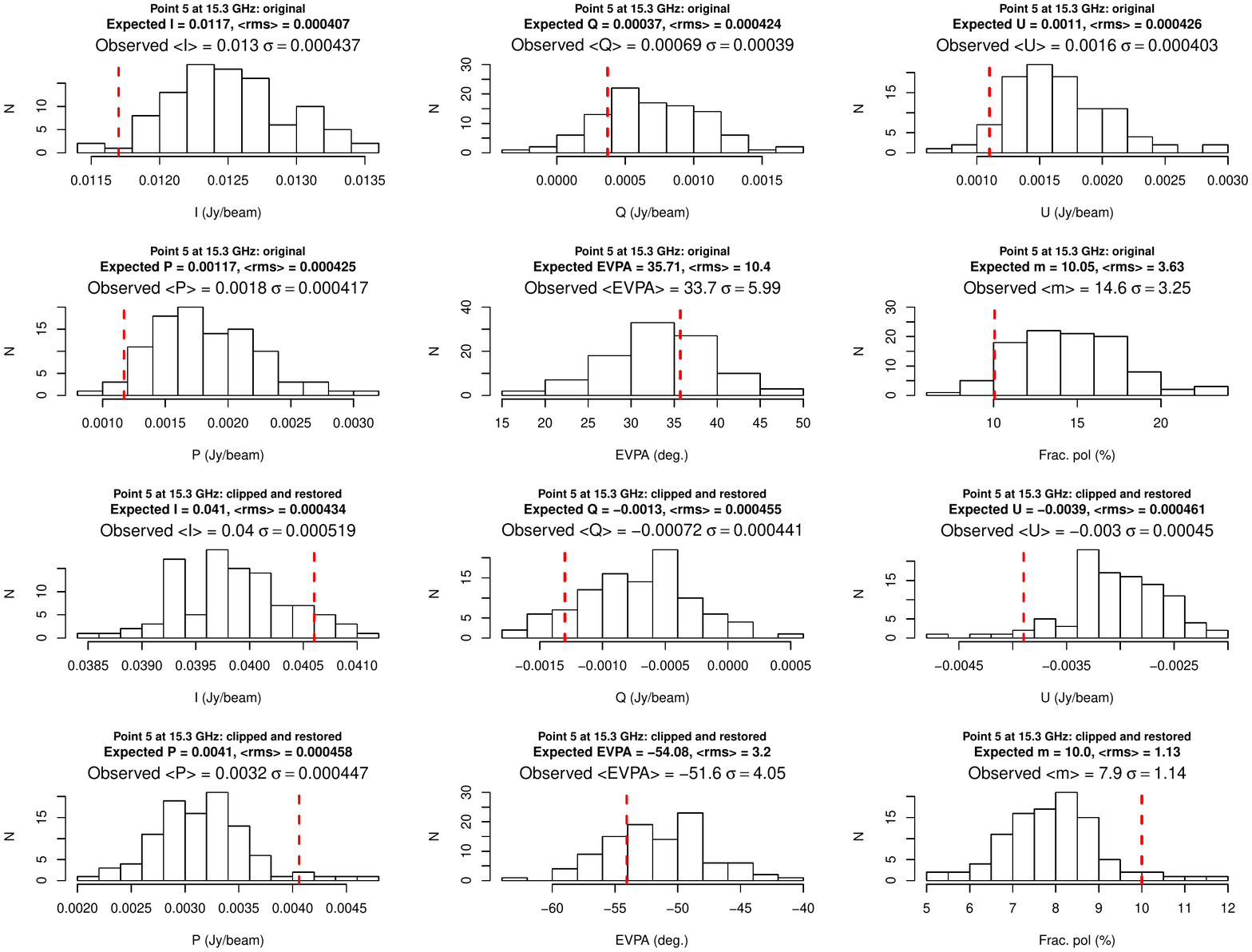}
\end{center}
\caption{Distributions of the 100 simulated values of total intensity, Stokes Q and U, polarized flux density, EVPA and fractional polarization for 0333+321 from jet location 5 shown in Fig.~\ref{fig:0333ipoints}. Top six panels show the original 15\,GHz data and bottom six panels data which has ({\it u,v})-range clipped as our real data and which were restored with the 8.1 GHz beam size. The expected value is shown by a dashed line (red in the online version). See text for information on the expected error values.}\label{fig:0333isim}
\end{figure*}

\begin{figure*}[htp]
\begin{center}
\includegraphics[scale=0.6]{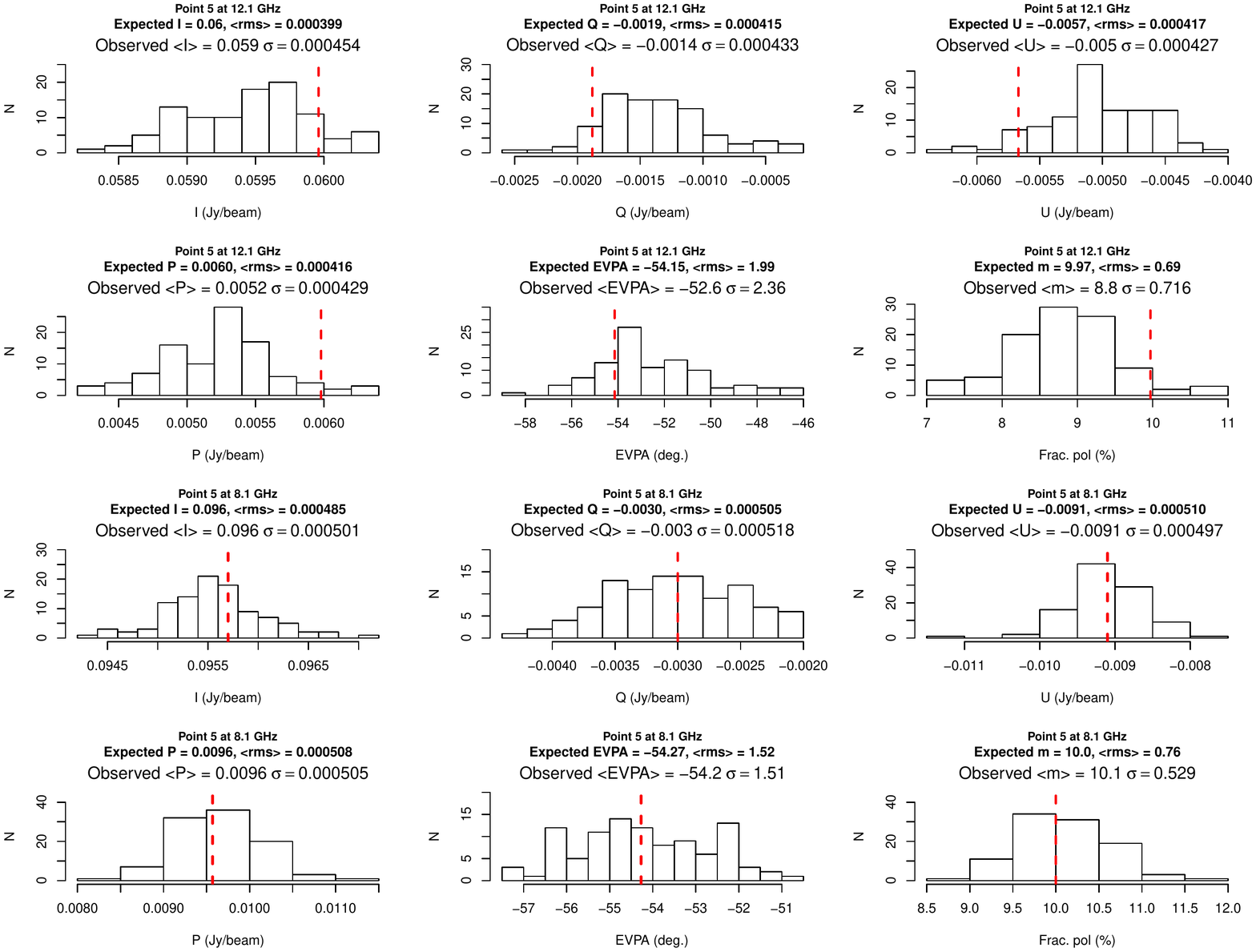}
\end{center}
\caption{Distributions of the 100 simulated values of total intensity, Stokes Q and U, polarized flux density, EVPA and fractional polarization for 0333+321 from jet location 5 shown in Fig.~\ref{fig:0333ipoints}. Top six panels show 12.1 GHz data and bottom six panels 8.1 GHz. The expected value is shown by a dashed line (red in the online version). See text for information on the expected error values.}\label{fig:0333isim_lowf}
\end{figure*}

In order to study the effects of ({\it u,v})-clipping and the CLEAN procedure, we simulated the original 15 GHz data of 0333+321 
and the ({\it u,v})-clipped data restored to the 8.1\,GHz beam. In these simulations we set Q = $-$0.031623 $\times$ I and U = $-$0.094869 $\times$ I,
which results in uniform fractional polarization $m=0.1$ and EVPA = $-$54.217 degrees. The expected values were determined as in the 
case of 1226+023. In Fig.~\ref{fig:0333isim} top six panels we show the distributions for the original images and in the bottom 
six panels the distributions for the images which have a reduced ({\it u,v})-coverage corresponding to the real data and restored 
with the 8.1\,GHz beam size. These are taken at location 6  (shown by a star) in Fig.~\ref{fig:0333ipoints} to show the effects on a faint edge 
region. Our equations for the error values work very well  showing that the CLEAN procedure does not affect the standard 
deviation of the distributions, but there is a large offset in the peaks of the distributions in both 
simulations. This is also seen in the lower frequency simulations, for which we show an example at 12.1 and 8.1\,GHz 
(8.4\,GHz is nearly identical to 8.1\,GHz and therefore not shown) in Fig.~\ref{fig:0333isim_lowf}.
This is due to the reduced ({\it u,v})-coverage, especially at 15.4 and 12.1\,GHz, and the CLEAN procedure. By looking at the offsets in 
all the component locations in the different sources and all the frequency bands, we find the error to vary between 1 and 
3 times the rms error and therefore we set it to be 1.5 times the rms error in our error estimates.

Based on all these simulations, we conclude that the errors in polarized flux density, EVPA and fractional polarization should be 
calculated using Eqs.~\ref{eq:p_err}, \ref{eq:evpa_err} and \ref{eq:m_err}, where $\sigma_Q$ and $\sigma_U$ should include 
contributions from the rms error, D-term error and CLEAN error so that they are defined as:
\begin{equation}
\sigma = \sqrt{\sigma_\mathrm{rms}^2 + \sigma_\mathrm{Dterm}^2 + (1.5 \times \sigma_\mathrm{rms})^2},
\end{equation}
where $\sigma_\mathrm{rms}$ is the corresponding rms error and $\sigma_\mathrm{Dterm}$ is calculated using Eq.~\ref{eq:dterm_err}.
The errors for our analysis are defined using the above criteria.

\section{Errors in RM maps}\label{app:rmerr}
Errors in the RM values can formally be obtained from the variance-covariance matrix of the 
linear fit as the error of the slope. This method is used in the RM task in AIPS. The resulting 
errors depend largely on the error of the EVPA, which is dominated by the absolute calibration error in 
the bright jet locations. As we show in Appendix \ref{app:polerr}, the rms contribution of the 
EVPA error is well-described by error propagation from Q and U rms values. It is crucial to obtain a 
correct error estimate for the RM values in order to distinguish between various depolarization 
models and to properly study gradients in the RM within a source. Using the same kind of approach as in Appendix 
\ref{app:polerr}, we simulated the effect of random noise and calibration error to see if the error 
estimate from the variance-covariance matrix is correct.

The simulations were done in the same manner as in Appendix \ref{app:polerr} but repeating the 
procedure for all the frequency bands. We then added a random error drawn from a Gaussian 
distribution with standard deviation corresponding to the calibration error at the given frequency band 
to each of the EVPA images.  This error is added systematically to each pixel in a given image but 
varies randomly from one epoch to another and needs to be accounted for in order to obtain correct 
error estimates for the RM. The simulated RM maps were then created using the same script 
as for the real data and 1000 RM maps were obtained. Without any added noise, the expected value 
for the RM in each map is 0 rad~m$^{-2}$ minus any Galactic Faraday rotation. This is because we created the 
simulated maps using the exact same procedure as for the real data and therefore also the Galactic 
Faraday rotation correction was taken into account. 

We repeated the simulations for three sources, 0333+321, 0735+178 and 1226+023. The source 0735+178 
was included so that we could study the effect on a faint source which does not have a well-resolved jet. 
The expected value of RM in each source is different because of the differing Galactic Faraday rotation. For 
0333+321 we expect a RM of $-$34.1 rad~m$^{-2}$, for 0735+178 a RM of $-$20.5 rad~m$^{-2}$ and for 1226+023 
a RM of $-$4.7 rad~m$^{-2}$.

\subsection{Errors in individual jet locations}
\begin{figure}[htp]
\begin{center}
\includegraphics[scale=0.5]{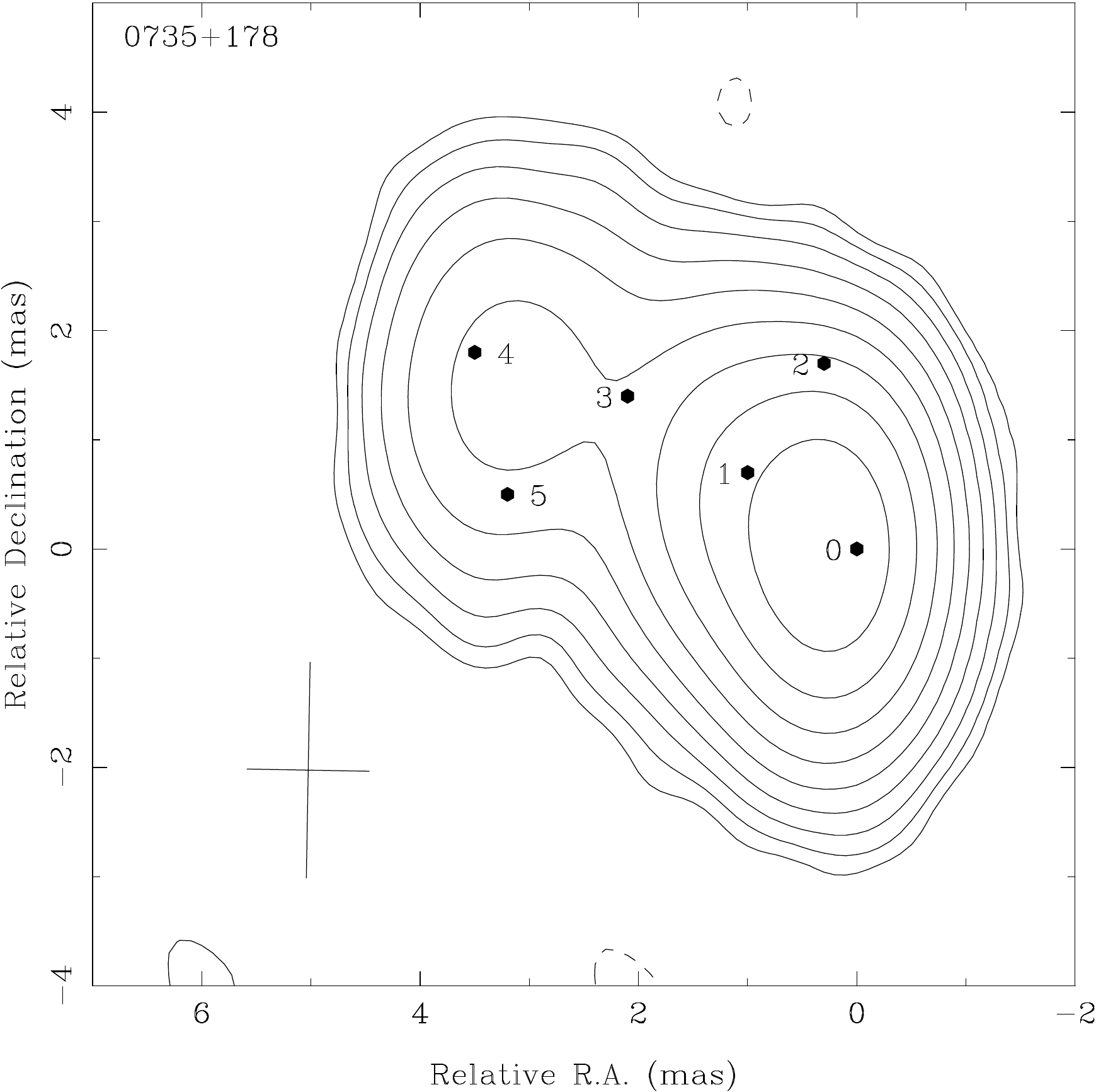}
\end{center}
\caption{Locations of the simulated RM distributions in 0735+178 overlaid on the 15 GHz total intensity contours.}\label{fig:0735_ipoints}
\end{figure}

\begin{figure*}[htp]
\begin{center}
\includegraphics[scale=0.6]{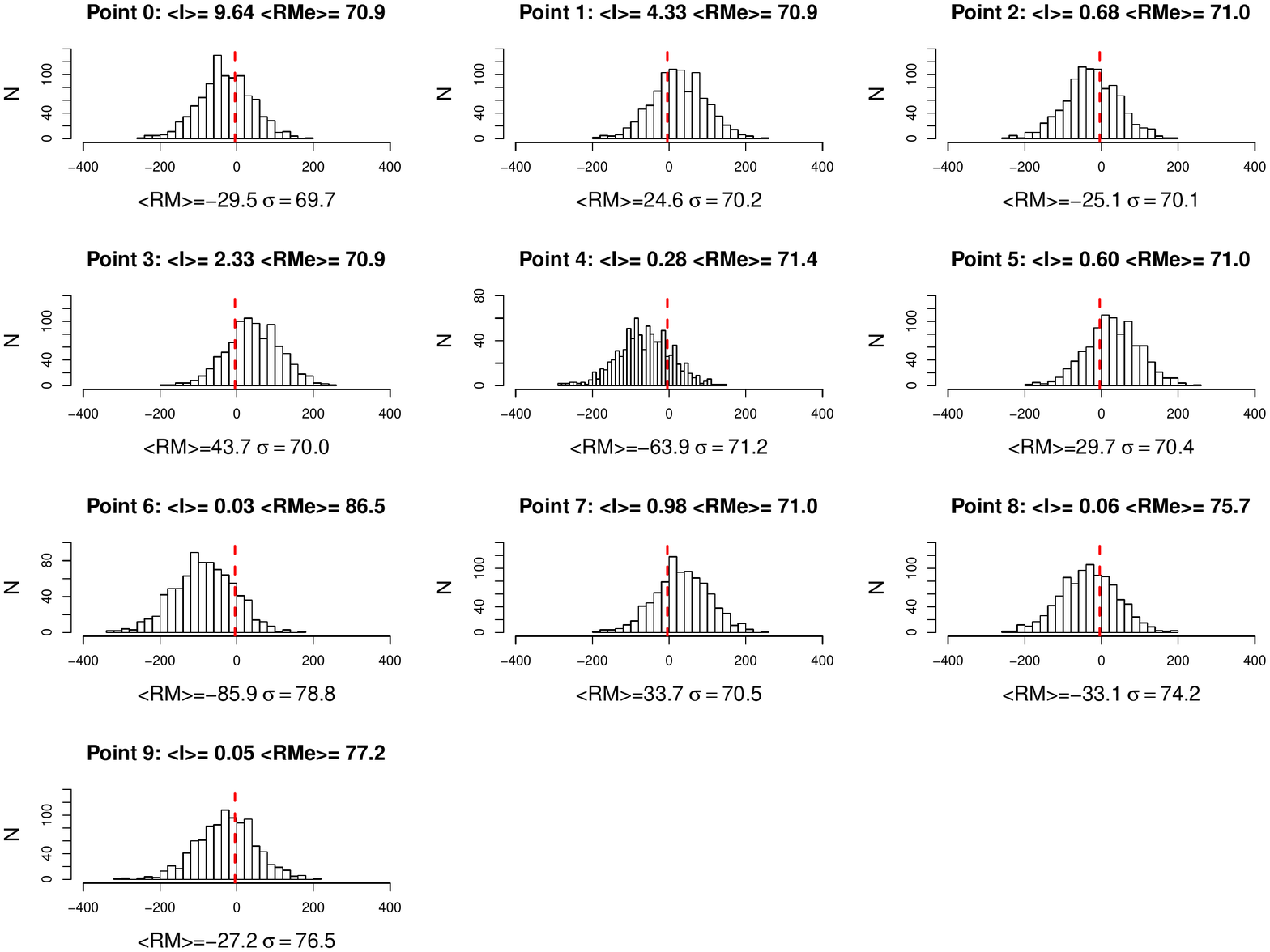}
\end{center}
\caption{Distributions of the 1000 simulated RM values for 1226+023 from jet locations shown in Fig.~\ref{fig:1226ipoints}. Average total intensity value $<$I$>$, average RM error from the variance-covariance matrix $<$RMe$>$, average RM value $<$RM$>$ from the simulated jets and the standard deviation of the distribution are given for each location. The expected RM value is shown by a dashed line (red in the online version).}\label{fig:1226rmhist}
\end{figure*}

\begin{figure*}[htp]
\begin{center}
\includegraphics[scale=0.6]{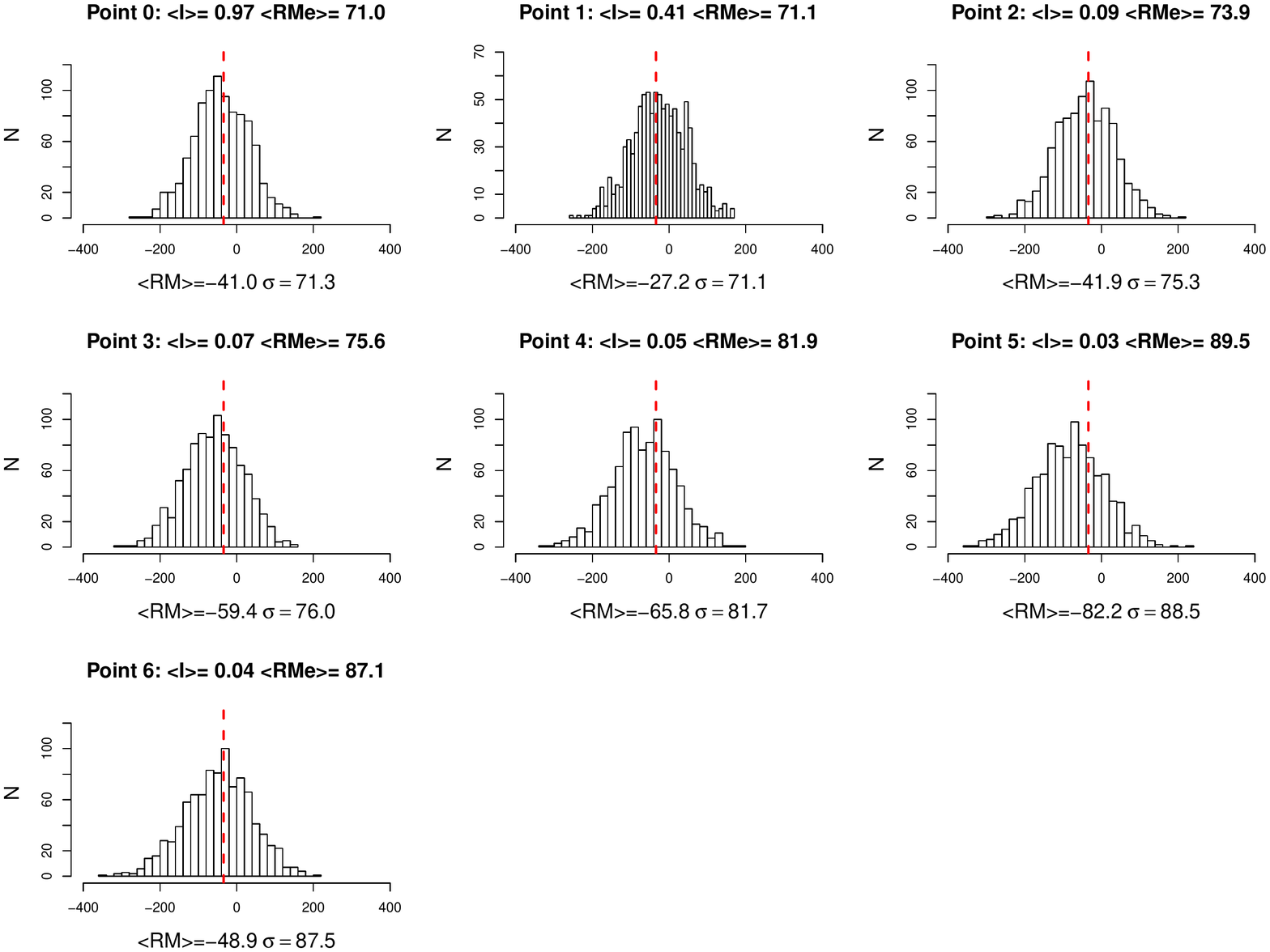}
\end{center}
\caption{Distributions of the 1000 simulated RM values for 0333+321 from jet locations shown in Fig.~\ref{fig:0333ipoints}. Average total intensity value $<$I$>$, average RM error from the variance-covariance matrix $<$RMe$>$, average RM value $<$RM$>$ from the simulated jets and the standard deviation of the distribution are given for each location. The expected RM value is shown by a dashed line (red in the online version).}\label{fig:0333rmhist}
\end{figure*}

\begin{figure*}[htp]
\begin{center}
\includegraphics[scale=0.6]{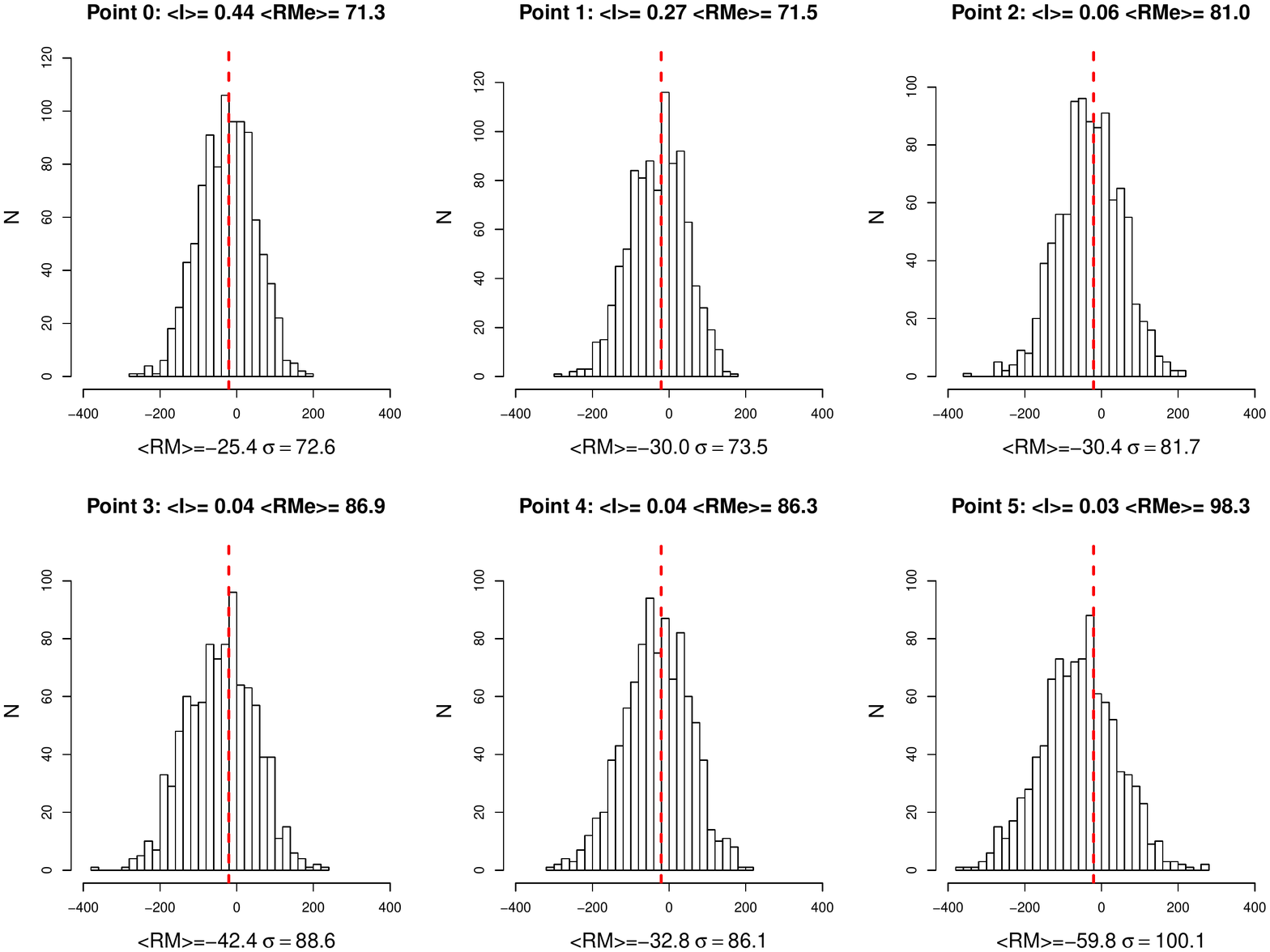}
\end{center}
\caption{Distributions of the 1000 simulated RM values for 0735+178 from jet locations shown in Fig.~\ref{fig:0735_ipoints}. Average total intensity value $<$I$>$, average RM error from the variance-covariance matrix $<$RMe$>$, average RM value $<$RM$>$ from the simulated jets and the standard deviation of the distribution are given for each location. The expected RM value is shown by a dashed line (red in the online version).}\label{fig:0735rmhist}
\end{figure*}

In order to study the RM error in individual jet locations, we chose the same jet locations for 0333+321 and 
1226+023 as in Appendix \ref{app:polerr} and shown in Figs.~\ref{fig:1226ipoints} and \ref{fig:0333ipoints}. For 0735+178 we chose the locations 
shown in Fig.~\ref{fig:0735_ipoints}. The distributions of the 1000 simulated values are shown in Figs.~\ref{fig:1226rmhist} - \ref{fig:0735rmhist}. 
Once again it is clear that the standard deviations $\sigma$ from our simulations, listed below each distribution, agree well with the 
error from the variance-covariance matrix $<$RMe$>$ (average value over the 1000 simulations), listed above each distribution.
There is, however, a visible offset in the peaks of the distributions to the left of the expected value.
This offset is due to small errors in the final CLEAN procedure as shown in Appendix \ref{app:polerr} and it is accounted for 
by the additional error due to the CLEAN procedure, which is taken into account in our final Q and U errors.

\begin{figure}[htp]
\begin{center}
\includegraphics[scale=0.5]{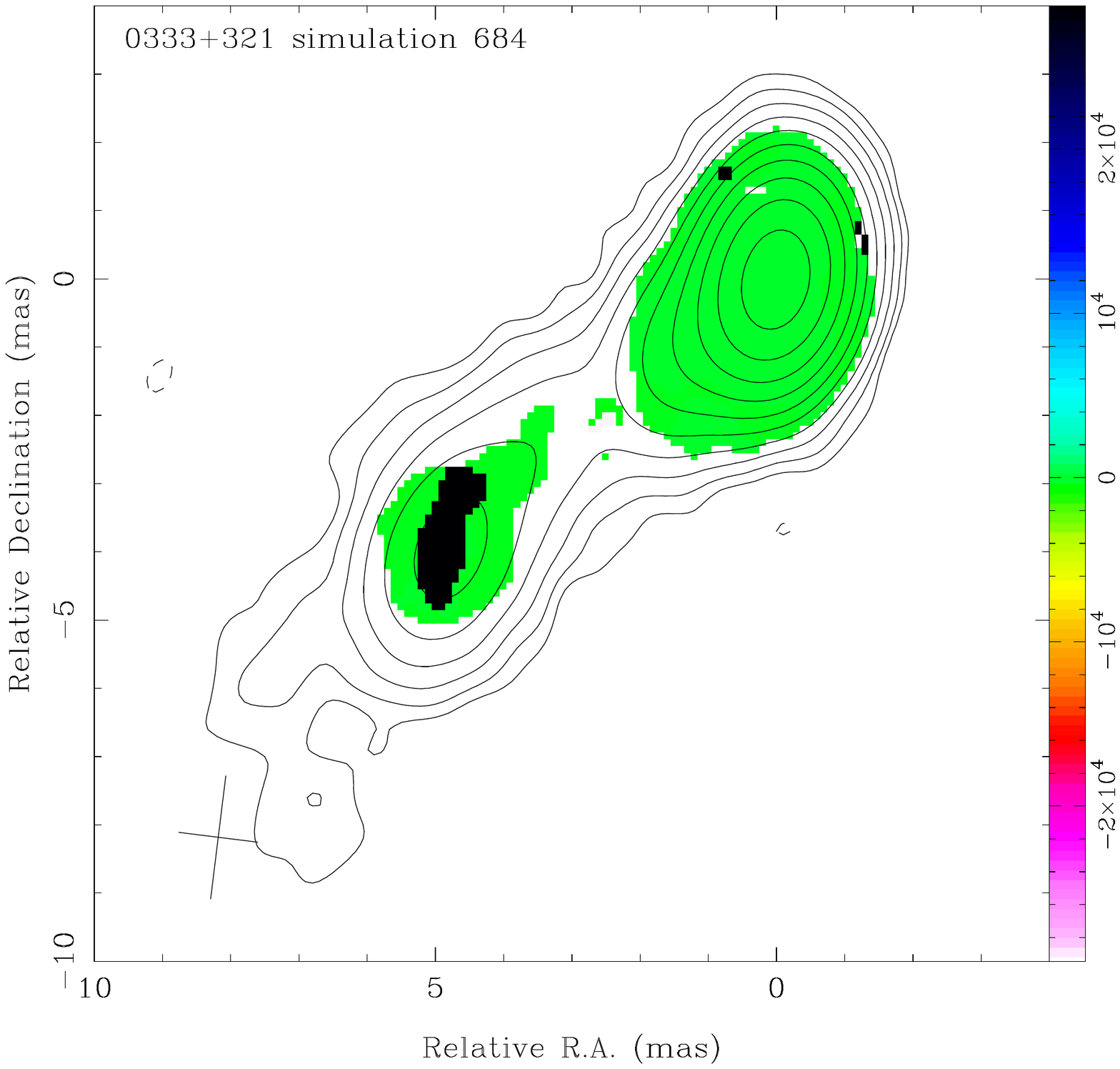}
\end{center}
\caption{Example of a simulation in which extreme RM values are generated due to noise in the data. Color version is available in the online edition of the Journal.}\label{fig:0333highRM}
\end{figure}

Our simulations also show that it is common to obtain very high RM values of 
$\sim \pm 2\times10^4$ rad~m$^{-2}$ purely due to noise in the data. An example of such simulated jet is shown in Fig.~\ref{fig:0333highRM}. 
We find at least one very high RM pixel in 67\% of the simulated maps in 0735+178, in 74\% in 0333+321 and in all the 
simulated maps of 1226+023. The number of the high-RM pixels depends on the total number of pixels in the simulated 
maps. In 1226+023 the fraction is much higher because it has 5 times more pixels than the two other sources. The median 
fraction of high-RM pixels in the simulated maps is 0.3 - 0.6 \% depending on the source but in 5.5 - 26\% of the maps 
there are more than 100 such pixels, resulting in patches of at least 1 times 1 mas. This happens because the pixels are not 
independent on scales smaller than the beam size. Therefore even fairly large very high-RM regions like these in the real RM maps are most likely 
spurious. For this reason, in our real RM maps, we have blanked the high-RM pixels because they are most likely due to 
noise in the data and do not represent real structure. 

Based on the simulations we conclude that the error from the variance-covariance matrix of the linear fit 
is a good estimate for the true error in the RM in individual jet locations.

\subsection{Errors in RM gradients}\label{app:grad}
\begin{figure}[htp]
\begin{center}
\includegraphics[scale=0.5]{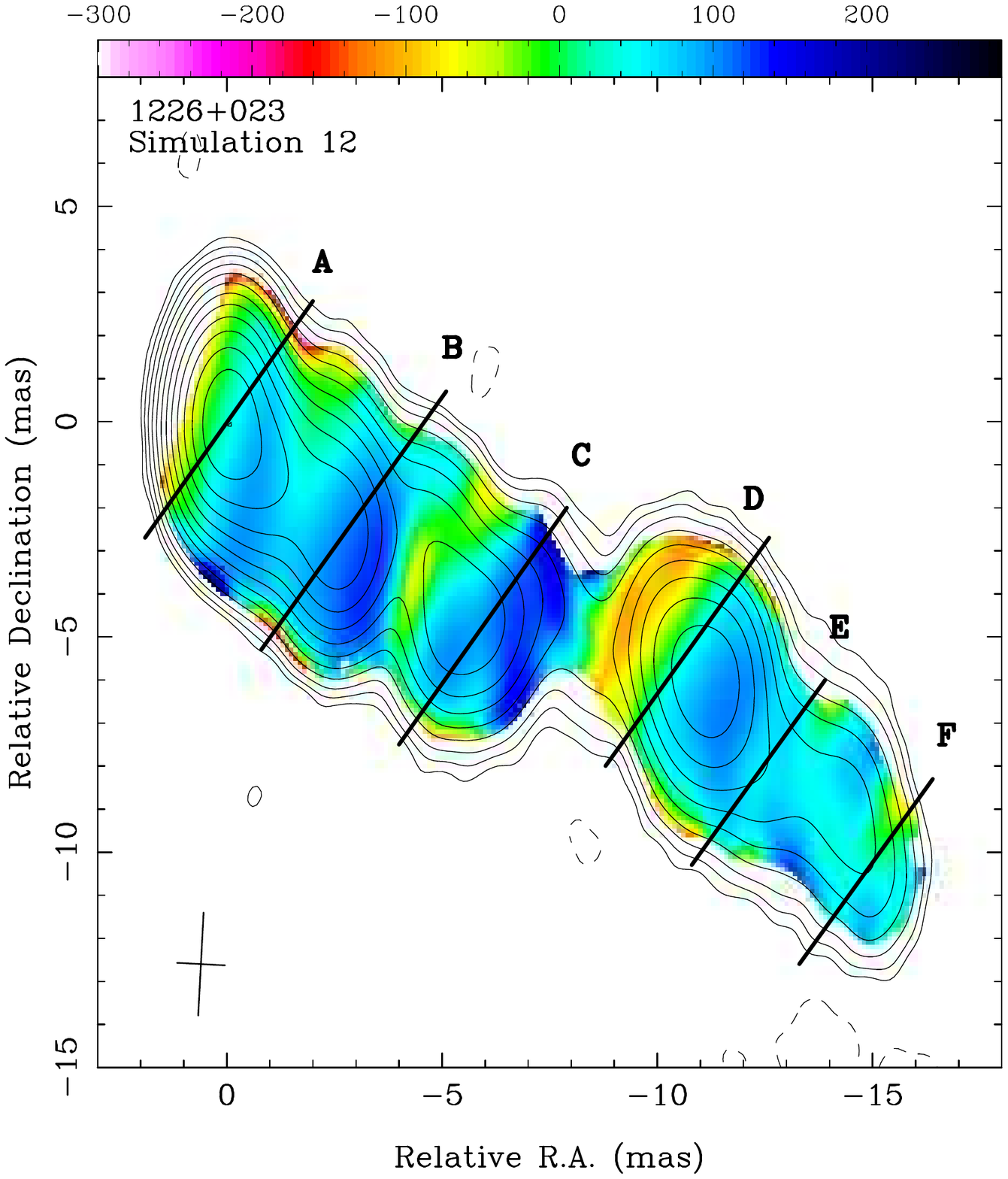}
\end{center}
\caption{Slices transverse to the jet in 1226+023 overlaid on a simulated RM map. Color version is available in the online edition of the Journal.}\label{fig:1226gradjet}
\end{figure}

\begin{figure}[htp]
\begin{center}
\includegraphics[scale=0.5]{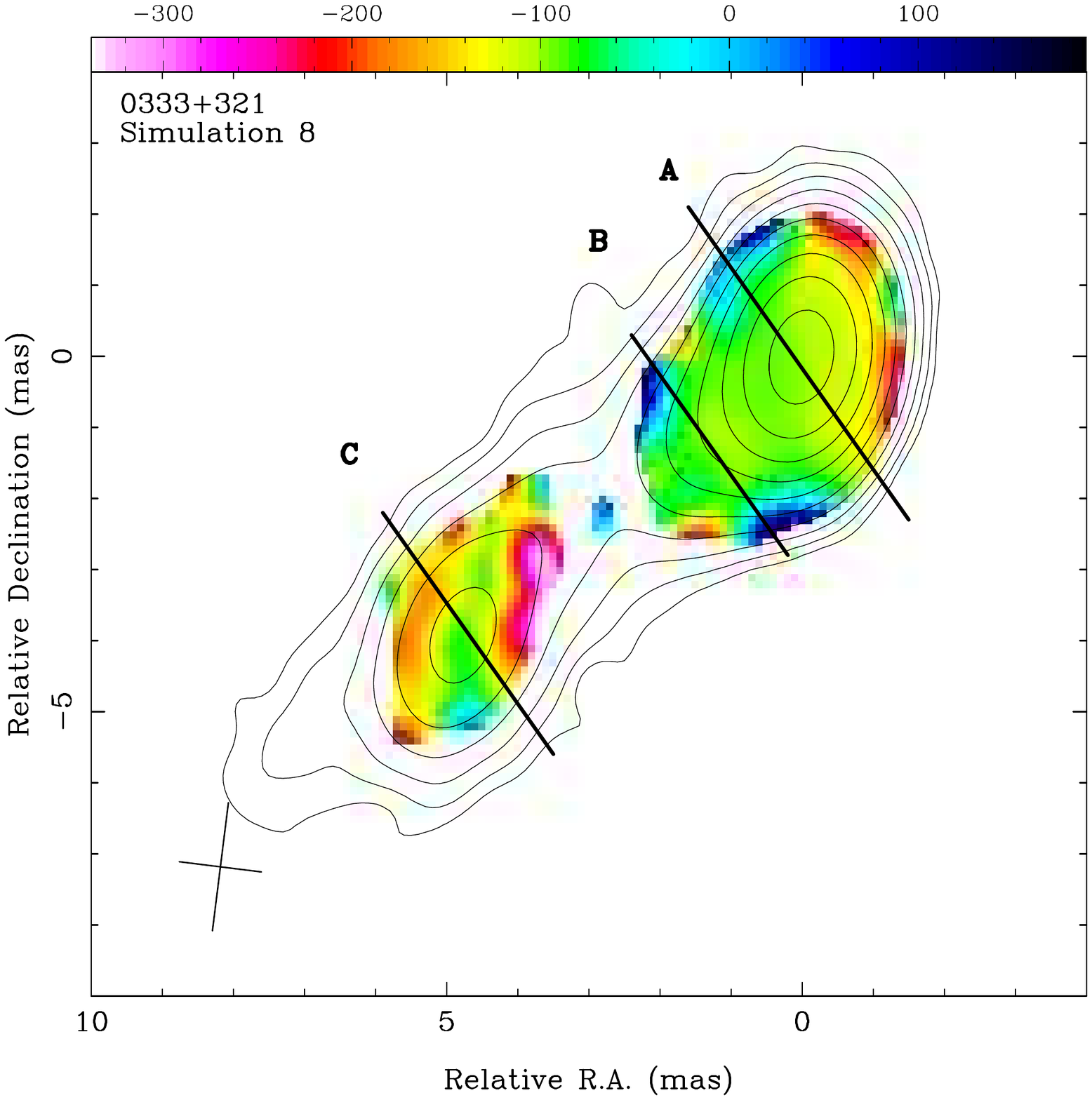}
\end{center}
\caption{Slices transverse to the jet in 0333+321 overlaid on a simulated RM map. Color version is available in the online edition of the Journal.}\label{fig:0333gradjet}
\end{figure}

\begin{figure}[htp]
\begin{center}
\includegraphics[scale=0.5]{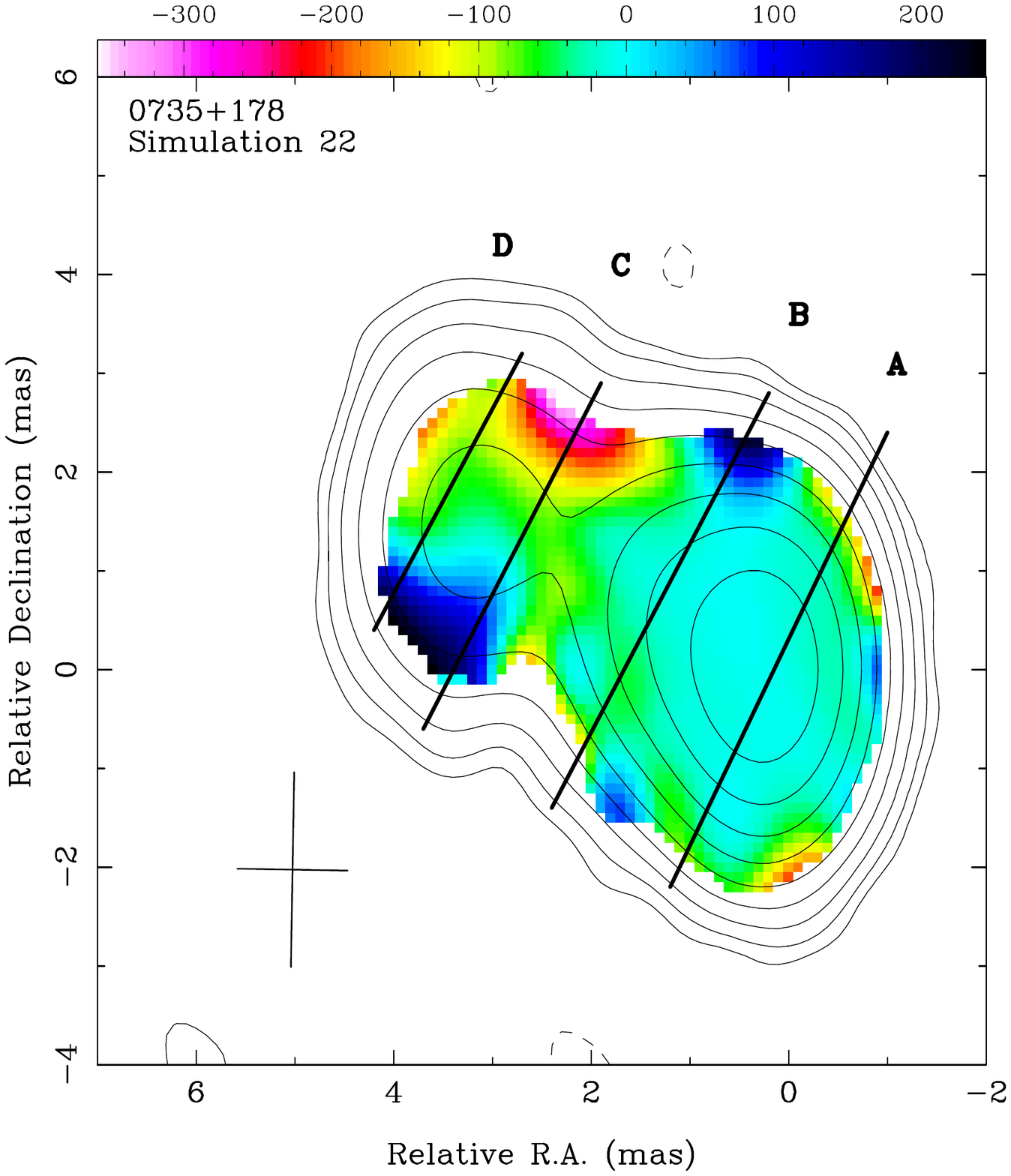}
\end{center}
\caption{Slices transverse to the jet in 0735+178 overlaid on a simulated RM map. Color version is available in the online edition of the Journal.}\label{fig:0735gradjet}
\end{figure}

Over the past decade there have been several claims of detections of RM gradients in many AGN 
\citep[e.g.,][for a compilation of results]{contopoulos09}. These results are still somewhat 
controversial due to the difficulty in addressing the errors in the gradients \citep{taylor10}. 
\cite{murphy11} convolve simulated gradients with different beam sizes and argue that a 
gradient can be significant even when the jet is not resolved. Their simulation does not, however, 
take into account noise in the data.
With simulations we have shown that the errors in individual pixels are well described by the 
error from the variance-covariance matrix of the linear fit. It is more complicated to study the 
errors in the RM gradients, or more precisely, the probability of detecting a spurious RM gradient 
due to noise in the data. Several of our simulated jets show structures resembling real gradients 
primarily due to effects of the finite beam size. To quantify this effect, we have taken several 
transverse slices across the simulated jets, shown in Figs.~\ref{fig:1226gradjet} - \ref{fig:0735gradjet}, and fitted a simple line 
to the data. We have then looked at the distributions of the slopes which give us the 
maximum spurious gradient in rad~m$^{-2}$~mas$^{-1}$, created by noise in the data.

\begin{figure*}[htp]
\begin{center}
\includegraphics[scale=0.55]{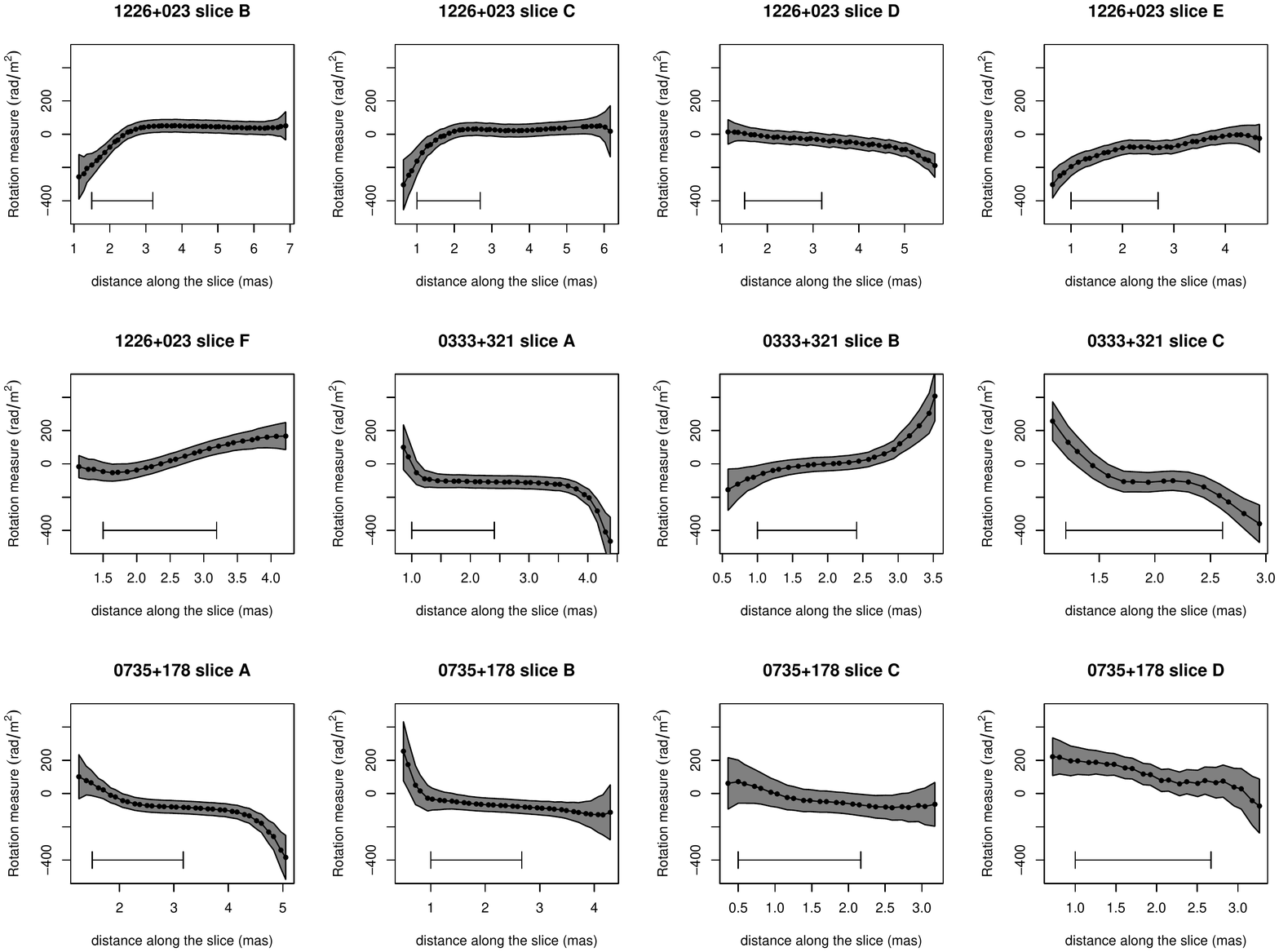}
\end{center}
\caption{The most extreme examples of simulated transverse gradients over the slices across the jet locations shown in Figs.~\ref{fig:1226gradjet} - \ref{fig:0735gradjet}. None of these simulated jets contain an actual RM gradient. The FWHM beam size along the slice direction is shown in each plot as a scale bar.}\label{fig:slices}
\end{figure*}

In Fig.~\ref{fig:slices} we show examples of the slices in the different sources. One thing to note is that 
when studying gradients, the absolute EVPA calibration error can be ignored because it affects each 
pixel in the same direction and thus will not affect a gradient across the same source 
\citep{mahmud09}. Therefore in each pixel we have subtracted in quadrature 
the amount of calibration error, $\sim$ 60 rad~m$^{-2}$, from the total RM error.
It is very important to note that the pixels are not independent but very much affected 
by the beam size. We have plotted the size of the beam along the transverse slice in 
each plot to show the scale in which the pixels are not independent. As can be seen from the plots, 
depending on the slice, there are only 1-4 independent points along each slice. 

\begin{figure*}[htp]
\begin{center}
\includegraphics[scale=0.6]{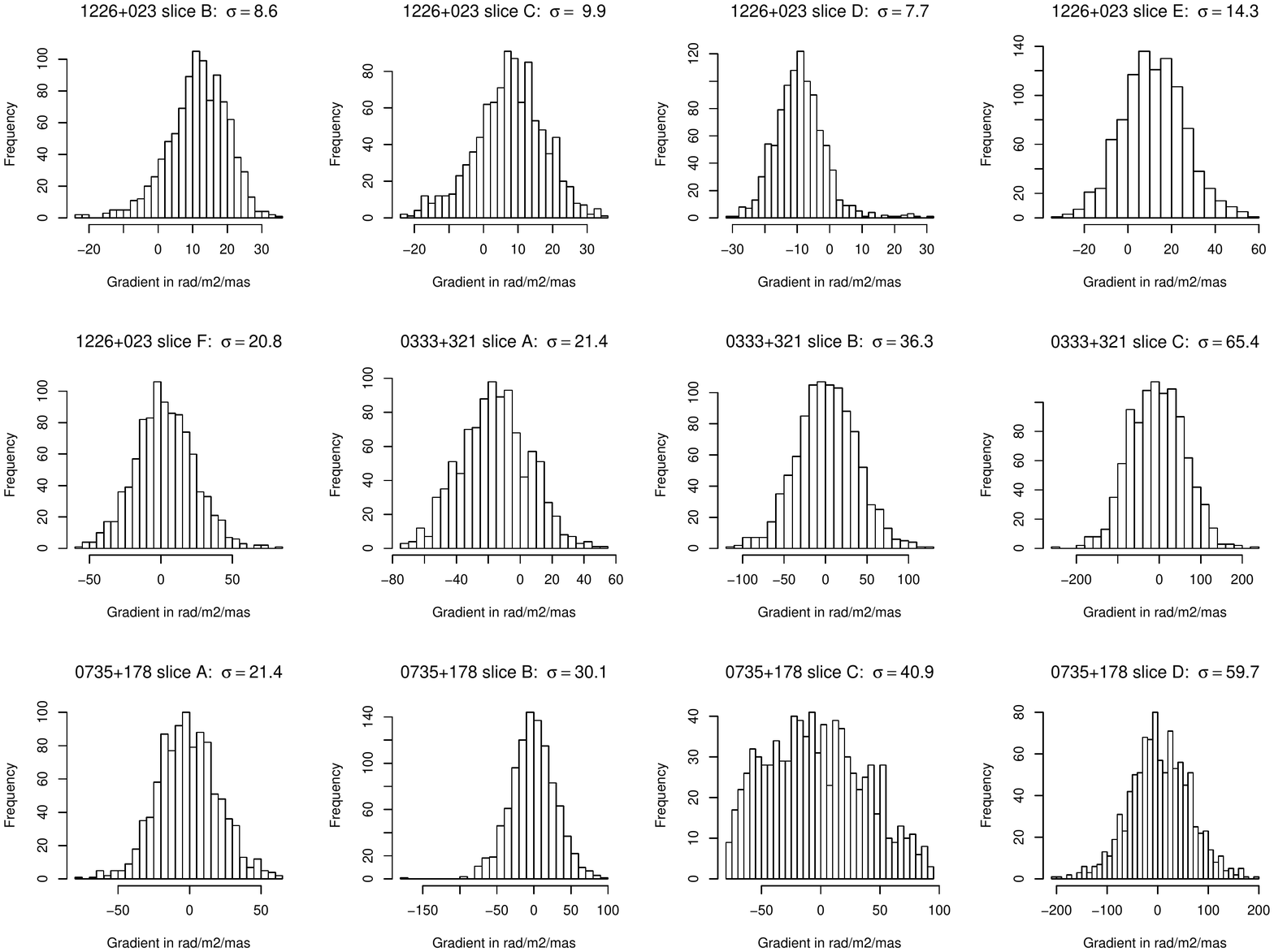}
\end{center}
\caption{Distributions of 1000 simulated gradients across the jet locations shown in Figs.~\ref{fig:1226gradjet} - \ref{fig:0735gradjet}. 1226+023 slice A is omitted from the plot because it is almost identical to slice B.} \label{fig:gradhist}
\end{figure*}

\begin{figure*}[htp]
\plottwo{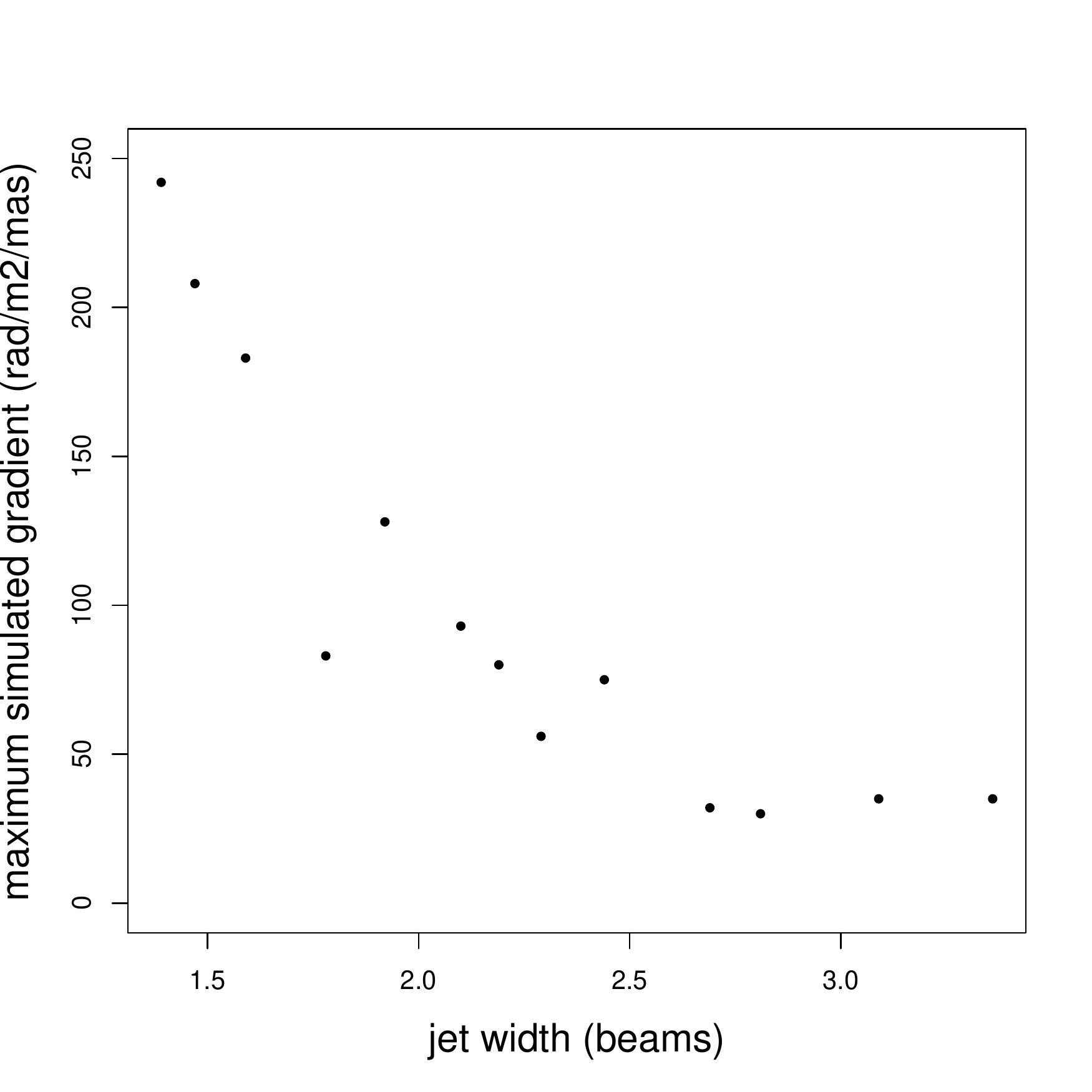}{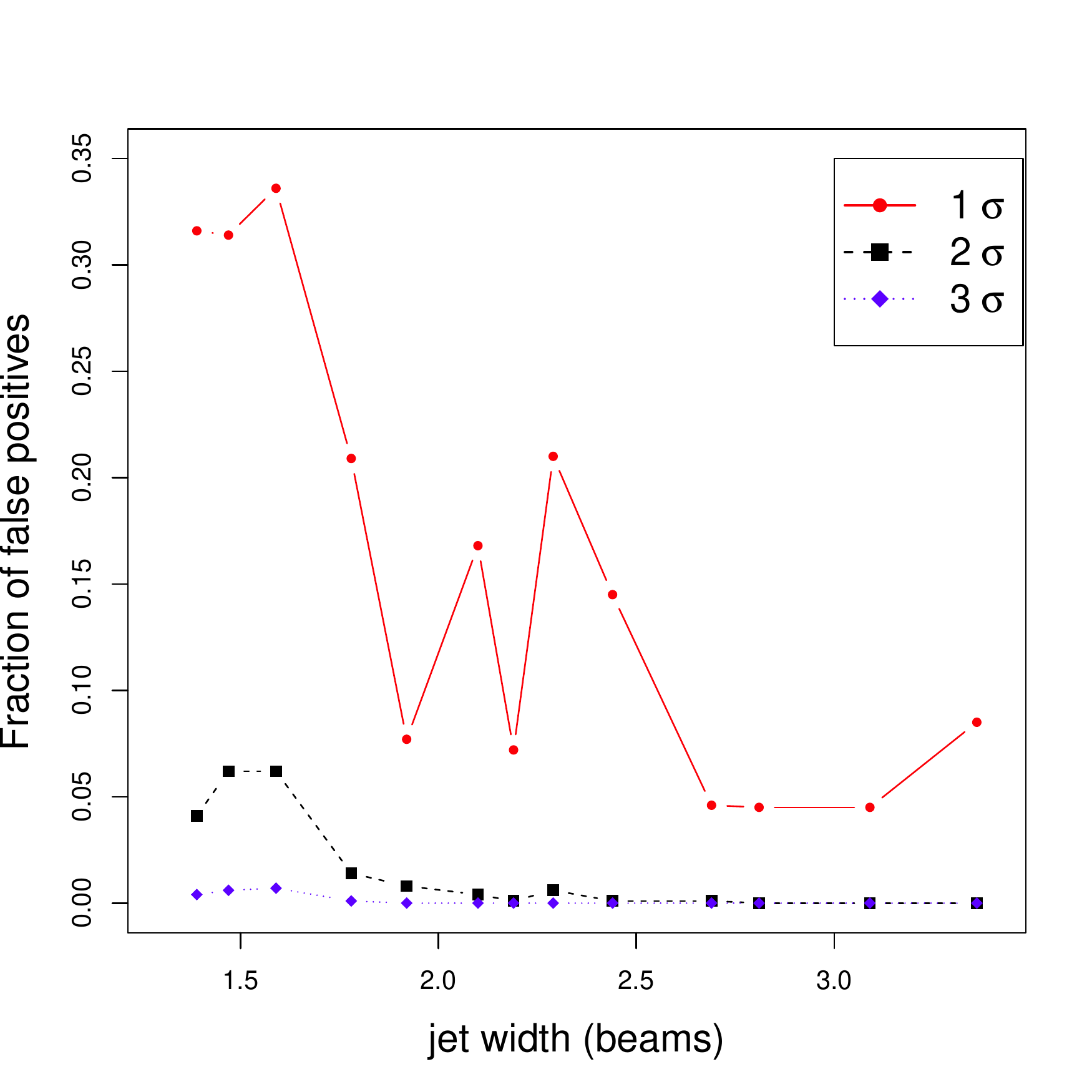}
\caption{Maximum gradient from the simulations against the jet width (left). Fraction of total gradients along the slice in our simulations exceeding 1 (filled circles, red in the online version), 2 (filled squares, black in the online version) or 3 (filled diamonds, blue in the online version) standard deviations against the jet width.\label{fig:gradbeam}}
\end{figure*}

The distributions of the measured slopes are shown in Fig.~\ref{fig:gradhist}. In the calculation of the slope we have 
ignored the very-high RM values. This resulted in slices with only a few pixels with measured RMs which 
usually resulted in steep slopes. These are ignored in the distribution plots.
The most notable are the distributions from slices C in 0333+321 and slices D in 0735+178 
which show gradients up to 200 rad~m$^{-2}$~mas$^{-1}$. The slices are about 1.5 beam sizes wide and 
with a typical beam width along the slice of 1.5 mas, this would result in a total gradient of 
450 rad~m$^{-2}$ over the slice. This is of the same order as most of the gradients in the literature. 
It is very close to the value we see in the jet of 3C 273. Reassuringly, the magnitude of the 
spurious gradient is strongly dependent on the width of the jet with respect to the beam size as 
shown in Fig.~\ref{fig:gradbeam} left panel. When the width of the polarized jet approaches three beam widths, the spurious gradient 
diminishes. To calculate the jet width in beams we first take the width of the slice in pixels and divide that 
by the beam size along the slice direction in pixels. This is because we want to ensure that we take into 
account that we do not detect polarization in as large area as total intensity and require the jet to be resolved in 
polarization as well.

 We also note that almost all the spurious gradients we see in the simulations are smaller than three times the RM errors at the location of the 
slice if using the errors as we have defined them and therefore would not be accepted as true gradients in our real data. 
This is shown in Fig.~\ref{fig:gradbeam} right panel, where we plot the fraction of 
false positives, i.e. number of spurious gradients which exceed the error limit, against the jet width 
for 1$\sigma$, 2$\sigma$ and 3$\sigma$ limits.  Here $\sigma$ is defined as the largest error bar in the end of a 
gradient slice. This plot clearly demonstrates how the fraction of false 
positives goes to zero if the jet is two beams wide and a 3$\sigma$ error limit is used. When the 
jet is more than 2.5 beams wide, even a 2$\sigma$ limit might be sufficient, if additionally taking into 
account the magnitude of the maximum possible spurious gradient shown in the left panel.
For our specific four-frequency setup between 8 and 15\,GHz using 128 Mbits s$^{-1}$ recording bit rate and $\sim$60 minute 
on-source time, we find the limit to be 200 rad~m$^{-2}$~mas$^{-1}$.

Our simulations also verify that it is impossible to get a persistent gradient over the whole jet length, 
as is seen in 3C 273, due to noise in the data. Therefore it 
is important that if gradients are detected, they are seen in multiple jet locations more than a beamwidth apart, or 
in the same jet location over multiple epochs, and that the jet is well-resolved in polarized flux density, preferably 
over two times the beam width. When studying jets that are not wide enough, our simulation 
results can be used as ``rules of thumb'' for estimating the magnitude of a reliable 
gradient. For example, in a jet that is two beams wide, it is possible to have spurious gradients up to 
100 rad~m$^{-2}$~mas$^{-1}$ which would result in a total gradient of 300 rad~m$^{-2}$ if the beam 
width is 1.5 mas.  We note that detection of three 2$\sigma$ gradients at different epochs corresponds to 
a 3$\sigma$ detection so that multi-epoch observations can help to determine if observed gradients in 
jets that are not well-resolved are real. However, this requires correct treatment of the RM errors and is not 
recommended for jets less than 1.5 beams wide due to the large fraction of false positives at these jet widths.

Our simulations overall agree well with the suggestions by \cite{taylor10} to consider a gradient 
reliable if the jet is more than three beams wide and use 3$\sigma$ error limits. Based on our 
simulations a jet which is two beams wide is already sufficient if the gradient exceeds the 3$\sigma$ limit. 
Attempting to detect RM gradients in jets less than two beams wide is discouraged 
as the probability to detect false positives exceeds 0.02 even when a 3$\sigma$ limit is used.
We emphasize that this approach requires the RM errors to be calculated from the variance-covariance matrix 
of the EVPA vs. $\lambda^2$ fit with appropriate EVPA errors taken into account; calibration errors 
can then be subtracted from the error bars. These simulations are applicable to the four-frequency configuration 
used in this paper and should be repeated if different frequency bands are used.

\clearpage
\begin{figure*}[htp]
\begin{center}
\includegraphics[scale=0.5]{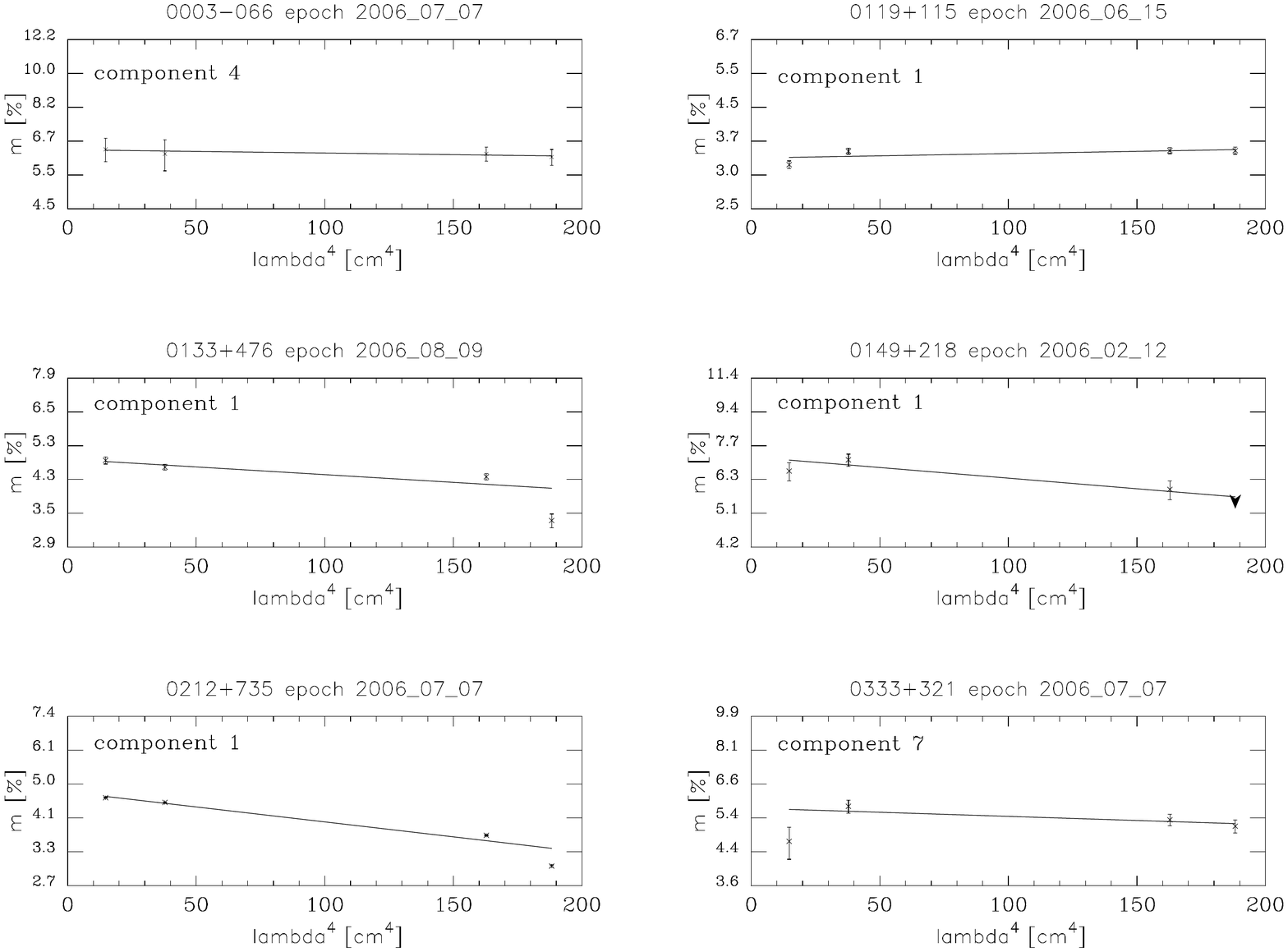}
\caption{Fits of depolarization curves to all isolated component. To appear Online only}
\end{center}
\end{figure*}

\begin{figure*}[htp]
\begin{center}
\includegraphics[scale=0.5]{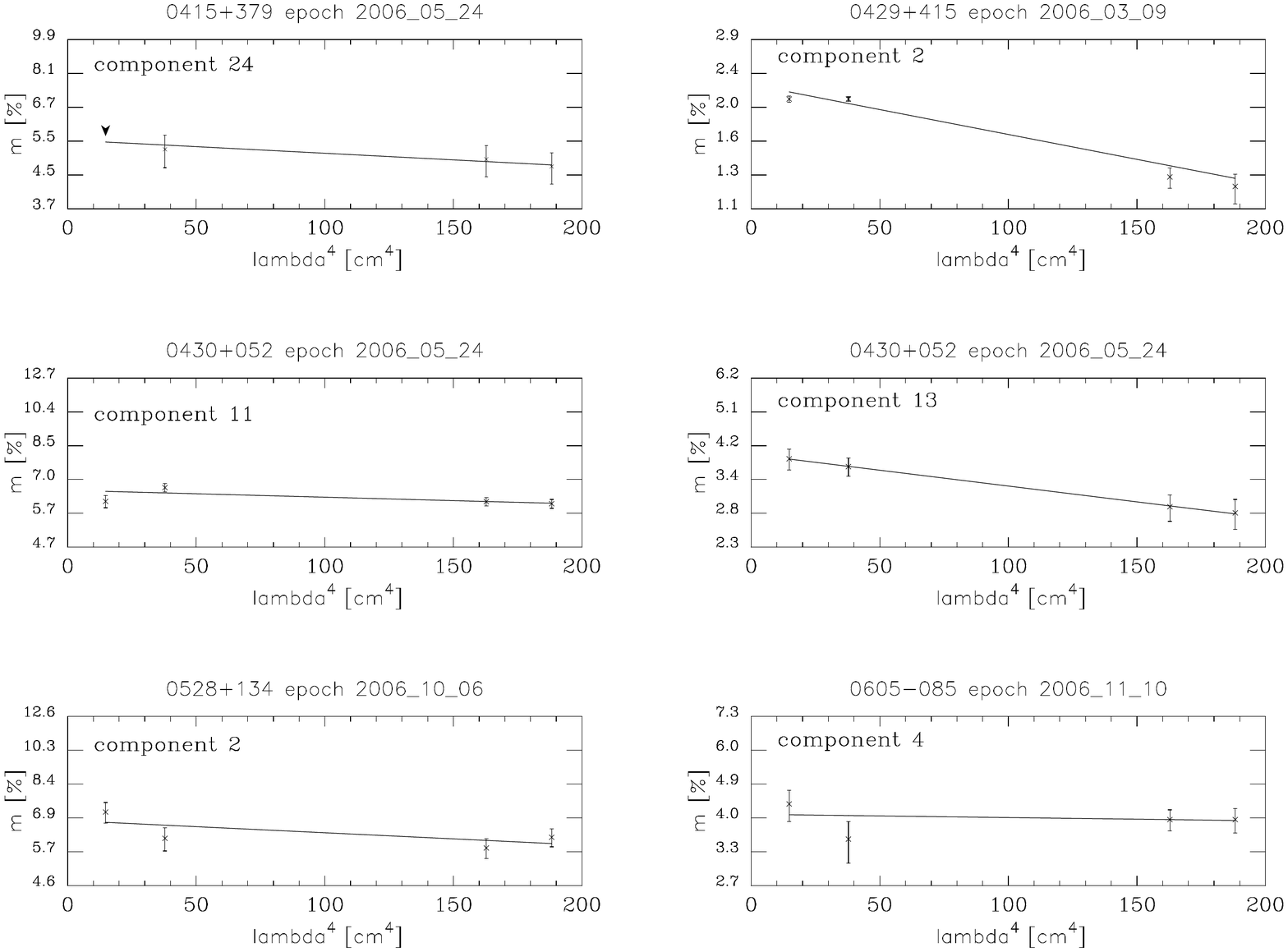}
\caption{Fits of depolarization curves to all isolated component. To appear Online only}
\end{center}
\end{figure*}

\begin{figure*}[htp]
\begin{center}
\includegraphics[scale=0.5]{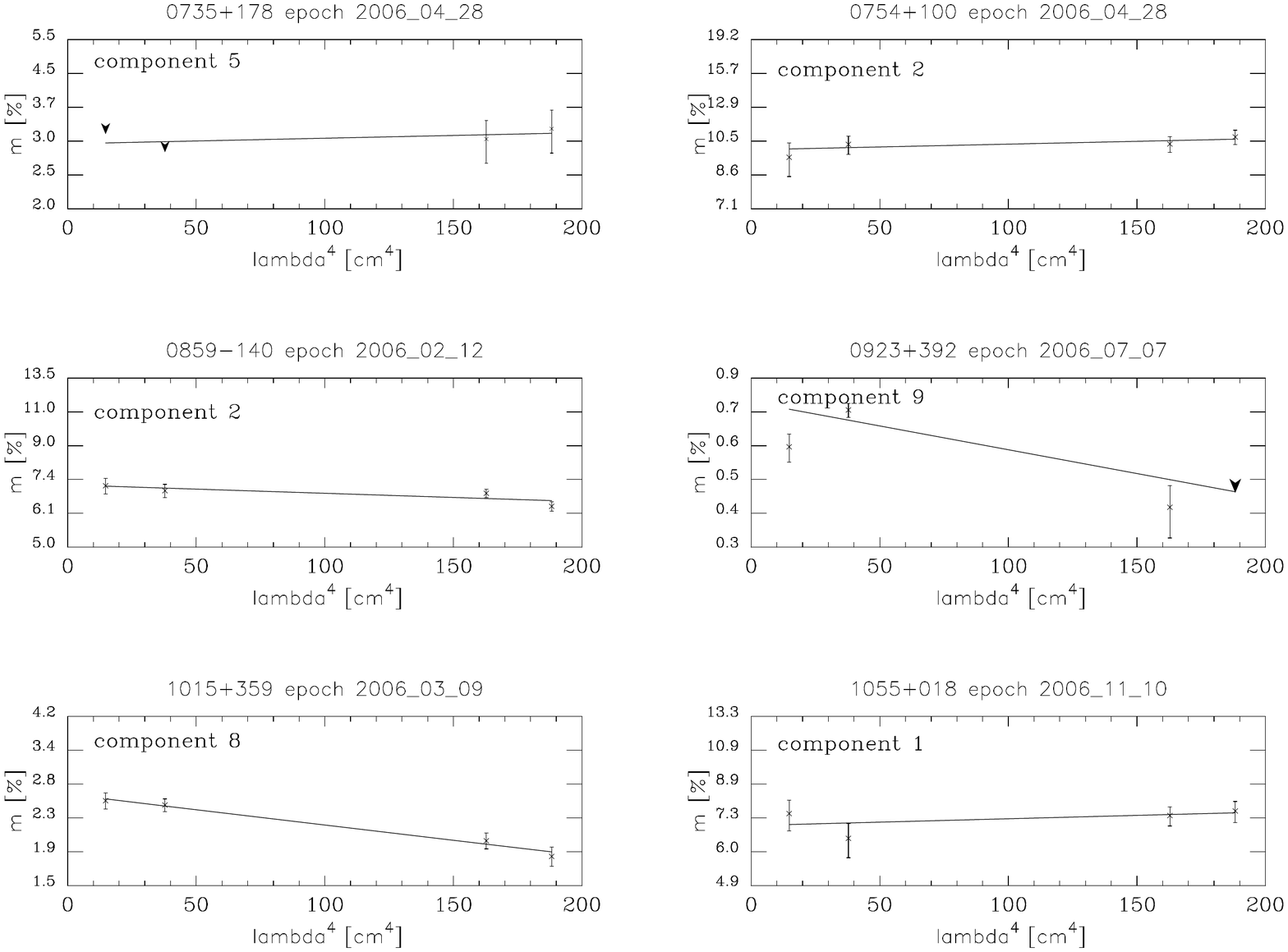}
\caption{Fits of depolarization curves to all isolated component. To appear Online only}
\end{center}
\end{figure*}

\begin{figure*}[htp]
\begin{center}
\includegraphics[scale=0.5]{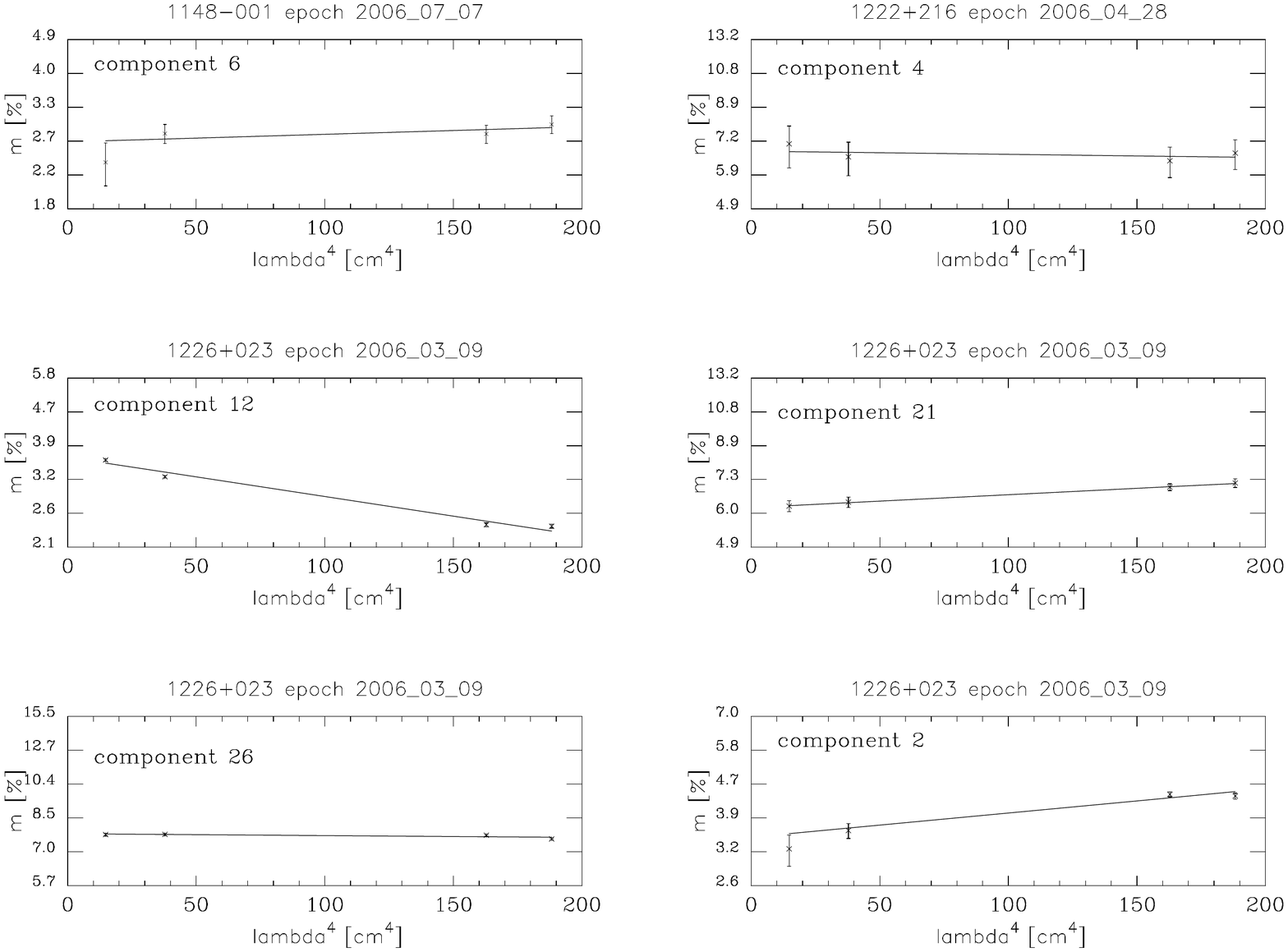}
\caption{Fits of depolarization curves to all isolated component. To appear Online only}
\end{center}
\end{figure*}

\begin{figure*}[htp]
\begin{center}
\includegraphics[scale=0.5]{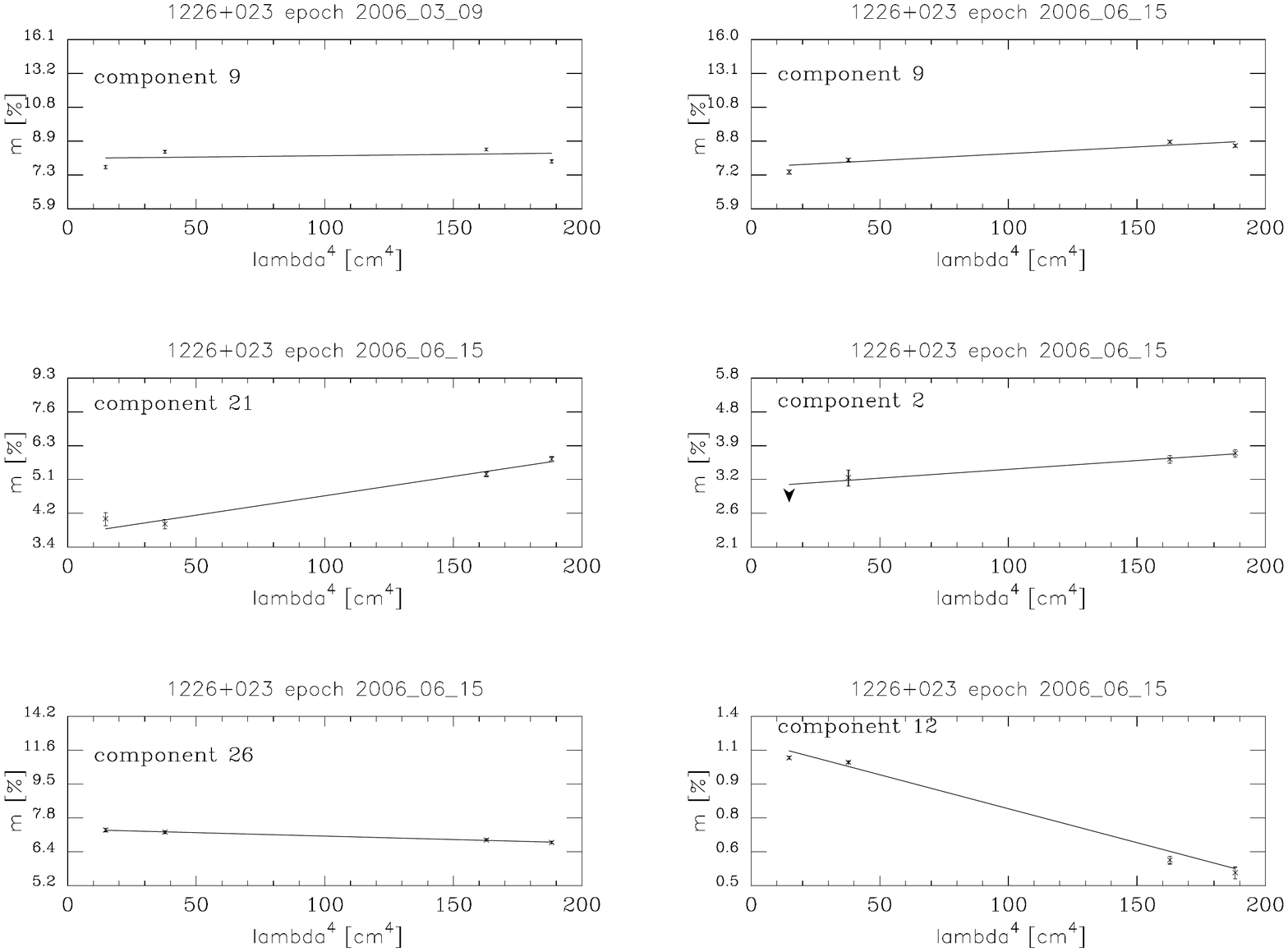}
\caption{Fits of depolarization curves to all isolated component. To appear Online only}
\end{center}
\end{figure*}

\begin{figure*}[htp]
\begin{center}
\includegraphics[scale=0.5]{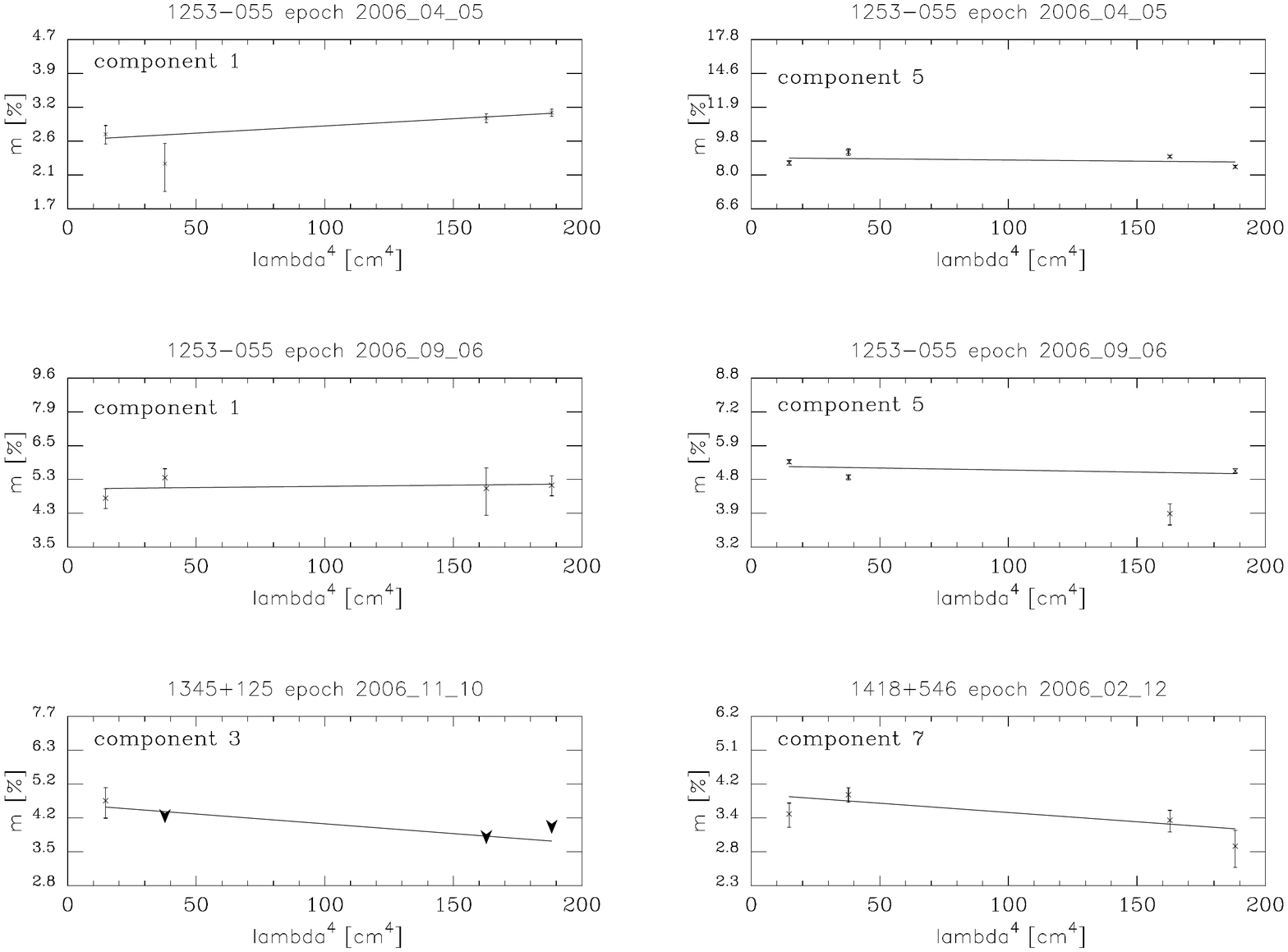}
\caption{Fits of depolarization curves to all isolated component. To appear Online only}
\end{center}
\end{figure*}

\begin{figure*}[htp]
\begin{center}
\includegraphics[scale=0.5]{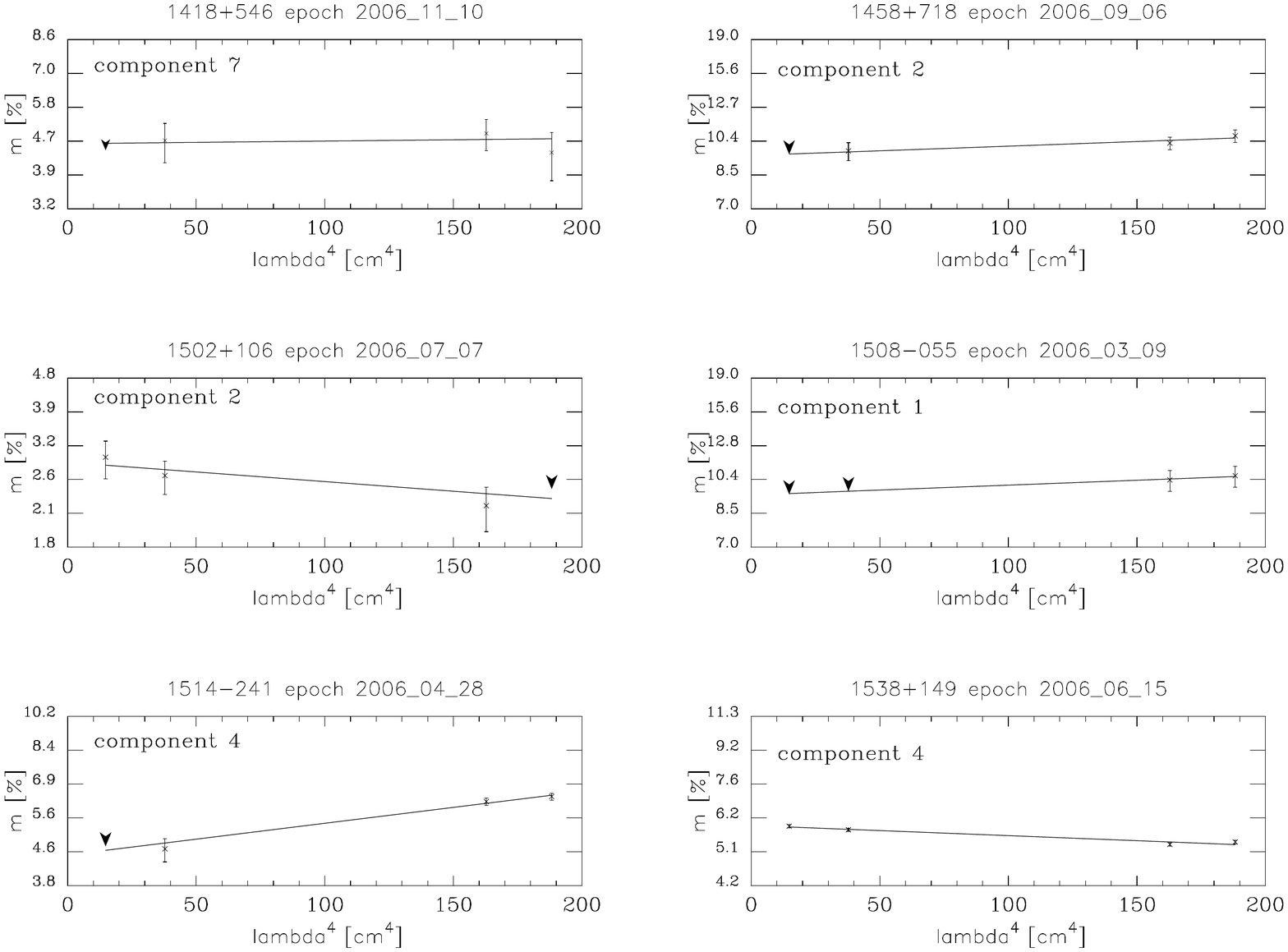}
\caption{Fits of depolarization curves to all isolated component. To appear Online only}
\end{center}
\end{figure*}

\begin{figure*}[htp]
\begin{center}
\includegraphics[scale=0.5]{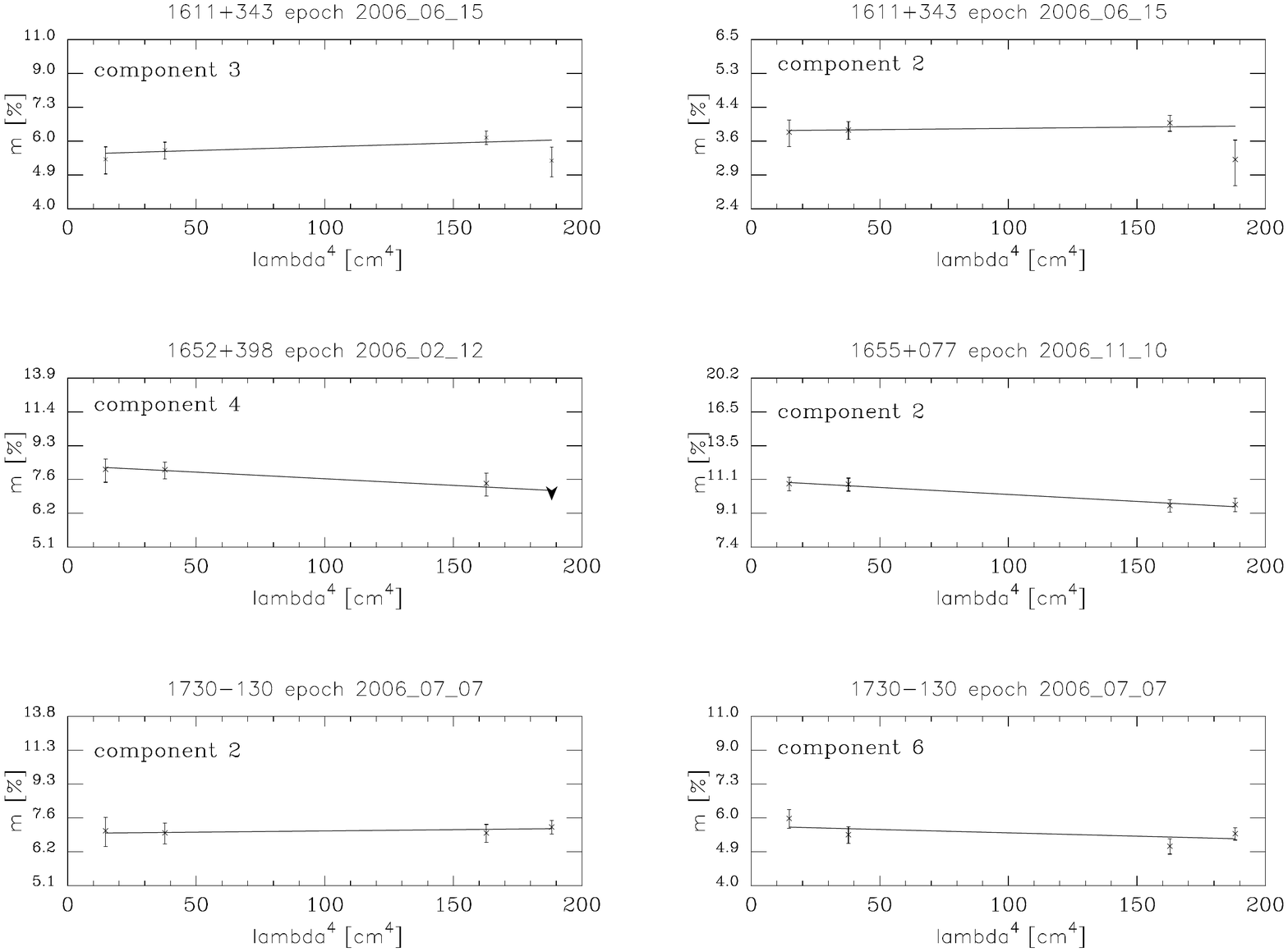}
\caption{Fits of depolarization curves to all isolated component. To appear Online only}
\end{center}
\end{figure*}

\begin{figure*}[htp]
\begin{center}
\includegraphics[scale=0.5]{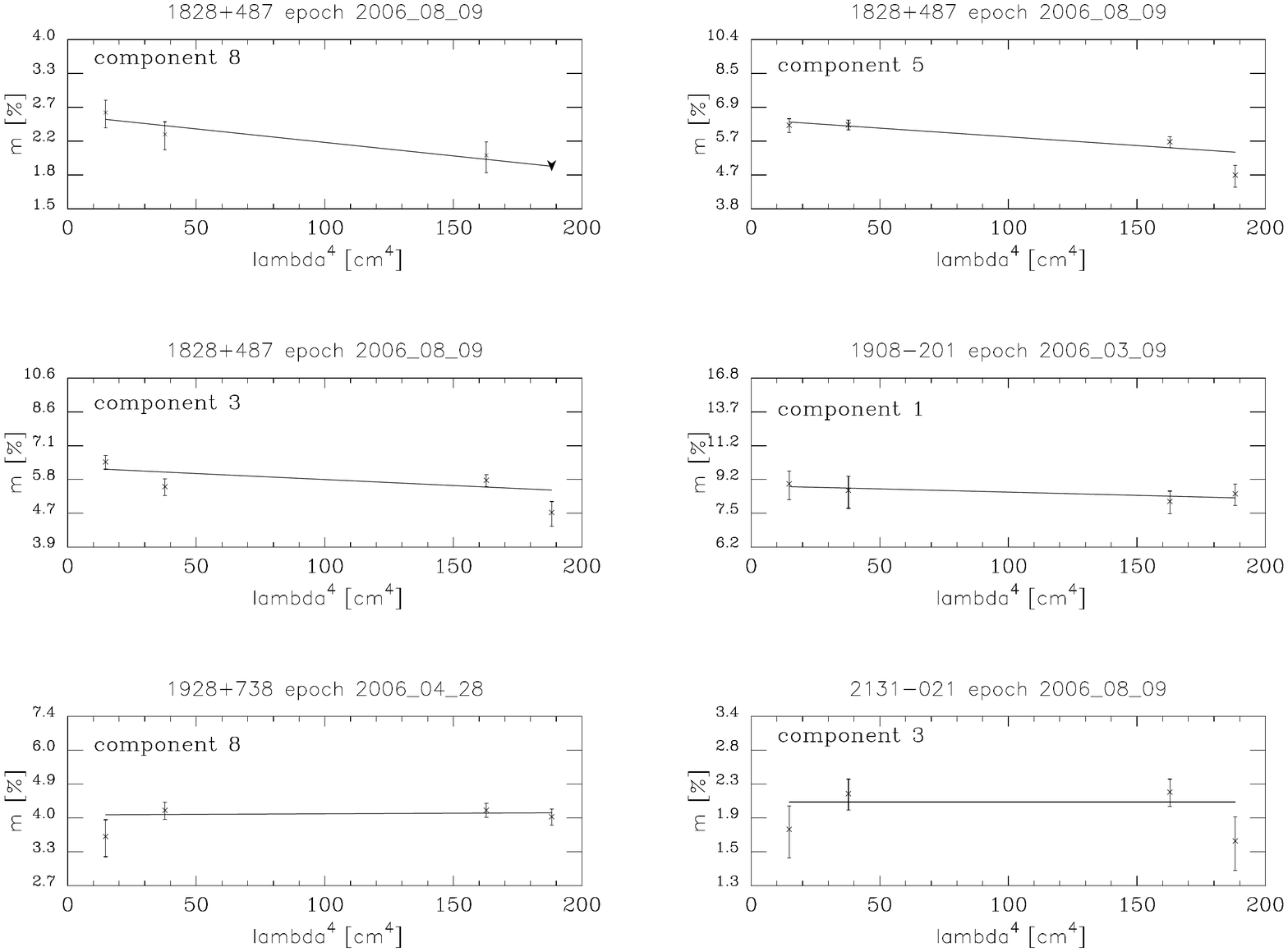}
\caption{Fits of depolarization curves to all isolated component. To appear Online only}
\end{center}
\end{figure*}

\begin{figure*}[htp]
\begin{center}
\includegraphics[scale=0.5]{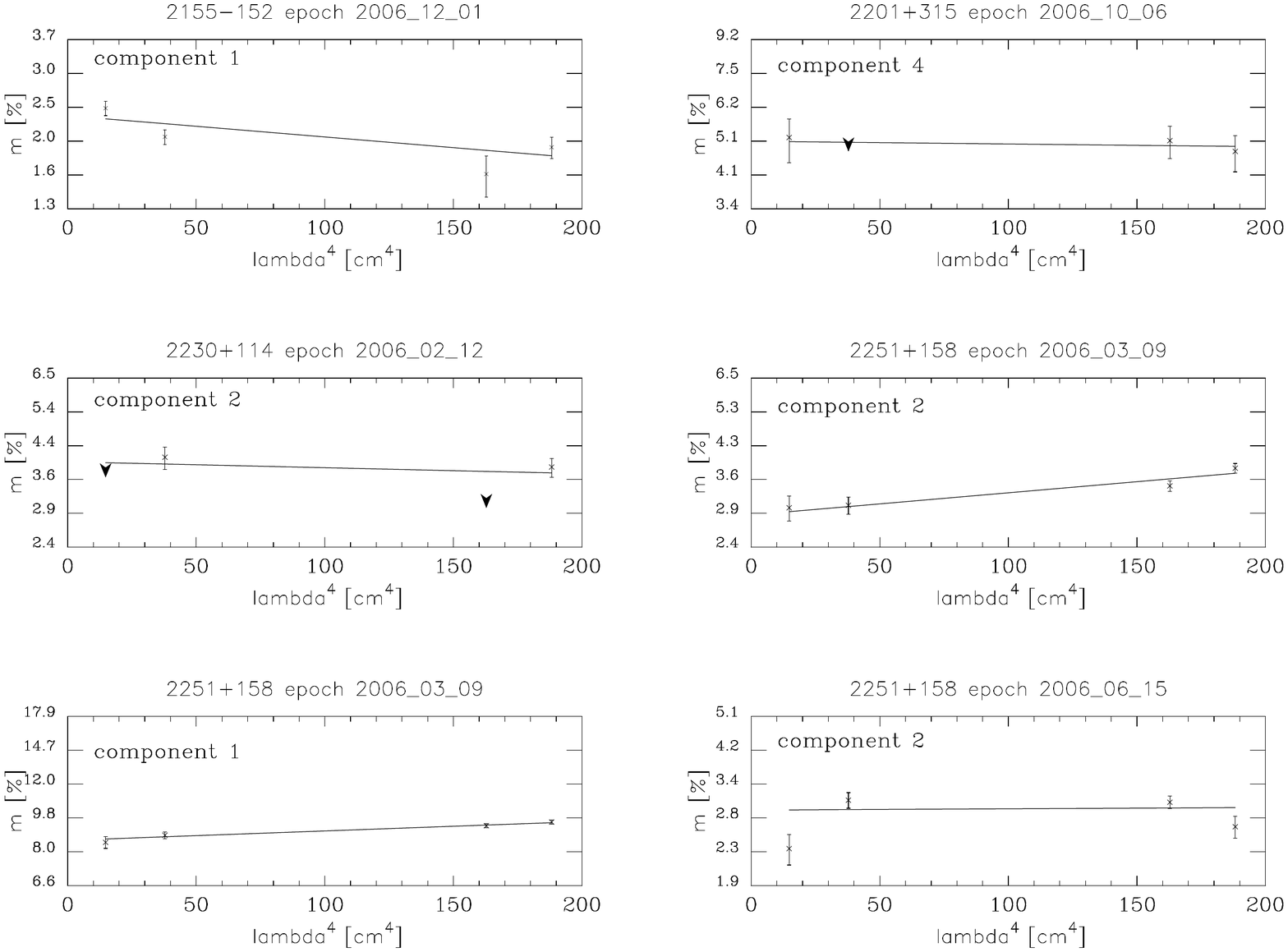}
\caption{Fits of depolarization curves to all isolated component. To appear Online only}
\end{center}
\end{figure*}

\begin{figure*}[htp]
\begin{center}
\includegraphics[scale=0.5]{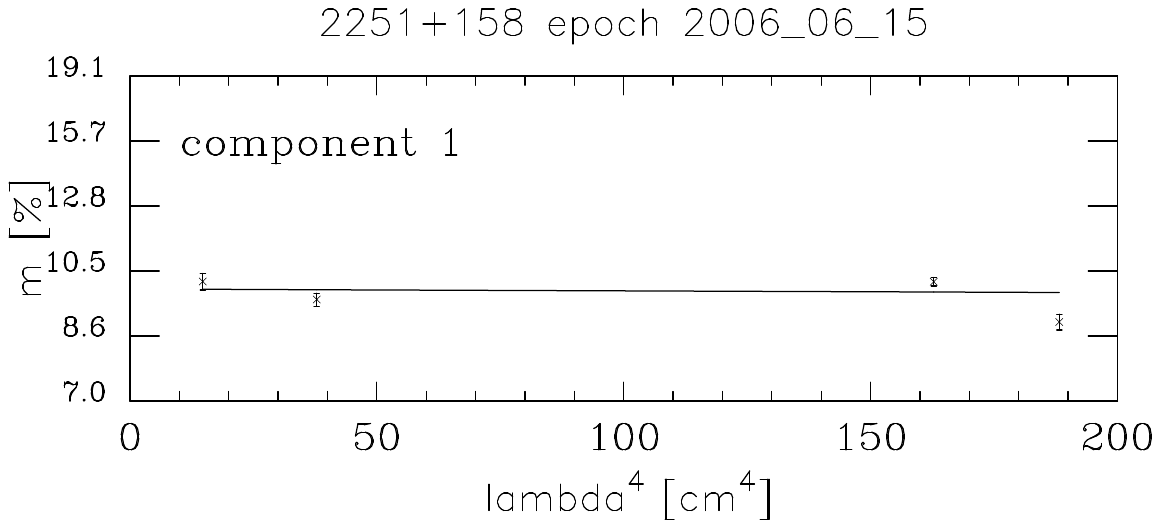}
\caption{Fits of depolarization curves to all isolated component. To appear Online only}
\end{center}
\end{figure*}

\clearpage
\LongTables
\tabletypesize{\scriptsize} 
\begin{deluxetable}{lccccccccccc} 
\tablecolumns{12} 
\tablewidth{0pt}  
\tablecaption{\label{depoltable}Isolated components at 15.3 GHz and their polarization values used in the depolarization analysis}  
\tablehead{\colhead{Source} &\colhead{I.D.} & \colhead{Epoch} & \colhead {r} &  
\colhead{P.A.} & \colhead{RM} & \colhead{m$_{15.3}$} & \colhead{m$_{12.1}$} & \colhead{m$_{8.4}$} & \colhead{m$_{8.1}$} & \colhead{b} & \colhead{signif. of b}\\  
\colhead{} & \colhead{} & \colhead{} & \colhead{(mas)} &  
\colhead{(deg)} & \colhead{(rad~m$^{-2}$)} & \colhead{(\%)} & \colhead{(\%)} & \colhead{(\%)} & \colhead{(\%)} & \colhead{($\times10^5 \mathrm{m}^{-4}$)} & \colhead{} \\ 
\colhead{(1)} & \colhead{(2)} & \colhead{(3)} & \colhead{(4)} &  
\colhead{(5)} & \colhead{(6)} & \colhead{(7)} & \colhead{(8)} & \colhead{(9)} & \colhead{(10)} & \colhead{(11)} & \colhead{(12)}} 
\startdata 
0003$-$066 & 4 & 2006-07-07 & 6.86 & $-$81.1 & 113.5 $\pm$ 9.6 & 8.6 $\pm$ 1.1 & 8.2 $\pm$ 1.4 & 8.2 $\pm$ 0.6 & 7.9 $\pm$ 0.7 & $-0.4 \pm 0.8$ & 0.4   \\ 
0119+115 & 1 & 2006-06-15 & 0.09 & 7.6 & $-$59.3 $\pm$ 8.7 & 3.2 $\pm$ 0.1 & 3.5 $\pm$ 0.1 & 3.5 $\pm$ 0.1 & 3.5 $\pm$ 0.1 & $+0.3 \pm 0.1$ & $2.0^c$   \\ 
0133+476 & 1 & 2006-08-09 & 2.89 & $-$26.6 & $-$153.5 $\pm$ 4.6 & 10.8 $\pm$ 0.6 & 9.7 $\pm$ 0.4 & 8.4 $\pm$ 0.4 & 4.4 $\pm$ 0.5 & $-2.3 \pm 0.4$ & $5.8^c$   \\ 
0149+218 & 1 & 2006-02-12 & 8.60 & $-$11.8 & \nodata & 22.7 $\pm$ 4.5 & 29.1 $\pm$ 3.9 & 15.1 $\pm$ 3.1 & $<10.0$ & $-4.7 \pm 1.7$ & 2.8  \\ 
0212+735 & 1 & 2006-07-07 & 0.02 & $-$127.4 & $-$284.9 $\pm$ 11.6 & 9.0 $\pm$ 0.1 & 8.5 $\pm$ 0.04 & 5.4 $\pm$ 0.04 & 3.6 $\pm$ 0.1 & $-4.1 \pm 0.1$ & $77.8^c$   \\ 
0333+321 & 7 & 2006-07-07 & 6.25 & 129.7 & $-$1.2 $\pm$ 8.0 & 7.0 $\pm$ 1.6 & 11.7 $\pm$ 1.1 & 9.6 $\pm$ 0.8 & 8.7 $\pm$ 0.9 & $-1.2 \pm 0.8$ & 1.5   \\ 
0415+379 & 24 & 2006-05-24 & 7.19 & 60.4 & \nodata & $<11.8$ & 9.4 $\pm$ 2.4 & 8.0 $\pm$ 1.9 & 7.2 $\pm$ 1.7 & $-2.1 \pm 2.1$ & 1.0  \\ 
0429+415 & 2 & 2006-03-09 & 20.54 & 59.4 & \nodata & 14.0 $\pm$ 1.0 & 14.0 $\pm$ 0.8 & 2.3 $\pm$ 0.5 & 1.8 $\pm$ 0.6 & $-11.6 \pm 1.5$ & $7.8^c$  \\
\enddata
\tablecomments{Columns are as follows: (1) IAU Name (B1950.0); (2) I.D. of the component (0 = core); (3) Epoch; (4) Component distance from the phase center of the I map; (5) Position angle of the component from the phase center; (6) RM$_\mathrm{obs}$ of the component (7) Fractional polarization at 15.3 GHz. Upper limits are 3$\sigma$ limits in fractional polarization; (8) Fractional polarization at 12.1 GHz; (9) Fractional polarization at 8.4 GHz; (10) Fractional polarization at 8.1 GHz; (11) Slope b from the depolarization fit in Sect. \ref{sect:depol}; (12) Significance of the slope. Sources marked with c have a $\chi^2 > 6$ indicating low probability that our simple model is accurate in explaining how the polarization changes with wavelength.}

\end{deluxetable}

\setcounter{table}{0}
\begin{deluxetable}{lccccrrrrr} 
\tablecolumns{10} 
\tabletypesize{\scriptsize} 
\tablewidth{0pt}  
\tablecaption{Sources and their rotation measure properties}  
\tablehead{\colhead{Source} &\colhead{Alias} & \colhead{z} & \colhead {Opt. Cl.} & \colhead{$\beta_{app}$} &  
\colhead{Epoch} & \colhead{Gal. RM} & \colhead{med. RM} & \colhead{med. core RM} & \colhead{med. jet RM}\\  
\colhead{} & \colhead{} & \colhead{} & \colhead{} & \colhead{(c)} & \colhead{} & \colhead{(rad~m$^{-2}$)} & \colhead{(rad~m$^{-2}$)} & \colhead{(rad~m$^{-2}$)} & \colhead{(rad~m$^{-2}$)} \\ 
\colhead{(1)} & \colhead{(2)} & \colhead{(3)} & \colhead{(4)} &  
\colhead{(5)} & \colhead{(6)} & \colhead{(7)} & \colhead{(8)} & \colhead{(9)} & \colhead{(10)}} 
\startdata 
0003$-$066 & NRAO 005 & 0.3467 & B & 8.4 & 2006-07-07 & 4.5 & -20.7 & -34.8 & 130.0  \\ 
0003+380 & S4 0003+38 & 0.229 & Q & \nodata & 2006-03-09 & -90.0 & \nodata & \nodata & \nodata  \\ 
0003+380 & S4 0003+38 & 0.229 & Q & \nodata & 2006-12-01 & -90.0 & 6053.3 & 6053.3 & \nodata  \\ 
0007+106 & III Zw 2 & 0.0893 & G & 1.2 & 2006-06-15 & -3.4 & 604.2 & 604.2 & \nodata  \\ 
0010+405 & 4C +40.01 & 0.256 & Q & \nodata & 2006-04-05 & -77.8 & \nodata & \nodata & \nodata  \\ 
0010+405 & 4C +40.01 & 0.256 & Q & \nodata & 2006-12-01 & -77.8 & \nodata & \nodata & \nodata  \\ 
0016+731 & S5 0016+73 & 1.781 & Q & 8.1 & 2006-08-09 & -9.1 & 264.9 & 264.9 & \nodata  \\ 
0048$-$097 & PKS 0048-09 & \nodata & B & \nodata & 2006-09-06 & -4.0 & -157.9 & -158.3 & -155.1  \\
\enddata 
\tablecomments{Columns are as follows: (1) IAU Name (B1950.0); (2) alternate name; (3) redshift; (4) optical classification where Q = quasar, B = BL Lac object, G = active galaxy, and U = unidentified; (5) Apparent speed used in viewing angle calculation of Fig. \ref{fig:rmdist}; (6) epoch of the RM observation; (7) Galactic RM correction taken from \cite{taylor09}; (8) median RM over the source; (9) median RM over the core area; (10) median RM over the jet. Table 1 is published in its entirety in the electronic edition of the Journal, a portion is shown here for guidance regarding its form and content.}

\end{deluxetable}

\setcounter{table}{2}
\begin{deluxetable}{lcccccc} 
\tablecolumns{7} 
\tabletypesize{\scriptsize} 
\tablewidth{0pt}  
\tablecaption{Modelfit components at 15.3 GHz and their RM values}  
\tablehead{\colhead{Source} &\colhead{I.D.} & \colhead{Epoch} & \colhead {r} &  
\colhead{P.A.} & \colhead{RM} & \colhead{Isolated}\\  
\colhead{} & \colhead{} & \colhead{} & \colhead{(mas)} &  
\colhead{(deg)} & \colhead{(rad~m$^{-2}$)} & \colhead{} \\ 
\colhead{(1)} & \colhead{(2)} & \colhead{(3)} & \colhead{(4)} &  
\colhead{(5)} & \colhead{(6)} & \colhead{(7)}} 
\startdata 
0003$-$066 & 0 & 2006-07-07 & 0.71 & $-$168.6 & $-$78 $\pm$ 73 &   \\ 
0003$-$066 & 1 & 2006-07-07 & 0.66 & $-$71.6 & $-$88 $\pm$ 75 &   \\ 
0003$-$066 & 4 & 2006-07-07 & 6.86 & $-$81.1 & 113 $\pm$ 98 & Y  \\ 
0003$-$066 & 5 & 2006-07-07 & 0.09 & 2.9 & $-$42 $\pm$ 72 &   \\ 
0003$-$066 & 6 & 2006-07-07 & 1.28 & $-$102.7 & $-$14 $\pm$ 77 &   \\ 
0007+106 & 1 & 2006-06-15 & 0.39 & $-$66.6 & 627 $\pm$ 130 &   \\ 
0016+731 & 0 & 2006-08-09 & 0.01 & $-$52.2 & 278 $\pm$ 74 &   \\ 
0016+731 & 2 & 2006-08-09 & 0.20 & 129.2 & 266 $\pm$ 74 &   \\
\enddata 
\tablecomments{Columns are as follows: (1) IAU Name (B1950.0); (2) I.D. of the component (0 = core); (3) Epoch; (4) Component distance from the phase center of the I map; (5) Position angle of the component from the phase center; (6) Component RM; (7) Flag for isolated jet components. Table 3 is published in its entirety in the electronic edition of the Journal; a portion is shown here for guidance regarding its form and content.}

\end{deluxetable}



\begin{thebibliography}{}
\bibitem[Alberdi et al.(2000)]{alberdi00} Alberdi, A., Gom\'ez, J.-L., Marcaide, J. M., Marscher, A. P. \& P\'erez-Torres, M. A. 2000, A \& A, 361, 529
\bibitem[Algaba et al.(2011)]{algaba11} Algaba, J. C., Gabuzda, D. C. \& Smith, P. S. 2011, arXiv:1110:5717
\bibitem[Asada et al.(2002)]{asada02} Asada, K., Inoue, M., Uchida, Y. et al. 2002, \pasj, 54, L39
\bibitem[Asada et al.(2008a)]{asada08a} Asada, K., Inoue, M., Kameno, S., \& Nagai, H. 2008, \apj, 675, 79
\bibitem[Asada et al.(2008b)]{asada08b} Asada, K., Inoue, M., Nakamura, M., Kameno, S., \& Nagai, H. 2008, \apj, 682, 798
\bibitem[Asada et al.(2010)]{asada10} Asada, K., Nakamura, M., Inoue, M., Kameno, S., \& Nagai, H. 2010, \apj, 720, 41
\bibitem[Attridge et al.(2005)]{attridge05} Attridge, J. M., Wardle, J. F. C. \& Homan, D. C. 2005, \apj, 633, 85
\bibitem[Blandford(1993)]{blandford93} Blandford, R. 1993, in Astrophysical Jets, Astrophysics and Space Science Library (Cambridge: Cambridge Univ. Press), Vol. 103, 15
\bibitem[Blandford \& Znajek(1977)]{blandford77} Blandford, R. D. \& Znajek, R. L. 1977, \mnras, 179, 433
\bibitem[Broderick \& McKinney(2010)]{broderick10} Broderick,, A. E. \& McKinney, J. C. 2010, \apj, 725, 750
\bibitem[Burn(1966)]{burn66} Burn, B. J. 1966, \mnras, 133, 67
\bibitem[Chen(2005)]{chen05} Chen, T. 2005, Ph.D. thesis, Brandeis Univ.
\bibitem[Clausen-Brown et al.(2011)]{clausen11} Clausen-Brown, E., Lyutikov, M. \& Kharb, P. 2011, \mnras, 415, 2081
\bibitem[Contopoulos et al.(2009)]{contopoulos09} Contopoulos, I., Christodoulou, D. M., Kazanas, D. \& Gabuzda, D. C. 2009, \apj, 702, L148
\bibitem[Croke \& Gabuzda(2008)]{croke08} Croke, S. \& Gabuzda, D. C. 2008, MNRAS, 386, 619
\bibitem[Croke et al.(2010)]{croke10} Croke, S. M., O'Sullivan, S. P. \& Gabuzda, D. C. 2010, \mnras, 402, 259
\bibitem[Gabuzda et al.(2004)]{gabuzda04} Gabuzda, D. C., Murray, E. \& Cronin, P. 2004, \mnras, L89
\bibitem[Gabuzda et al.(2008)]{gabuzda08} Gabuzda, D. C., Vitrishchak, V. M., Mahmud, M. \& O'Sullivan, S. P. 2008, \mnras, 384, 1003
\bibitem[G\'omez et al.(2002)]{gomez02} G\'omez, J.-L., Marscher, A. P., Alberdi, A., Jorstad, S. G. \& Agudo, I. 2002, VLBA Scientific Memo No. 30
\bibitem[G\'omez et al.(2008)]{gomez08} G\'omez, J.-L., Marscher, A. P., Jorstad, S. G., Agudo, I \& Roca-Sogorb, M. 2008, \apj, 681, L69
\bibitem[G\'omez et al.(2011)]{gomez11} G\'omez, J.-L., Roca-Sogorb, M., Agudo, I, Marscher, A. P. \& Jorstad, S. G. 2011, \apj, 733, 11
\bibitem[Homan et al.(2001)]{homan01} Homan, D. C., Attridge, J. M. \& Wardle, J. F. C. 2001, \apj, 556, 113
\bibitem[Homan \& Lister(2006)]{homan06} Homan, D. C. \& Lister, M. L. 2006, \aj, 131, 1262
\bibitem[Homan et al.(2009)]{homan09} Homan, D. C., Lister, M. L., Aller, H. D., Aller, M. F. \& Wardle, J. F. C. 2009, \apj, 696, 328
\bibitem[Homan(2012)]{homan12} Homan, D. C. 2012, ApJL, 747, L24
\bibitem[Hovatta et al.(2009)]{hovatta09} Hovatta, T., Valtaoja, E., Tornikoski, M. \& L\"ahteenm\"aki, A. 2009, \aa, 494, 527
\bibitem[Hughes et al.(1989)]{hughes89} Hughes, P. A, Aller, H. D. \& Aller, M. F. 1989, \apj, 341, 68
\bibitem[Jones \& O'Dell(1977)]{jones77} Jones, T. W. \& O'Dell, S. L. 1977, \apj, 214, 522
\bibitem[Kellermann et al.(2004)]{kellermann04} Kellermann, K. I., Lister, M. L., Homan, D. C., et al. 2004, \apj, 609, 539
\bibitem[Komatsu et al.(2009)]{komatsu09} Komatsu, E., Dunkley, J., Nolta, M. R. et al. 2009, \apjs, 180, 330
\bibitem[Kovalev et al.(2008)]{kovalev08} Kovalev, Y.Y., Lobanov, A. P., Pushkarev, A. B. \& Zensus, J. A. 2008, A\&A, 483, 759
\bibitem[Laing(1980)]{laing80} Laing, R. 1980, \mnras, 193, 439
\bibitem[Lister et al.(2009a)]{lister09} Lister, M. L., Aller, H. D., Aller, M. F., et al. 2009a, \aj, 137, 3718
\bibitem[Lister et al.(2009b)]{lister09b} Lister, M. L., Cohen, M. H., Homan, D. C. et al. 2009b, \aj, 138, 1874 
\bibitem[Lister \& Homan(2005)]{lister05} Lister, M. L. \& Homan, D. C. 2005, \aj, 130, 1389
\bibitem[Lister et al.(2001)]{lister01} Lister, M. L., Tingay, S. J., Murphy, D. W., Piner, B. G., Jones, D. L., Preston, R. A. 2001, \apj, 554, 948
\bibitem[Lobanov(1998)]{lobanov98} Lobanov, A. P. 1998, A\&A, 330, 79
\bibitem[Lyutikov et al.(2005)]{lyutikov05} Lyutikov, M., Pariev, V. I. \& Gabuzda, D. C. 2005, \mnras, 360, 869
\bibitem[Mahmud et al.(2009)]{mahmud09} Mahmud, M., Gabuzda, D. \& Bezrukovs, V. 2009, \mnras, 400, 2
\bibitem[Mantovani et al.(2010)]{mantovani10} Mantovani, F., Rossetti, A., Junor, W., Saikia, D. J. \& Salter, C. J. 2010, A\&A, 518, A33
\bibitem[Marscher et al.(2008)]{marscher08} Marscher, A. P., Jorstad, S. G., D'Arcangelo, F. D. et al. 2008, Nature, 452, 966
\bibitem[Marr et al.(2001)]{marr01} Marr, J. M., Taylor, G. B. \& Crawford III, F. 2001 \apj, 550, 160
\bibitem[Meier et al.(2001)]{meier01} Meier, D. L., Koide, S. \& Uchida, Y. 2001, Science, 291, 84
\bibitem[McKinney \& Narayan(2007)]{mckinney07} McKinney, J. C. \& Narayan, R. 2007, \mnras, 375, 531
\bibitem[Murphy \& Gabuzda (2011)]{murphy11} Murphy, E. \& Gabuzda, D. C. 2011, arXiv:1109.4778
\bibitem[Mutel et al.(2005)]{mutel05} Mutel, R. L., Denn, G. R. \& Dreier, C. 2005, ASPC, 340, 155
\bibitem[O'Sullivan \& Gabuzda(2009)]{osullivan09a} O'Sullivan, S. P. \& Gabuzda, D. C. 2009, \mnras, 393, 429
\bibitem[Pacholczyk(1970)]{pacholczyk70} Pacholczyk, A. G. 1970, Radio Astophysics, Nonthermal Processes in Galactic and Extragalactic Sources, (W. H. Freeman and Company)
\bibitem[Press(1992)]{press92} Press, W. H., Flannery, B. P., Teukolsky, S. A., \& Vetterling, W. T. 1992, Numerical Recipes in FORTRAN 77, 2nd edn. (Cambridge University Press)
\bibitem[Pushkarev(2001)]{pushkarev01} Pushkarev, A. B. 2001, ARep, 45, 667 
\bibitem[Pushkarev(2012)]{pushkarev12} Pushkarev, A. B., Hovatta, T., Kovalev, Y. Y. et al. 2012, A\&A, submitted
\bibitem[Reichstein \& Gabuzda(2011)]{reichstein11} Reichstein, A. \& Gabuzda, D. 2011, arXiv:1102.0702
\bibitem[Reichstein \& Gabuzda(2012)]{reichstein12} Reichstein, A. \& Gabuzda, D. 2012, J. Phys.: Conf. Ser. 355, 012021
\bibitem[Roberts, Wardle \& Brown(1994)]{roberts94} Roberts, D. H., Wardle, J. F. C. \& Brown, L. F. 1994, \apj, 427, 718
\bibitem[Rudnick \& Jones(1983)]{rudnick83} Rudnick, L. \& Jones, T. W. 1983, \aj, 88 518
\bibitem[Rusk(1988)]{rusk88} Rusk, R. E. 1988 Ph.D. thesis, Univ. Toronto
\bibitem[Savolainen et al.(2008)]{savolainen08} Savolainen, T., Wiik, K., Valtaoja, E. \& Tornikoski, M. 2008, ASPC, 386, 451
\bibitem[Shepherd(1997)]{shepherd97} Shepherd, M. C. 1997 ASPC 125, 77
\bibitem[Sokolovsky et al.(2011)]{sokolovsky11} Sokolovsky, K. V,. Kovalev, Y. Y., Pushkarev, A. B. \& Lobanov, A. P. 2011, A\&A, 532, 38
\bibitem[Taylor(1998)]{taylor98} Taylor, G. B. 1998, \apj, 506, 637
\bibitem[Taylor(2000)]{taylor00} Taylor, G. B. 2000, \apj, 533, 95
\bibitem[Taylor et al.(2009)]{taylor09} Taylor, A. R., Stil, J. M. \& Sunstrum, C. 2009, \apj, 702, 1230
\bibitem[Taylor \& Zavala(2010)]{taylor10} Taylor, G, B. \& Zavala, R. T. 2010, \apj, 722, L183
\bibitem[Tribble(1991)]{tribble91} Tribble, P. C. 1991, \mnras, 250, 726
\bibitem[Udomprasert et al.(1997)]{udomprasert97} Udomprasert, P. S., Taylor, G. B., Pearson, T. J. \& Roberts, D. H. 1997, \apj, 483, L9
\bibitem[Vlahakis \& K\"onigl(2004)]{vlahakis04} Vlahakis, N. \& K\"onigl, A. 2004, \apj, 605, 656
\bibitem[Walker et al.(2000)]{walker00} Walker, R. C., Dhawan, V., Romney, J. D., Kellermann, K. I. \& Vermeulen, R. C. 2000, \apj, 530, 233
\bibitem[Wardle \& Homan(2003)]{wardle03} Wardle, J. F. C. \& Homan, D. C. 2003, Ap\&SS, 288, 143
\bibitem[Wrobel(1993)]{wrobel93} Wrobel, J. M. 1993, \aj, 106, 444
\bibitem[Zavala \& Taylor(2002)]{zavala02} Zavala, R. T. \& Taylor, G. B. 2003, \apj, 566, L9
\bibitem[Zavala \& Taylor(2003)]{zavala03} Zavala, R. T. \& Taylor, G. B. 2003, \apj, 589, 126
\bibitem[Zavala \& Taylor(2004)]{zavala04} Zavala, R. T. \& Taylor, G. B. 2004, \apj, 612, 749
\bibitem[Zavala \& Taylor(2005)]{zavala05} Zavala, R. T. \& Taylor, G. B. 2005, \apj, 626, L73
\end{thebibliography}
\end{document}